\documentclass[11pt, a4paper]{article}
\pdfoutput=1

\usepackage{amsmath}
\usepackage{amsfonts}
\usepackage{amssymb}
\usepackage{bm}
\usepackage{bbm}
\usepackage{verbatim}
\usepackage{dsfont}
\usepackage{booktabs}
\usepackage{slashed}
\usepackage{nicefrac}
\usepackage{mathtools}
\usepackage{physics}
\usepackage{array}
\usepackage{adjustbox}
\usepackage{rotating}
\usepackage{float}

\usepackage{epsfig}
\usepackage{xcolor}
\usepackage{graphicx}
\usepackage[font=small]{caption}
\usepackage[listofformat=empty,subrefformat=empty]{subfig}

\usepackage{float}
\usepackage{overpic}
\usepackage{tikz}
\usepackage{tabularx}
\usepackage{pdflscape}

\usepackage{enumerate}
\usepackage{hhline}
\usepackage{multirow}

\usepackage[nosort]{cite}
\usepackage{xspace}
\usepackage{setspace}

\usepackage[pdftitle={},
  pdfauthor={},
  pdfsubject={},
  bookmarksopen, bookmarksnumbered, bookmarksopenlevel=2, colorlinks=false, linkcolor=blue, citecolor=blue, urlcolor=blue]{hyperref}

\usepackage[nameinlink,capitalize]{cleveref}

\makeatletter{}
\g@addto@macro\bfseries{\boldmath}
\makeatother

\usepackage[normalem]{ulem}

\newcommand{\be}{\begin{equation}}
\newcommand{\ee}{\end{equation}}
\newcommand{\ba}{\begin{aligned}}
\newcommand{\ea}{\end{aligned}}

\def\Z{{\mathbb{Z}}}

\newcommand{\rr}{\rightarrow}

\begin{document}

\baselineskip=18pt  
\numberwithin{equation}{section}  
\allowdisplaybreaks  


\thispagestyle{empty}

\vspace*{-2cm}

\vspace*{2.6cm}
\begin{center}
{\Large{\textbf{SymTFT actions, condensable algebras and categorical anomaly resolutions 
}}} 
\\
\vspace*{1.5cm}
Daniel Robbins and Subham Roy\\

{\it Department of Physics, \\
University at Albany, SUNY\\
Albany, NY 12222, USA}

\vspace*{0.8cm}
\end{center}
\vspace*{.5cm}

\noindent

We investigate symmetry topological field theories (SymTFTs) of non-abelian and non-invertible symmetries and the different Lagrangian algebras associated with a given Drinfeld center.  For several examples we analyze the condensable algebras of the Drinfeld center to identify the intrinsically gapless symmetry protected topological (igSPT) phases. In the previous work, the relation between igSPT phases and resolving anomalies by embedding an anomalous symmetry inside a larger fusion category was demonstrated. Here we present more examples of this mechanism that involve both group-like and categorical symmetries.

\newpage
\setcounter{tocdepth}{2}
\tableofcontents

\newpage

\section{Introduction}
One of the most powerful tools used to analyze quantum field theories is provided by symmetries. From Noether's theorem \cite{Noether_1971} to constraining renormalization group flows, symmetry principles have provided a plethora of insights that consolidated our understanding of quantum field theories. Thanks to the seminal work of \cite{Gaiotto:2014kfa}, the notion of symmetry has been reformulated in terms of the extended topological operators of the QFT. Since then, the notion of symmetry has been generalized in many directions --- higher form symmetries, higher group symmetries, non-invertible symmetries and $(-1)$-form symmetries \cite[and references therein]{Freed:2022iao,Cordova:2022ruw,Bah:2022xfv,Schafer-Nameki:2023jdn,Bhardwaj:2023kri,Shao:2023gho,Gomes:2023ahz,Brennan:2023mmt,McGreevy:2022oyu,Chang:2018iay,Bhardwaj:2022yxj,Iqbal:2024pee,Costa:2024wks,Cordova:2019jnf, Cordova:2019uob,Aloni:2024jpb,Lin:2025oml,Yu:2024jtk,Najjar:2024vmm, Heidenreich:2020pkc, Robbins:2025apg,Heckman:2022muc,Apruzzi:2022rei,Apruzzi:2021nmk,Albertini:2020mdx, GarciaEtxebarria:2022vzq}.   

These generalized symmetries can be studied under the framework Symmetry TFT \cite{Freed:2012bs,Freed:2022qnc,Gaiotto:2020iye,Kaidi:2022cpf,Kaidi:2023maf,Apruzzi:2021nmk,Bhardwaj:2023ayw,Heckman:2022xgu,Baume:2023kkf,Heckman:2024oot,Heckman:2024zdo,Heckman:2025lmw,Cvetic:2024dzu}.  The SymTFT is a $(d+1)$-dimensional topological field theory we can define to probe the global symmetries of $d$-dimensional quantum field theory. The sandwich construction puts the $(d+1)$-dimensional theory on the $d$-dimensional spacetime cross a compact interval, with topological boundary conditions on one end (the symmetry boundary), and non-topological degrees of freedom encoding the remaining dynamics of the QFT at the other boundary (the physical boundary). The topological operators that appear in the bulk of SymTFT can capture the generalized charges\cite{Bhardwaj:2023wzd} of the QFT. Roughly speaking, there are two types of topological operators in the bulk of SymTFT. First there are the operators that remain parallel to the symmetry boundary and encode the symmetry generators of the absolute theory. Conversely, the set of operators that extend from one boundary to the other correspond to operators charged under the symmetry. As a concrete example, the SymTFT for a $2d$ theory with a $\Z^{(0)}_2$ symmetry is captured by the $3d~\Z_2$ Dijkgraaf-Witten (DW) theory\cite{dijkgraaf1990topological}. The topological operators of the DW theory encodes the information of the symmetry generator and the operators charged under the symmetry. The presence of the 't Hooft anomaly in $2d$ theory can be incorporated by adding twist terms or additional couplings in the DW action.  

The topological operator of the SymTFT, anchored between the two boundaries gives rise to the charged local operators the theory. If for some reason this operator is not allowed to \textit{`end'} on the physical boundary of the SymTFT, then the corresponding charge goes missing. In other words, it can result in a trivially acting symmetry. The notion of missing charges were central to the extension of SymTFT into a Club Sandwich \cite{Bhardwaj:2023bbf,Bhardwaj:2024qrf}. Trivially acting symmetries have been previously constructed independently using the idea that a topological line operator which can terminate on a topological point operator will not link non-trivially with any $(d-2)$-dimensional operator \cite{Robbins:2022wlr}. Trivially acting symmetries have their own interesting applications, one of them is that gauging a trivially acting $(d-2)$ form symmetry leads us to a theory with $(d-1)$-form symmetry --- decomposition\cite{Hellerman:2006zs, Sharpe:2014tca, Sharpe:2019ddn, Tanizaki:2019rbk, Komargodski:2020mxz, Cherman:2020cvw, Robbins:2020msp, Nguyen:2021naa, Sharpe:2021srf, Cherman:2021nox, Sharpe:2022ene,Lin:2022xod, Bhardwaj:2023bbf}. 
This is a phenomenon that occurs whenever a quantum field theory carries $(d-1)$-form symmetry; essentially, our QFT breaks down into several universes.  In doing so, the partition function of the QFT splits into the sum of partition functions of the individual universes. Furthermore, trivially acting symmetries can be used to resolve anomalies \cite{Wang:2017loc, Tachikawa:2017gyf, Pantev:2005rh, Pantev:2022kpl, Pantev:2022pbf,Robbins:2020msp, Robbins:2021lry, Robbins:2021ibx,Robbins:2021xce,Robbins:2022wlr}. The 't Hooft anomaly of a smaller symmetry was resolved by introducing additional trivial symmetry generators, by embedding the smaller anomalous symmetry inside a larger symmetry.

In this work, we investigate the trivially acting symmetries and anomaly resolutions using the framework of SymTFT and the club sandwich. The foundations for this work were already established in \cite{Perez-Lona:2025ncg}, where it was pointed out that intrinsically gapless SPT phases \cite{Scaffidi_2017,Bhardwaj:2024qrf,Thorngren:2020wet,Wen:2022tkg,Li:2023knf} can potentially allow us to identify examples of anomaly resolutions. In this work, we extend this, identifying the anomalous symmetries that can be resolved by embedding them inside our examples of $D_4,Q_8$ or non-invertible $\text{Rep}(D_4)$ or $\text{Rep}(Q_8)$. Unlike the previous work, which only focused on the anomaly resolution by extending the anomalous symmetry into a non-invertible symmetry, here we stay democratic and discuss anomaly resolutions with both group-like and categorical symmetry. In doing so, we give a complete SymTFT interpretation of the previous works on anomaly resolution in the context of $2d$ orbifolds \cite{Robbins:2020msp, Robbins:2021lry, Robbins:2021ibx,Robbins:2021xce,Robbins:2022wlr}. 

Moreover, the short exact sequences related to the anomaly resolutions can be used to model renormalization group flows \cite{Antinucci:2025fjp}. Take any IR theory with an anomalous 0-form symmetry $\mathcal{G}^\omega_{IR}$, which can be embedded into a larger UV symmetry group $\mathcal{G}_{UV}$ without that anomaly. This defines a homomorphism, $\Phi: ~ \mathcal{G}_{UV} \longrightarrow \mathcal{G}^\omega_{IR}$. In other words, the UV symmetry group has a part that acts trivially in the IR.  It can be detected by the nontrivial kernel of homomorphism,
\begin{equation}
    \mathcal{G}_{triv} = \text{ker} (\Phi)
\end{equation}
We can summarize this with the following short exact sequence:
\begin{equation}
    0 \longrightarrow \mathcal{G}_{triv} \longrightarrow \mathcal{G}_{UV} \stackrel{\Phi}{\longrightarrow} {\mathcal{G}^\omega_{IR}} \longrightarrow 0
\end{equation}
where we can say that the pullback of the anomaly to $\mathcal{G}_{UV}$ is trivial. For categorical short exact sequences, $\Phi$ is more like a tensor functor. In this work, using the club sandwich we have identified categorical short exact sequences, which can be used to identify such tensor functors between non-invertible and group-like symmetries.  

The paper is organized as follows. In order to achieve our goal, we need to identify the igSPT phases and to do that we need to identify the condensable algebras associated with the Drinfeld center. In section \ref{sec:condensable_algebra}, we review the methods to identify the condensable algebras associated with the Drinfeld center. Moreover, we exemplify this with a detailed calculation of the condensable algebras of $\mathcal{Z}(Q_8)$. The condensable algebras play the central role in identification of the igSPT phases later. We hope that our treatment will provide a useful reference for future work.

We then move onto discussing the SymTFT for $D_4$ in \ref{sec:symtft_for_d4}. We have briefly reviewed the SymTFT action, different Lagrangian algebras of this SymTFT and how different lagrangian algebras are related with each other via some discrete gauging with or without discrete torsion. The same section also presents a symmetry web of $D_4$ in the space of $c=1$. We closed this section with a discussion of anomaly resolutions involving $D_4/\text{Rep}(D_4)$ symmetry. Section \ref{sec:symtft_q4} presents a discussion of the SymTFT that encodes $Q_8/\text{Rep}(Q_8)$ symmetry and its boundary conditions. Moreover, we have identified the igSPT phases associated with this SymTFT which would eventually allow us to identify the anomalous symmetries that can be resolved by 
$Q_8/\text{Rep}(Q_8)$. We conclude in section \ref{sec:conclusion} with a list of future directions.

\section{Enumerating condensable algebras for $\mathcal{Z}(G)$} \label{sec:condensable_algebra}

Let $G$ be a finite group with identity element $e$.  The Drinfeld center $\mathcal{Z}(G)$ is\footnote{Throughout this work we will use $\mathcal{Z}(G)$ to refer to this construction which can serve as the bulk SymTFT for either anomaly-free global $G$ symmetry or $\operatorname{Rep}(G)$ symmetry, depending on the choice of boundary condition at the symmetry boundary.  Later we will, somewhat imprecisely, use notations $\mathcal{Z}(\operatorname{Vec}_G)$ or $\mathcal{Z}(\operatorname{Rep}(G))$ to emphasize which boundary condition has been chosen.  We will also $\mathcal{Z}(\Z_2^\omega)$ for the double semion theory which can serve as the SymTFT for an anomalous $\Z_2$ symmetry, and $\mathcal{Z}(\Z_2^\omega\times\Z_2^\omega)$ for the modular tensor category $\mathcal{Z}(\Z_2^\omega)\boxtimes\mathcal{Z}(\Z_2^\omega)$ which is the SymTFT for a product of two anomalous $\Z_2$ symmetries.} a three-dimensional topological field theory whose operators are line operators called anyons.  Formally $\mathcal{Z}(G)$ is the category of finite-dimensional complex representations of a certain Hopf algebra $D(G)$, the Drinfeld double, and the anyons discussed below are the objects in this category, but for our purposes it will suffice to simply quote some results\cite{Beigi:2010htr,beigi2012indistinguishablechargeonfluxionpairsquantum,Kitaev:2005hzj}.  The anyons (i.e.~the simple objects) are labeled by a conjugacy class $[a]=\{gag^{-1}|g\in G\}$ and an irreducible representation $\pi:Z(a)\rr\operatorname{GL}(V_\pi)$ of $Z(a)=\{g\in G|ag=ga\}$, the centralizer of some representative $a$ of the conjugacy class $[a]$.  The corresponding anyon will be labeled $([a],\pi)$.  To ensure uniqueness, we will pick a fixed representative $a$ for each conjugacy class, and for some formulae below we also need to pick a conjugating element $\kappa_b$ for every $b\in[a]$, that is an element $\kappa_b$ such that $b=\kappa_ba\kappa_b^{-1}$, with the convention that for the distinguished representative $a$ we have $\kappa_a=1$.

The {\it{dimension}} $d_{([a],\pi)}$ and {\it{topological spin}} $s_{([a],\pi)}$ are given by
\begin{equation}
    d_{([a],\pi)}=\left|[a]\right|\,d_\pi,\qquad s_{([a],\pi)}=\frac{\chi_\pi(a)}{d_\pi},
\end{equation}
where $|[a]|$ is the number of elements in the conjugacy class $[a]$, $\chi_\pi$ is the character of the representation $\pi$, i.e.~$\chi_\pi(g)=\operatorname{Tr}_{V_\pi}(\pi(g))$, and $d_\pi=\chi_\pi(1)$ is the dimension of the representation $\pi$.  Note that $d_\pi$ here is a positive integer, and $s_{([a],\pi)}$ is a phase (this follows since $a$ is in the center of $Z(a)$ and so by Schur's lemma $\pi(a)$ is simply a phase times the identity operator).  Anyons  for which $s_{([a],\pi)}=1$ are called {\it{bosons}}, those with $s_{([a],\pi)}=-1$ are {\it{fermions}}.

We have explicit formulae \cite{Beigi:2010htr,beigi2012indistinguishablechargeonfluxionpairsquantum}for the S-matrix between anyons,
\begin{equation}
    S_{([a],\pi),([a'],\pi')}=\frac{1}{|Z(a)||Z(a')|}\sum_{\substack{h\in G \\ ha'h^{-1}\in Z(a)}}\chi_\pi(h(a')^{-1}h^{-1})\chi_{\pi'}(h^{-1}a^{-1}h),
\end{equation}
and the fusion coefficients for fusion of anyons (Verlinde formula)
\begin{equation}
    N_{([a_1],\pi_1),([a_2],\pi_2)}^{([a_3],\pi_3)}=\sum_{([a_4],\pi_4)}\frac{S_{([a_1],\pi_1),([a_4],\pi_4)}S_{([a_2],\pi_2),([a_4],\pi_4)}\overline{S_{([a_3],\pi_3),([a_4],\pi_4)}}}{S_{([e],\mathbf{1}),([a_4],\pi_4)}}.
\end{equation}
We can often compute the fusion coefficients more efficiently using
\begin{equation}\label{eq:FusionComputation}
    N_{([a_1],\pi_1),([a_2],\pi_2)}^{([a_3],\pi_3)}=\frac{1}{|G|}\sum_{\substack{b_1\in[a_1],\ b_2\in[a_2], \\ b_1b_2\in[a_3]}}\sum_{g\in Z(b_1)\cap Z(b_2)}\overline{\chi_{\pi_3}(\kappa_{b_1b_2}^{-1}g\kappa_{b_1b_2})}\chi_{\pi_1}(\kappa_{b_1}^{-1}g\kappa_{b_1})\chi_{\pi_2}(\kappa_{b_2}^{-1}g\kappa_{b_2}),
\end{equation}
which follows from orthogonality of characters of the underlying Hopf algebra representations.

A condensable algebra is a formal linear combination of anyons, $A=\sum n_{([a],\pi)}\,([a],\pi)$, with $n_{([a],\pi)}\in\Z_{\ge 0}$, satisfying several conditions\footnote{The conditions we have specified here are necessary but not sufficient. In general, we also need to specify a morphism $m: A \times A \rightarrow A
$. The definition of condensable alegbra is given in terms of several conditions involving $m$, see \cite{Fuchs:2002cm, kong2021anyoncondensationtensorcategories}. See~\cite{Wen:2023otf} for an analysis of condensable algebras of $\mathcal{Z}(G)$, particularly those that lead to gapless SPT phases.  For a physics-friendly reference, see \cite{Benini:2022hzx}.}~\cite{Chatterjee:2022tyg}.  Of these, the ones relevant for us are that
\begin{enumerate}
\item $n_{([e],\mathbf{1})}=1$.
\item If $n_{([a],\pi)}\ne 0$, then $s_{([a],\pi)}=1$, i.e.~only bosons appear.
\item $0\le n_{([a],\pi)}\le d_{([a],\pi)}=|[a]|d_\pi$,
\item $n_{([a],\pi)}n_{([a'],\pi')}\le \sum_{([a''],\pi'')}N_{([a],\pi),([a'],\pi')}^{([a''],\pi'')}n_{([a''],\pi'')}$.
\item If $d_A=\sum n_{([a],\pi)}d_{([a],\pi)}=|G|$, then $A$ is called a Lagrangian algebra and it should be S-transformation invariant, 
\be
n_{([a],\pi)}=\sum_{([a'],\pi')}S_{([a],\pi),([a'],\pi')}n_{([a'],\pi')}.
\ee
If $d_A<|G|$, then for every $([a],\pi)$, the quantities
\be
\zeta_{([a],\pi)}:=\frac{\left|G\right|}{d_A}\sum_{([a'],\pi')}S_{([a],\pi),([a'],\pi')}n_{([a'],\pi')}
\ee
must be cyclotomic integers, meaning they can be written as integer linear combinations of $N$th roots of unity for some $N$.  A particularly useful fact is that a rational number is a cyclotomic integer if and only if it is actually an integer.
\end{enumerate}

There are three algebras that are always present, namely
\begin{itemize}
\item $A_{triv}=1$,
\item $A_{el}=\sum_\pi d_\pi ([e],\pi)$,
\item $A_{mag}=\sum_{[a]} ([a],\mathbf{1})$.
\end{itemize}

The algebras with $d_A=|G|$ are called Lagrangian algebras, and can be used to pick consistent topological boundary conditions for the TFT. In the sandwich construction, we put the SymTFT on a two-dimensional manifold cross an interval, with topological boundary conditions (corresponding to a choice of Lagrangian algebra) on one of the boundaries (the {\it{symmetry boundary}}).  The two-dimensional absolute theory obtained by shrinking the interval to a point will have global symmetries determined by the choice of Lagrangian algebra, which for $\mathcal{Z}(G)$ is determined by a choice of a subgroup $H$ (up to conjugation) and a cocycle $\nu\in H^2(H,U(1))$ (i.e.~the data needed to gauge a subgroup $H$)\footnote{For a general discussion on Lagrangian algebra, see \cite{Zhang:2023wlu, davydov2011wittgroupnondegeneratebraided,kong2014anyon,Cheng:2020rpl,Cong:2017ffh}.}.  $A_{el}$ corresponds to ordinary non-anomalous group-like symmetries $\operatorname{Vec}(G)$ (or $\operatorname{Vec}(G,1)$ if we want to emphasize that the anomaly class is trivial), while $A_{mag}$ corresponds to a theory with symmetries described by the fusion category $\operatorname{Rep}(G)$.  For given $G$, other Lagrangian algebras may exist corresponding to other possible gaugings of the group $G$. 

\subsection{Condensable algebras for $\mathcal{Z}(Q_8)$}\label{sec:condensablealgebraq8}
Now we are in a position to derive the condensable algebras for $\mathcal{Z}(Q_8)$ as an example for the general discussion in the previous section. The group $Q_8=\{1,-1,i,-i,j,-i,k,-k\}$ has multiplications $ij=k$, $ji=-k$, $ki=j$, $ik=-j$, $jk=i$, $kj=-1$, $i^2=j^2=k^2=-1$, and multiplication of the central elements $\pm 1$ is hopefully obvious.  The conjugacy classes are
\begin{equation}
    [1]=\{1\},\quad [-1]=\{-1\},\quad [i]=\{i,-i\},\quad [j]=\{j,-j\},\quad\text{and}\ [k]=\{k,-k\}.
\end{equation}
We'll choose conjugacy class representatives $1$, $-1$, $i$, $j$, and $k$, and will set $\kappa_{-i}=j$, $\kappa_{-j}=k$ and $\kappa_{-k}=i$.  For $1$ and $-1$ the centralizer is the entire group $Q_8$, and its irreducible representations will be labeled $\mathbf{1}$, $\pi_a$, $\pi_b$, $\pi_c$, and $\pi_m$.  The first four of these are one-dimensional irreps, while $\pi_m$ is a two-dimensional irrep.  The action on generators $i$ and $j$ is given by
\begin{equation}
    \pi_a(i)=1,\qquad\pi_a(j)=-1,
\end{equation}
\begin{equation}
    \pi_b(i)=-1,\qquad\pi_b(j)=1,
\end{equation}
\begin{equation}
    \pi_c(i)=-1,\qquad\pi_c(j)=-1,
\end{equation}
\begin{equation}
    \pi_m(i)=\left(\begin{matrix} i & 0 \\ 0 & -i \end{matrix}\right),\qquad\pi_m(j)=\left(\begin{matrix} 0 & 1 \\ -1 & 0 \end{matrix}\right).
\end{equation}
For $g=i,j,k$, the centralizer $Z(g)$ is isomorphic to $\Z_4$ and the four irreducible representations will be labeled $\rho_p$, $p=0,1,2,3$, that acts as
\begin{equation}
    \rho_p(g)=i^p.
\end{equation}
With this information, we can list the anyons of the theory, along with their dimensions and topological spins.
\begin{center}
\begin{minipage}{0.45\textwidth}
\centering
\begin{tabular}{|c|c|c|}
\hline
anyon & $d$ & $s$ \\
\hline
$([1],\mathbf{1})$ & 1 & 1 \\
\hline
$([1],\pi_a)$ & 1 & 1 \\
\hline
$([1],\pi_b)$ & 1 & 1 \\
\hline
$([1],\pi_c)$ & 1 & 1 \\
\hline
$([1],\pi_m)$ & 2 & 1 \\
\hline
$([-1],\mathbf{1})$ & 1 & 1 \\
\hline
$([-1],\pi_a)$ & 1 & 1 \\
\hline
$([-1],\pi_b)$ & 1 & 1 \\
\hline
$([-1],\pi_c)$ & 1 & 1 \\
\hline
$([-1],\pi_m)$ & 2 & -1 \\
\hline
$([i],\rho_0)$ & 2 & 1 \\
\hline
$([i],\rho_1)$ & 2 & i \\
\hline
$([i],\rho_2)$ & 2 & -1 \\
\hline
$([i],\rho_3)$ & 2 & -i \\
\hline
\end{tabular}
\end{minipage}
\hspace{-2mm}
\begin{minipage}{0.45\textwidth}
\centering
\begin{tabular}{|c|c|c|}
\hline
anyon & $d$ & $s$ \\
\hline
$([j],\rho_0)$ & 2 & 1 \\
\hline
$([j],\rho_1)$ & 2 & i \\
\hline
$([j],\rho_2)$ & 2 & -1 \\
\hline
$([j],\rho_3)$ & 2 & -i \\
\hline
$([k],\rho_0)$ & 2 & 1 \\
\hline
$([k],\rho_1)$ & 2 & i \\
\hline
$([k],\rho_2)$ & 2 & -1 \\
\hline
$([k],\rho_3)$ & 2 & -i \\
\hline
\end{tabular}
\end{minipage}
\end{center}

Of the twenty-two anyons, twelve are bosons, which we will list in the order $([1],\mathbf{1})$, $([1],\pi_a)$, $([1],\pi_b)$, $([1],\pi_c)$, $([1],\pi_m)$, $([-1],\mathbf{1})$, $([-1],\pi_a)$, $([-1],\pi_b)$, $([-1],\pi_c)$, $([i],\rho_0)$, $([j],\rho_0)$, $([k],\rho_0)$.  The boson-boson block of the S-matrix is then computed to be
\setcounter{MaxMatrixCols}{20}
\begin{equation}
    S|_{\text{bosons}}=\frac{1}{8}\left(\begin{matrix} 1 & 1 & 1 & 1 & 2 & 1 & 1 & 1 & 1 & 2 & 2 & 2 \\ 1 & 1 & 1 & 1 & 2 & 1 & 1 & 1 & 1 & 2 & -2 & -2 \\    1 & 1 & 1 & 1 & 2 & 1 & 1 & 1 & 1 & -2 & 2 & -2 \\ 1 & 1 & 1 & 1 & 2 & 1 & 1 & 1 & 1 & -2 & -2 & 2 \\ 2 & 2 & 2 & 2 & 4 & -2 & -2 & -2 & -2 & 0 & 0 & 0 \\ 1 & 1 & 1 & 1 & -2 & 1 & 1 & 1 & 1 & 2 & 2 & 2 \\ 1 & 1 & 1 & 1 & -2 & 1 & 1 & 1 & 1 & 2 & -2 & -2 \\ 1 & 1 & 1 & 1 & -2 & 1 & 1 & 1 & 1 & -2 & 2 & -2 \\ 1 & 1 & 1 & 1 & -2 & 1 & 1 & 1 & 1 & -2 & -2 & 2 \\ 2 & 2 & -2 & -2 & 0 & 2 & 2 & -2 & -2 & 4 & 0 & 0 \\ 2 & -2 & 2 & -2 & 0 & 2 & -2 & 2 & -2 & 0 & 4 & 0 \\ 2 & -2 & -2 & 2 & 0 & 2 & -2 & -2 & 2 & 0 & 0 & 4 \end{matrix}\right).
\end{equation}
Note that since this is just one block of the full anyon S-matrix it does not need to square to the identity.

We can compute the fusions of bosons as well.  Of course $([1],\mathbf{1})$ acts as the identity under fusion.  To write the remaining fusions we will take advantage of the cyclic symmetry that permutes $i$, $j$, and $k$ and simultaneously permutes $\pi_a$, $\pi_b$, and $\pi_c$.  Then we have
\begin{align}
\textbf{Fusion with } ([1],\pi_a):\qquad
& ([1],\pi_a)^2 = ([1],\mathbf{1}),  \\
& ([1],\pi_a)\cdot([1],\pi_b) = ([1],\pi_c), \quad
  ([1],\pi_a)\cdot([1],\pi_m) = ([1],\pi_m), \nonumber \\
& ([1],\pi_a)\cdot([-1],\mathbf{1}) = ([-1],\pi_a), \quad
  ([1],\pi_a)\cdot([-1],\pi_a) = ([-1],\mathbf{1}), \nonumber \\
& ([1],\pi_a)\cdot([-1],\pi_b) = ([-1],\pi_c), \nonumber \\
& ([1],\pi_a)\cdot([i],\rho_0) = ([i],\rho_0), \quad
  ([1],\pi_a)\cdot([j],\rho_0) = ([j],\rho_2),
\nonumber \\[6pt]
\textbf{Fusion with } ([1],\pi_m):\qquad
& ([1],\pi_m)^2 = ([1],\mathbf{1}) \oplus ([1],\pi_a)\oplus ([1],\pi_b)\oplus([1],\pi_c), \\
& ([1],\pi_m)\cdot([-1],\mathbf{1}) = ([-1],\pi_m), \quad
  ([1],\pi_m)\cdot([-1],\pi_a) = ([-1],\pi_m), \nonumber \\
& ([1],\pi_m)\cdot([i],\rho_0) = ([i],\rho_1)\oplus([i],\rho_3),
 \nonumber \\[6pt]
\textbf{Fusion with } ([-1],\mathbf{1}):\qquad
& ([-1],\mathbf{1})^2 = ([1],\mathbf{1}),  \\
& ([-1],\mathbf{1})\cdot([-1],\pi_a) = ([1],\pi_a), \quad
  ([-1],\mathbf{1})\cdot([i],\rho_0) = ([i],\rho_0), \nonumber
\\[6pt]
\textbf{Fusion with } ([-1],\pi_a):\qquad
& ([-1],\pi_a)^2 = ([1],\mathbf{1}),  \\
& ([-1],\pi_a)\cdot([-1],\pi_b) = ([1],\pi_c), \nonumber \\
& ([-1],\pi_a)\cdot([i],\rho_0) = ([i],\rho_0), \quad
  ([-1],\pi_a)\cdot([j],\rho_0) = ([j],\rho_2), \nonumber
\\[6pt]
\textbf{Fusion with } ([i],\rho_0):\qquad
& ([i],\rho_0)^2 = ([1],\mathbf{1})\oplus([1],\pi_a)\oplus([-1],\mathbf{1})\oplus([-1],\pi_a), \\
& ([i],\rho_0)\cdot([j],\rho_0) = ([k],\rho_0)\oplus([k],\rho_2). \nonumber
\end{align}
as well as fusions related to these by cyclic permutations. Note that the right-hand sides of these fusions can have non-bosonic anyons. Moreover, we have also computed fusion of non-bosonic objects, which will be useful to identify the reduced topological order. These can also be computed using the formula (\ref{eq:FusionComputation}), with the result
\begin{align}
\textbf{Fusion with } ([1],\pi_a):\qquad
& ([1],\pi_a)\cdot([-1],\pi_m) = ([-1],\pi_m), \\
& ([1],\pi_a)\cdot([i],\rho_r) = ([i],\rho_r), \nonumber \\
& ([1],\pi_a)\cdot([j],\rho_r) = ([j],\rho_{r+2}),
\nonumber \\[6pt]
\textbf{Fusion with } ([1],\pi_m):\qquad
& ([1],\pi_m)\cdot([i],\rho_r) = ([i],\rho_{r+1})\oplus ([i],\rho_{r+3}) \\
& ([1],\pi_m)\cdot([-1],\pi_m) =([-1],\mathbf{1})\oplus([-1],\pi_a)\oplus([-1],\pi_b)\oplus([-1],\pi_c), 
\nonumber \\[6pt]
\textbf{Fusion with } ([-1],\mathbf{1}):\qquad
& ([-1],\mathbf{1})\cdot([-1],\pi_m) = ([1],\pi_m), \\ 
& ([-1],\mathbf{1}) \cdot ([i],\rho_r)=([i],\rho_{-r}),  
\nonumber
\\[6pt]
\textbf{Fusion with } ([-1],\pi_a):\qquad
&  ([-1],\pi_a)\cdot([-1],\pi_m) = ([1],\pi_m),\\ 
&  ([-1],\pi_a)\cdot([i],\rho_r) = ([i],\rho_{-r}), \nonumber \\
&  ([-1],\pi_a)\cdot([j],\rho_r) = ([j],\rho_{2-r}), 
\nonumber \\[6pt]
\textbf{Fusion with } ([-1],\pi_m):\qquad
& ([-1],\pi_m)\cdot([i],\rho_r) =([i],\rho_{r+1})\oplus([i],\rho_{r+3}), \\
& ([-1],\pi_m)^2 =([1],\mathbf{1})\oplus([1],\pi_a)\oplus([1],\pi_b)\oplus([1],\pi_c),
\nonumber
\\[6pt]
\textbf{Fusion with } ([i],\rho_0):\qquad
& ([i],\rho_0) \cdot ([i],\rho_1) = ([1],\pi_m)\oplus([-1],\pi_m), \\
& ([i],\rho_0) \cdot ([i],\rho_3) = ([1],\pi_m)\oplus([-1],\pi_m), \nonumber \\
& ([i],\rho_0) \cdot ([i],\rho_2) = ([1],\pi_b)\oplus([1],\pi_c)\oplus([-1],\pi_b)\oplus([-1],\pi_c),\nonumber
\end{align}
Then we have the cross terms, 
\begin{equation}
    \begin{aligned}
        ([i],\rho_1) \cdot ([i],\rho_2) &=([1],\pi_m) \oplus ([-1],\pi_m), \\ 
        ([i],\rho_2) \cdot ([i],\rho_3) &= ([1],\pi_m) \oplus ([-1],\pi_m), \\
        ([i],\rho_r)\cdot([j],\rho_s) &=([k],\rho_{r+s})\oplus([k],\rho_{r+s+2}) , \\
        ([i],\rho_1) \cdot ([i],\rho_3) &= ([1],\pi_b) \oplus ([1],\pi_c) \oplus ([-1],\mathbf{1}) \oplus ([-1],\pi_a), \\
    \end{aligned}
\end{equation}
Finally, 
\begin{equation}
    \begin{aligned}
        ([i],\rho_1)^2 &=([1],\mathbf{1})\oplus([1],\pi_a)\oplus([-1],\pi_b)\oplus([-1],\pi_c), \\
        ([i],\rho_2)^2 &=([1],\mathbf{1})\oplus([1],\pi_a)\oplus([-1],\mathbf{1})\oplus([-1],\pi_a), \\
        ([i],\rho_3)^2 &=([1],\mathbf{1})\oplus([1],\pi_a)\oplus([-1],\pi_b)\oplus([-1],\pi_c).
    \end{aligned}
\end{equation}
Here the indices $r$ and $s$ are taken mod 4, and we should also include everything related to the above by simultaneously permuting $\{\pi_a,\pi_b,\pi_c\}$ and $\{i,j,k\}$.

Now we attempt to find all condensable algebras.  We note that the fusions above lead to nontrivial constraints (using condition 4 for condensable algebras),
\begin{equation}
    \begin{aligned}
        n_{([1],\pi_a)}n_{([1],\pi_b)} &\le n_{([1],\pi_c)}, \qquad n_{([1],\pi_a)}n_{([j],\rho_0)}=0, \\
        n_{([1],\pi_a)}n_{([-1],\mathbf{1})} &\le n_{([-1],\pi_a)}, \qquad n_{([1],\pi_m)}n_{([-1],\mathbf{1})}=0, \\
        n_{([1],\pi_a)}n_{([-1],\pi_a)} &\le n_{([-1],\mathbf{1})}, \qquad n_{([1],\pi_m)}n_{([-1],\pi_a)}=0, \\
        n_{([1],\pi_a)}n_{([-1],\pi_b)} &\le n_{([-1],\pi_c)}, \qquad n_{([1],\pi_m)}n_{([i],\rho_0)}=0, \\
        n_{([-1],\mathbf{1})}n_{([-1],\pi_a)} &\le n_{([1],\pi_a)}, \qquad n_{([-1],\pi_a)}n_{([j],\rho_0)}=0, \\
        n_{([-1],\pi_a)}n_{([-1],\pi_b)} &\le n_{([1],\pi_c)}, \qquad n_{([i],\rho_0)}n_{([j],\rho_0)}\le n_{([k],\rho_0)} \\
        n_{([1],\pi_m)}^2 &\le 1+n_{([1],\pi_a)}+n_{([1],\pi_b)}+n_{([1],\pi_c)}, \\
        n_{([i],\rho_0)}^2 &\le 1+n_{([1],\pi_a)}+n_{([-1],\mathbf{1})}+n_{([-1],\pi_a)},
    \end{aligned}
\end{equation}
and cyclic permutations.

It's easiest to first constrain the coefficients of the two-dimensional bosons.  Suppose first that $n_{([1],\pi_m)}=2$.  Then the constraints above immediately imply that $n_{([-1],\mathbf{1})}=n_{([-1],\pi_a)}=n_{([-1],\pi_b)}=n_{([-1],\pi_c)}=n_{([i],\rho_0)}=n_{[j],\rho_0)}=n_{([k],\rho_0)}=0$ and (combined with the condensable algebra condition 3) $n_{[1],\pi_a)}=n_{([1],\pi_b)}=n_{([1],\pi_c)}=1$.  This algebra has total dimension $8$ and is S-transformation invariant as required by condition 5, so we do have the algebra
\begin{equation}
    \mathcal{A}_{8,1}=([1],\mathbf{1}) \oplus ([1],\pi_a) \oplus ([1],\pi_b) \oplus ([1],\pi_c) \oplus 2([1],\pi_m).
\end{equation}
We recognize this as the electric algebra, $\mathcal{A}_{8,1}=A_{el}$.

Suppose instead that $n_{([1],\pi_m)}=1$.  We still have $n_{([-1],\mathbf{1})}=n_{([-1],\pi_a)}=n_{([-1],\pi_b)}=n_{([-1],\pi_c)}=n_{([i],\rho_0)}=n_{([j],\rho_0)}=n_{([k],\rho_0)}=0$, and of $x=n_{([1],\pi_a)}$, $y=n_{([1],\pi_b)}$, and $z=n_{([1],\pi_c)}$, they are either all zero, all one, or one of them is one and the other two are zero (having two nonzero would violate the inequality $n_{([1],\pi_a)}n_{([1],\pi_b)}\le n_{([1],\pi_c)}$ or one of its cyclic partners).  Plugging this into condition 5, we have
\begin{equation*}
    \zeta_{([1],\pi)}=d_\pi,\qquad\zeta_{([-1],\pi)}=\frac{x+y+z-1}{3+x+y+z},\qquad\zeta_{([i],\rho_0)}=\frac{2+2x-2y-2z}{3+x+y+z},
\end{equation*}
\begin{equation}
    \zeta_{([j],\rho_0)}=\frac{2-2x+2y-2z}{3+x+y+z},\qquad\zeta_{([k],\rho_0)}=\frac{2-2x-2y+2z}{3+x+y+z}.
\end{equation}
These are all manifestly rational numbers, so the condition is that they must all be integers\footnote{In principle we should also check $\zeta_{([a],\pi)}$ for non-bosonic $([a],\pi)$ as well.  This can be done and doesn't lead to any additional constraints.}.  This rules out the cases where $x=y=z$, but allows all three of the cases where one of $x$, $y$, and $z$ is equal to one.  Thus we get three more algebras of dimension four,
\begin{equation}
\begin{aligned}
    \mathcal{A}_{4,1}=\ & ([1],\mathbf{1}) \oplus ([1],\pi_a) \oplus ([1],\pi_m),\\
    \mathcal{A}_{4,2}=\ & ([1],\mathbf{1}) \oplus ([1],\pi_b) \oplus ([1],\pi_m),\\
    \mathcal{A}_{4,3}=\ & ([1],\mathbf{1}) \oplus ([1],\pi_c) \oplus ([1],\pi_m).
\end{aligned}
\end{equation}

Any remaining possibilities will have $n_{([1],\pi_m)}=0$.  Now suppose $n_{([i],\rho_0)}=2$.  Then our inequalities immediately require $n_{([1],\pi_b)}=n_{([1],\pi_c)}=n_{([-1],\pi_b)}=n_{([-1],\pi_c)}=0$ and $n_{([1],\pi_a)}=n_{([-1],\mathbf{1})}=n_{([-1],\pi_a)}=1$.  We also have inequalities $2n_{([j],\rho_0)}\le n_{([k],\rho_0)}$ and $2n_{([k],\rho_0)}\le n_{([j],\rho_0)}$ which can only be satisfied if $n_{([j],\rho_0)}=n_{([k],\rho_0)}=0$.  The resulting algebra has dimension 8 and is S-transformation invariant, so we get
\begin{equation}
    \mathcal{A}_{8,2}=([1],\mathbf{1}) \oplus ([1],\pi_a) \oplus ([-1],\mathbf{1}) \oplus ([-1],\pi_a) \oplus 2([i],\rho_0),
\end{equation}
and by cyclic symmetry we also have
\begin{equation}
\begin{aligned}
    \mathcal{A}_{8,3}=\ & ([1],\mathbf{1}) \oplus ([1],\pi_b) \oplus ([-1],\mathbf{1}) \oplus ([-1],\pi_b) \oplus 2([j],\rho_0),\\
    \mathcal{A}_{8,4}=\ & ([1],\mathbf{1}) \oplus ([1],\pi_c) \oplus ([-1],\mathbf{1}) \oplus ([-1],\pi_c)+2([k],\rho_0).
\end{aligned}
\end{equation}

So we can assume that $n_{([i],\rho_0)}$, $n_{([j],\rho_0)}$, and $n_{([k],\rho_0)}$ are all less than two.  The inequalities imply that they are either all equal to one, exactly one of them is equal to one and the other two are zero, or they are all equal to zero.  For the case $n_{([i],\rho_0)}=n_{([j],\rho_0)}=n_{([k],\rho_0)}=1$, the inequalities force $n_{([1],\pi_a)}=n_{([1],\pi_b)}=n_{([1],\pi_c)}=n_{([-1],\pi_a)}=n_{([-1],\pi_b)}=n_{([-1],\pi_c)}=0$.  Let $x=n_{([-1],\mathbf{1})}$.  Then we have
\begin{equation*}
    \zeta_{([1],\mathbf{1})}=\zeta_{([1],\pi_a)}=\zeta_{([1],\pi_b)}=\zeta_{([1],\pi_c)}=\zeta_{([-1],\mathbf{1})}=1,\qquad\zeta_{([1],\pi_m)}=\frac{2-2x}{7+x},
\end{equation*}
\begin{equation}
    \zeta_{([-1],\pi_a)}=\zeta_{([-1],\pi_b)}=\zeta_{([-1],\pi_c)}=\frac{x-1}{7+x},\qquad\zeta_{([i],\rho_0)}=\zeta_{([j],\rho_0)}=\zeta_{([k],\rho_0)}=\frac{6+2x}{7+x}.
\end{equation}
These will all be integers if $x=1$ and not if $x=0$.  Thus we get another dimension eight algebra,
\begin{equation}
    \mathcal{A}_{8,5}=([1],\mathbf{1}) \oplus ([-1],\mathbf{1}) \oplus ([i],\rho_0) \oplus ([j],\rho_0) \oplus \oplus ([k],\rho_0),
\end{equation}
which we recognize as the magnetic algebra, $\mathcal{A}_{8,5}=A_{mag}$.

We move on to the case where one of $n_{([i],\rho_0)}$, $n_{([j],\rho_0)}$, and $n_{([k],\rho_0)}$ is one and the other two vanish.  Without loss of generality, set $n_{([i],\rho_0)}=1$, $n_{([j],\rho_0)}=n_{([k],\rho_0)}=0$.  The inequalities require $n_{([1],\pi_b)}=n_{([1],\pi_c)}=n_{[-1],\pi_b)}=n_{([-1],\pi_c)}=0$.  Setting $x=n_{([1],\pi_a)}$, $y=n_{([-1],\mathbf{1})}$, and $z=n_{([-1],\pi_a)}$, the inequalities force either $x=y=z=1$, $x=y=z=0$, or one of the three is equal to one with the other two equal to zero.  We also have
\begin{equation*}
    \zeta_{([1],\mathbf{1})}=\zeta_{([1],\pi_a)}=\zeta_{([-1],\mathbf{1})}=\zeta_{([-1],\pi_a)}=1,
\end{equation*}
\begin{equation*}
    \zeta_{([1],\pi_b)}=\zeta_{([1],\pi_c)}=\zeta_{([-1],\pi_b)}=\zeta_{([-1],\pi_c)}=\frac{x+y+z-1}{3+x+y+z},\qquad\zeta_{([1],\pi_m)}=\frac{2+2x-2y-2z}{3+x+y+z},
\end{equation*}
\begin{equation}
    \zeta_{([i],\rho_0)}=2,\qquad\zeta_{([j],\rho_0)}=\zeta_{([k],\rho_0)}=\frac{2-2x+2y-2z}{3+x+y+z}.
\end{equation}
This rules out the cases $x=y=z=1$ and $x=y=z=0$, but allows all the cases where exactly one of them is one.  So we get three new dimension four algebras,
\begin{equation}
\begin{aligned}
    \mathcal{A}_{4,4}=\ & ([1],\mathbf{1})\oplus ([1],\pi_a)\oplus ([i],\rho_0),\\
    \mathcal{A}_{4,5}=\ & ([1],\mathbf{1})\oplus ([-1],\mathbf{1})\oplus ([i],\rho_0),\\
    \mathcal{A}_{4,6}=\ & ([1],\mathbf{1})\oplus ([-1],\pi_a)\oplus ([i],\rho_0),
\end{aligned}
\end{equation}
and by cyclic symmetry we get six more,
\begin{equation}
\begin{aligned}
    \mathcal{A}_{4,7}=\ & ([1],\mathbf{1})\oplus([1],\pi_b)\oplus([j],\rho_0),\\
    \mathcal{A}_{4,8}=\ & ([1],\mathbf{1})\oplus([-1],\mathbf{1})\oplus([j],\rho_0),\\
    \mathcal{A}_{4,9}=\ & ([1],\mathbf{1})\oplus([-1],\pi_b)\oplus([j],\rho_0),\\
    \mathcal{A}_{4,10}=\ & ([1],\mathbf{1})\oplus([1],\pi_c)\oplus([k],\rho_0),\\
    \mathcal{A}_{4,11}=\ & ([1],\mathbf{1})\oplus([-1],\mathbf{1})\oplus([k],\rho_0),\\
    \mathcal{A}_{4,12}=\ & ([1],\mathbf{1})\oplus([-1],\pi_c)\oplus([k],\rho_0),
\end{aligned}
\end{equation}
So we have now reduced down to the case where all the coefficients of dimension two bosons vanish.  The dimension one bosons form a group under fusion ($\Z_2^3$ in this case).  It turns out that for an algebra built purely from dimension one bosons, the inequalities in condition 4 are equivalent to the statement that the bosons with coefficient one must form a subgroup.  It can be checked that each such subgroup also satisfies condition 5, so we end up with
\begin{equation}
\begin{aligned}
     \mathcal{A}_1=\ & ([1],\mathbf{1}), \\
    \mathcal{A}_{2,1}=\ & ([1],\mathbf{1})\oplus([1],\pi_a),\\
    \mathcal{A}_{2,2}=\ & ([1],\mathbf{1})\oplus([1],\pi_b),\\
    \mathcal{A}_{2,3}=\ & ([1],\mathbf{1})\oplus([1],\pi_c),\\
    \mathcal{A}_{2,4}=\ & ([1],\mathbf{1})\oplus([-1],\mathbf{1}),\\
    \mathcal{A}_{2,5}=\ & ([1],\mathbf{1})\oplus([-1],\pi_a),\\
    \mathcal{A}_{2,6}=\ & ([1],\mathbf{1})\oplus([-1],\pi_b),\\
    \mathcal{A}_{2,7}=\ & ([1],\mathbf{1})\oplus([-1],\pi_c),\\
    \mathcal{A}_{4,13}=\ & ([1],\mathbf{1})\oplus([1],\pi_a)\oplus([1],\pi_b)\oplus([1],\pi_c),\\
    \mathcal{A}_{4,14}=\ & ([1],\mathbf{1})\oplus([1],\pi_a)\oplus([-1],\mathbf{1})\oplus([-1],\pi_a),\\
    \mathcal{A}_{4,15}=\ & ([1],\mathbf{1})\oplus([1],\pi_b)\oplus([-1],\mathbf{1})\oplus([-1],\pi_b),\\
    \mathcal{A}_{4,16}=\ & ([1],\mathbf{1})\oplus([1],\pi_c)\oplus([-1],\mathbf{1})\oplus([-1],\pi_c),\\
    \mathcal{A}_{4,17}=\ & ([1],\mathbf{1})\oplus([1],\pi_a)\oplus([-1],\pi_b)\oplus([-1],\pi_c),\\
    \mathcal{A}_{4,18}=\ & ([1],\mathbf{1})\oplus([1],\pi_b)\oplus([-1],\pi_a)\oplus([-1]\,\pi_c),\\
    \mathcal{A}_{4,19}=\ & ([1],\mathbf{1})\oplus([1],\pi_c)\oplus([-1],\pi_a)\oplus([-1],\pi_b),\\
\end{aligned}
\end{equation}
Finally, we have another eight dimensional one, 
\begin{equation}
    \mathcal{A}_{8,6}=\  ([1],\mathbf{1})\oplus([1],\pi_a)\oplus([1],\pi_b)\oplus([1],\pi_c)\oplus([-1],\mathbf{1})\oplus([-1],\pi_a)\oplus([-1],\pi_b)\oplus([-1],\pi_c).
\end{equation}
We note that a few of the condensable algebras (including all of the Lagrangian algebras) were guessed in~\cite{Perez-Lona:2025ncg}, but we have derived them here rigorously.

\section{SymTFT for $\text{Rep}(D_4)$}\label{sec:symtft_for_d4}

In this section, we will review the $3d$ SymTFT for $(D_4)$ symmetry and discuss all possible gaugings associated with this symmetry in terms of Lagrangian algebras of the Drinfeld center and the corresponding symmetry boundary conditions. We then address, igSPT phases and their connection to anomaly resolution for $D_4/\text{Rep}(D_4)$.

\subsection{SymTFT Action}

The $3d$ SymTFT for a $2d$ QFT with a $D_4$ or non-invertible $\text{Rep}(D_4)$ symmetry can be constructed using three $\Z_2$-valued fields. This follows from the fact that $D_4$ can be realized as an extension of $\Z^{(a)}_2 \times \Z^{(b)}_2$ by $\Z^{(c)}_2$. The SymTFT action has previously appeared in \cite{He:2016xpi, Yu:2023nyn, Franco:2024mxa,Kaidi:2023maf,deWildPropitius:1995cf}, we are going to review that construction, and we will largely follow the presentation of \cite{Kaidi:2023maf}. The SymTFT action takes the following form: 
\begin{equation}\label{eqn:d4_symtft}
 S_{3d} = {\pi} \int_{X_3} \left(\hat{a}\delta a + \hat{b} \delta b + \hat{c} \delta c + abc\right).
\end{equation}
This action consists of three BF pairs, each corresponding to a distinct $\Z_2$ symmetry. The cubic interaction term encodes a mixed anomaly involving all three $\Z_2$ factors, enforcing the group extension criteria. The equation of motion of $c$ reads: $\delta \hat{c} = ab$, which is the cocyle condition for a $D_4$ extension \cite{Kapustin:2014lwa,Kapustin:2014zva}. This action is invariant under the following gauge transformations: 
\begin{equation}
    \begin{aligned}
         a \rightarrow a+ \delta \alpha, \qquad \hat{a} \rightarrow \hat{a} +\delta\hat{\alpha} - \left( \beta c - \gamma b + \beta \delta\gamma \right) \\
        b \rightarrow b+ \delta\beta, \qquad \hat{b} \rightarrow \hat{b} + \delta\hat{\beta} - \left( \gamma a - \alpha c + \gamma \delta \alpha \right)\\
        c \rightarrow c+ \delta\gamma, \qquad \hat{c} \rightarrow \hat{c} + \delta\hat{\gamma} - \left( \alpha b - \beta a + \alpha \delta\beta \right)
    \end{aligned}
\end{equation}
We verify this explicitly in appendix \ref{app:gauge_inv_d4}. Similar gauge transformations were also discussed in \cite{Kaidi:2023maf,He:2016xpi}, though our gauge transformations differ by an overall minus sign, which is immaterial for $\Z_2$ valued fields. One may also add a self-anomaly term in the action for any of the symmetries, as this action treats each $\Z_2$-valued field on the same footing and such a modification will not alter the gauge transformations\footnote{The action with a similar self-anomaly term indeed appears in \cite{Kaidi:2023maf}}. The fact that we can add three separate self-anomaly couplings to this action is in coherence with $H^3(D_4, U(1)) = \Z_2 \times \Z_2$.

\subsubsection*{Topological Operators in the bulk}

The bulk of the SymTFT supports in total 22 line operators \footnote{In addition to topological line operators, SymTFT can support surface operators in the bulk. In this paper, we are only focusing on the topological line operators of the bulk. Hence wherever we use the phrase topological operator, it implies line operators.}, among which 8 are invertible and remaining 14 are non invertible. The invertible line operators arise as combinations of the following operators: 
\begin{equation}\label{eqn:inv_op_d4}
    U_a[\gamma_1] = \exp\left(i \pi\int_{\gamma_1} a \right), \quad U_b[\gamma_1] = \exp\left(i \pi\int_{\gamma_1} b \right), \quad U_c[\gamma_1] = \exp\left(i \pi\int_{\gamma_1} c \right)
\end{equation} 
where $\gamma_1$ is a closed loop. Moreover, there are 14 non-invertible line operators arising from the dual gauge fields. we list a few of them\footnote{ for further details check \cite{Kaidi:2023maf}, also check \cite{Yu:2023nyn,Franco:2024mxa} for a top down discussion of this SymTFT.}. 
\begin{equation}\label{eqn:non_inv_op_d4}
    \begin{aligned}
        \hat{U}_a[\delta_1] &= \int D\phi_0 D\tilde{\phi}_0 \exp \left[ i \pi \left(  \int_{\delta_1} \hat{a} + \int_{\delta_1} \left( -\phi_0 \delta\tilde{\phi}_0  + \phi_0 c - \tilde{\phi}_0 b \right) \right) \right] \\
         \hat{U}_b[\rho_1] &= \int D\tilde{\phi}_0 D\hat{\phi}_0 \exp \left[ i \pi \left( \int_{\rho_1} \hat{b} + \int_{\rho_1} \left( -\tilde{\phi}_0 \delta\hat{\phi}_0 + \tilde{\phi}_0 a - \hat{\phi}_0 c \right) \right) \right] \\
        \hat{U}_c[\sigma_1] &=  \int D\phi_0 D\hat{\phi}_0 \exp \left[ i \pi \left(  \int_{\sigma_1} \hat{c} + \int_{\sigma_1} \left( -\hat{\phi}_0 \delta\phi_0 + \hat{\phi}_0 b - \phi_0 a \right) \right) \right]
    \end{aligned}
\end{equation}
where $\phi_0, \hat{\phi}_0 $ and $\tilde{\phi}_0$ are gauge fields living on the defect coupled to the bulk fields, which transform under the gauge transformations of the bulk fields, 
\begin{equation}
    \phi_0 \rightarrow \phi_0 + \beta, \qquad \tilde{\phi}_0 \rightarrow \tilde{\phi}_0 + \gamma, \qquad \hat{\phi} \rightarrow \hat{\phi}_0 + \alpha .
\end{equation}

\subsubsection*{Lagrangian Algebras for $D_4$}

The choice of boundary condition determines the symmetry category of the resulting absolute theory. Different choice of boundary conditions will lead to different symmetry categories. The realization of $D_4$ symmetry from this SymTFT depends on selecting boundary conditions consistent with the group extension condition. Moreover, there are Lagrangian algebras associated with various symmetries, related to $D_4$ through discrete gaugings, such as $\text{Rep}(D_4)$. Before analyzing the SymTFT and its boundary conditions, we tabulate the Lagrangian algebras associated with this SymTFT as presented in \cite{Bhardwaj:2024qrf, Bhardwaj:2023idu}. In that work, the algebras were presented in Drinfeld double notation. Appendix \ref{app:dictionary} provides a dictionary translating between that notation and ours.
\begin{table}[H]
    \centering
    \begin{tabular}{|c|c|}
    \hline
     & Lagrangian Algebras   \\
    \hline
    1 & $1 \oplus U_a \oplus U_b \oplus U_c \oplus U_{ab} \oplus U_{bc} \oplus U_{ac} \oplus U_{abc}$  \\
    \hline
    2 &   $1 \oplus U_a \oplus U_b \oplus U_{ab} \oplus 2\hat{U}_{c}$    \\
    \hline
    3 &   $1 \oplus U_b \oplus U_c \oplus U_{bc} \oplus 2\hat{U}_a$   \\
    \hline
    4 &   $ 1\oplus U_a \oplus U_c \oplus U_{ac} \oplus 2 \hat{U}_{b}$    \\
    \hline
    5 &  $1 \oplus U_c\oplus \hat{U}_a \oplus \hat{U}_b \oplus \hat{U}_{ab} $    \\
    \hline  
    6 &   $1 \oplus U_a \oplus \hat{U}_{b} \oplus \hat{U}_{c} \oplus \hat{U}_{bc}$   \\
    \hline 
    7 &   $1 \oplus U_b \oplus \hat{U}_{a} \oplus \hat{U}_{c} \oplus \hat{U}_{ac}$  \\
    \hline  
    8 &   $1 \oplus U_a \oplus U_{bc} \oplus U_{abc} \oplus 2 \hat{U}_{bc}$    \\
    \hline 
    9 &   $1 \oplus U_b \oplus U_{ac} \oplus U_{abc} \oplus 2\hat{U}_{ac}$    \\
    \hline 
    10 &   $1 \oplus U_c \oplus U_{ab} \oplus U_{abc} \oplus 2\hat{U}_{ab}$     \\
    \hline
     11 &  $1 \oplus U_{abc} \oplus \hat{U}_{ab} \oplus \hat{U}_{ac} \oplus \hat{U}_{bc}$   \\
    \hline
    \end{tabular}
    \caption{Lagrangian algebras of $D_4$ SymTFT in the $\Z^3_2$ notation}
    \label{tab:lag_alg_d4}
\end{table}

\subsection{Discrete gaugings and quantum symmetry}

We now turn to identifying the symmetry in the $2d$ absolute theory associated with each Lagrangian algebra. Some of these are straightforward; while others require more care. To begin with, imposing Dirichlet boundary condition on the gauge fields $a,b,c$, terminates all invertible lines of the SymTFT and yields a theory with $\Z^{(a)}_2 \times \Z^{(b)}_2 \times \Z^{(c)}_2$ symmetry, subject to a non-trivial mixed anomaly. In terms of boundary Lagrangian algebra, this corresponds to,
\begin{equation}\label{eqn:lag_z2_cube}
    \mathcal{L}_{\Z^3_2} = 1 \oplus U_a \oplus U_b \oplus U_c \oplus U_{ab} \oplus U_{bc} \oplus U_{ac} \oplus U_{abc}   
\end{equation}
To realize a $D_4$ symmetry, we instead impose Neumann boundary condition on one of $a,b$ or $c$, which are all equivalent at the level of action. This amounts to starting from the anomalous $\Z^{(a)}_2 \times \Z^{(b)}_2 \times \Z^{(c)}_2$ theory and gauging a $\Z_2$ subgroup, say $\Z_2^{(c)}$, after which the quantum dual $\hat{\Z}_2$ extends the remaining $\Z_2 \times \Z_2$ to $D_4$: 
\begin{equation}\label{eqn:d4_ses}
    0 \rightarrow \hat{\Z}^{(c)}_2 \rightarrow D_4 \rightarrow \Z^{(a)}_2 \times \Z^{(b)}_2 \rightarrow 0
\end{equation}
In terms of boundary Lagrangian algebra, gauging $\Z^{(c)}_2$ subgroup corresponds to modifying the boundary condition on the corresponding field, $c$. Imposing Neumann boundary condition on $c$ should correspond roughly to exchanging $c$ with $\hat{c}$ in the Lagrangian algebra, and so we conjecture that it should correspond to:
\begin{equation}\label{eqn:lag_1_d4}
    \mathcal{L}^1_{D_4} = 1 \oplus U_a \oplus U_b \oplus U_{ab} \oplus 2\hat{U}_{c}
\end{equation}
By symmetry, imposing Neumann boundary condition on $a$ or $b$ instead of $c$ yields the following Lagrangian algebras: 
\begin{equation}\label{eqn:lag_2_d4}
    \begin{aligned}
        \mathcal{L}^2_{D_4} = 1 \oplus U_b \oplus U_c \oplus U_{bc} \oplus 2\hat{U}_a \\
        \mathcal{L}^3_{D_4} = 1 \oplus U_a \oplus U_c \oplus U_{ac} \oplus 2 \hat{U}_{b}
    \end{aligned}
\end{equation}
This change of boundary condition from \eqref{eqn:lag_1_d4} to \eqref{eqn:lag_2_d4} can be interpreted as a discrete gauging of a subgroup of $D_4$, specifically $\Z^{(a)}_2 \times \hat{\Z}^{(c)}_2$ or $\Z^{(b)}_2 \times \hat{\Z}^{(c)}_2$. Conversely, starting from a theory with $D_4$ symmetry, one can obtain a theory with an anomalous $\Z^{3}_2$ symmetry by gauging the central $\Z_2$ subgroup (e.g.~$\hat{\Z}_2^{(c)}$) of $D_4$.
Furthermore, gauging the full $D_4$ symmetry instead produces a non-invertible $\text{Rep}(D_4)$ symmetry, corresponds to trading all the endable electric lines with magnetic line operators and vice versa (on the symmetry boundary). The associated Lagrangian algebra is,  
\begin{equation}\label{eqn:lag_rep_d4}
    \mathcal{L}^1_{\text{Rep}(D_4)} = 1 \oplus U_c\oplus \hat{U}_a \oplus \hat{U}_b \oplus \hat{U}_{ab} 
\end{equation}
As evident from \eqref{eqn:lag_1_d4}, this amounts to exchanging Dirichlet with Neumann boundary conditions for $a,b$ and $\hat{c}$. Alternatively, $\text{Rep}(D_4)$ can also be obtained from \eqref{eqn:lag_2_d4}, which yields the following lagrangian algebras:
\begin{equation}
    \begin{aligned}
      \mathcal{L}^2_{\text{Rep}(D_4)} = 1 \oplus U_a \oplus \hat{U}_{b} \oplus \hat{U}_{c} \oplus \hat{U}_{bc} \\
      \mathcal{L}^3_{\text{Rep}(D_4)} = 1 \oplus U_b \oplus \hat{U}_{a} \oplus \hat{U}_{c} \oplus \hat{U}_{ac}
    \end{aligned}
\end{equation}
We can obtain these directly from the Lagrangian algebra in \eqref{eqn:lag_rep_d4} by a discrete gauging of $\hat{\Z}^{(a)}_2 \times \Z^{(c)}_2$ or $\hat{\Z}^{(b)}_2 \times \Z^{(c)}_2$ subgroup of $\text{Rep}(D_4)$ or equivalently by gauging either $\Z^{(b)}_2$ or $\Z^{(a)}_2$ subgroup of $D_4$. This provides an alternate construction of $\text{Rep}(D_4)$, consistent with the fact that we can obtain a theory with a non-invertible $\text{Rep}(D_4)$ symmetry by gauging the $\Z^2_2 \subset \Z^3_2$ \cite{Kaidi:2023maf,Yu:2023nyn}.

At this stage, we have identified the corresponding symmetry associated with seven of the Lagrangian algebras in table \ref{tab:lag_alg_d4}. The remaining four are less obvious.
\begin{equation}\label{eqn:spcl_lag_alg}
    \begin{aligned}
        \mathcal{L} = 1 \oplus U_a \oplus U_{bc} \oplus U_{abc} \oplus 2 \hat{U}_{bc} \\
        \mathcal{L} = 1 \oplus U_b \oplus U_{ac} \oplus U_{abc} \oplus 2\hat{U}_{ac} \\
        \mathcal{L} = 1 \oplus U_c \oplus U_{ab} \oplus U_{abc} \oplus 2\hat{U}_{ab} \\
        \mathcal{L} = 1 \oplus U_{abc} \oplus \hat{U}_{ab} \oplus \hat{U}_{ac} \oplus \hat{U}_{bc}
    \end{aligned}
\end{equation}
These can be related with the rest through discrete gauging of a subgroup of $D_4$, potentially with discrete torsion (since $H^2(\Z_2 \times \Z_2, U(1)) = \Z_2$). Two of the algebras in \eqref{eqn:spcl_lag_alg} must correspond to gauging $\hat{\Z}^{(c)}_2 \times \Z^{(i)}_2 \subset D_4$ with a discrete torsion. Finally, gauging full $D_4$ with discrete torsion must produce another one, while gauging $\Z_4 \subset D_4$ should complete the list\footnote{For a list of possible gaugings of $D_4$, see \cite{Perez-Lona:2023djo,Putrov:2024uor}}. In order to identify these systematically, we check for boundary variation of the action and identify the conditions under which it vanishes.

\subsubsection*{Variation of the action} 

We take the $3d$ Dijkgraaf-Witten action, vary it and analyze the boundary variation. 
\begin{equation*}
    \begin{aligned}
         S_{3d} &= \pi \int_{X_3} \left(\hat{a}\delta a + \hat{b} \delta b + \hat{c} \delta c + abc\right), \\
        \Delta S_{3d} &= \pi \int_{X_{3}}  \left(\Delta \hat{a} \delta a + \hat{a} \delta \Delta a  + \Delta \hat{b} \delta b + \hat{b} \delta \Delta b + \Delta \hat{c} \delta c + \hat{c} \delta \Delta c + \left(\Delta a bc+ a \Delta b c + ab \Delta c \right)\right). \\
    \end{aligned}
\end{equation*}
We have used the symbol $\Delta$ to denote variation. We perform an integration by parts, 
\begin{equation*}
    \begin{aligned}
         \Delta S_{3d} =  {\pi} \int_{X_3} \left(\Delta \hat{a} \delta a +\Delta \hat{b} \delta b + \Delta \hat{c} \delta c - \delta \hat{a} \Delta a - \delta \hat{b}\Delta b - \delta \hat{c}\Delta c + \left( \Delta a bc+ a \Delta b c + ab \Delta c \right)\right) \\ +\pi\int_{\partial X_{3}} \left(\hat{a}\Delta a + \hat{b} \Delta b + \hat{c} \Delta c\right),
    \end{aligned}
\end{equation*}
\begin{equation}
\begin{aligned}
    \Delta S_{3d} =  {\pi} \int_{X_3} \left(\Delta \hat{a} \delta a +\Delta \hat{b} \delta b + \Delta \hat{c} \delta c - \Delta a \left(\delta \hat{a} -  bc\right) -\Delta b \left(\delta \hat{b} -  ac \right) - \Delta c \left( \delta \hat{c}-  ab \right) \right)\\ 
    +\pi\int_{\partial X_3} \left(\hat{a}\Delta a + \hat{b} \Delta b + \hat{c} \Delta c\right).
\end{aligned}    
\end{equation}
We start with $\mathcal{L} = 1 \oplus U_a \oplus U_b \oplus U_{ab} \oplus 2 \hat{U}_c$ to have a $D_4$ symmetry. The boundary variation vanishes provided we add a boundary coupling $\pi\int_{\partial X_3}c\hat{c}$ and impose $\Delta a =0,\Delta b=0$ and $\Delta \hat{c} = 0$ at the boundary $\partial X_3$. In order to gauge $\hat{\Z}^{(c)}_2 \times \Z^{(i)}_2$ with discrete torsion, we stack the $2d$ $\hat{\Z}^{(c)}_2 \times \Z^{(i)}_2$-SPT phase on the symmetry boundary. This is equivalent, for $i=a$, to adding a boundary coupling of $\pi\int_{\partial X_3}\hat{c}a$~\cite{Kaidi:2023maf,Perez-Lona:2025ncg}. The new boundary variation including this SPT, 
\begin{equation}
    \int_{\partial X_3} \left(\hat{a}\Delta a + \hat{b} \Delta b - c \Delta \hat{c} + \hat{c}\Delta a + \Delta \hat{c} a\right) = \int_{\partial X_3} \left(\hat{a}\Delta a + \hat{b} \Delta b + (a - c) \Delta \hat{c} + \hat{c}\Delta a\right).
\end{equation}
Imposing the condition, $\Delta b = 0$ and $a = c$ at the boundary leaves us with, 
\begin{equation*}
    \int_{\partial M} \left(\hat{a} \Delta a +  \hat{c} \Delta a \right).
\end{equation*}
Now, if we impose, $\hat{a} + \hat{c} = 0 $, the boundary variation cancels (note that we are just working modulo 2, so many of the signs are inessential). Collectively, these conditions translate to the fact that we need to terminate $U_b,U_{ac}, \hat{U}_{ac}$ on the symmetry boundary, leading us to conjecture that the Lagrangian algebra is
\begin{equation}
    \mathcal{L} = 1 \oplus U_b \oplus U_{ac} \oplus U_{abc} \oplus 2\hat{U}_{ac}.
\end{equation}
In the same manner, we can associate the following Lagrangian algebra, 
\begin{equation}
\label{eq:LagrangianForZ2^2WithDT}
    \mathcal{L} = 1 \oplus U_a \oplus U_{bc} \oplus U_{abc} \oplus 2\hat{U}_{bc}
\end{equation}
with gauging a $\Z^{(c)}_2 \times \Z^{(b)}_2$ subgroup of $D_4$ with discrete torsion. Here we impose the following conditions to ensure the boundary variations vanish, 
\begin{equation}
\label{eq:BCsForZ2^2WithDT}
    \hat{b} + \hat{c} =0, \quad \Delta a = 0, \quad b - c = 0,
\end{equation}
which in turn allows us to identify the Lagrangian algebra. That leaves us with two Lagrangian algebras: 
\begin{equation}
    \begin{aligned}
        \mathcal{L} = 1 \oplus U_c \oplus U_{ab} \oplus U_{abc} \oplus 2\hat{U}_{ab}, \\
        \mathcal{L} = 1 \oplus U_{abc} \oplus \hat{U}_{ab} \oplus \hat{U}_{ac} \oplus \hat{U}_{bc}.
    \end{aligned}
\end{equation}
We can identify the former with a discrete gauging of $\Z_4 \subset D_4$. Since we are starting from $\mathcal{L} = 1 \oplus U_a \oplus U_b \oplus U_{ab} \oplus 2 \hat{U}_c$, in order to obtain $\mathcal{L} = 1 \oplus U_c \oplus U_{ab} \oplus U_{abc} \oplus 2\hat{U}_{ab}$, we must terminate or change the boundary condition on $U_c$ and $\hat{U}_{ab}$. The fusion relations,
\begin{equation}
  \hat{U}^2_{ab} = 1\oplus U_c \oplus U_{ab}\oplus U_{abc}, \qquad U^2_c = 1  .
\end{equation}
These fusions could be consistent with either a $\Z_4$ subgroup on the symmetry boundary (if $\hat{U}_{ab}$ decomposes into the lines for the $\Z_4$ generator and its inverse) or with a $\Z_2^2$ subgroup (in which case $\hat{U}_{ab}$ breaks into lines for two of the nontrivial elements).  However, since we have already identified the only $\Z_2^2$ gaugings of $D_4$, we conclude that this Lagrangian algebra must correspond to gauging the $\Z_4$ subgroup.  We can also check that the boundary variation cancels provided 
\begin{equation}
    \hat{a} + \hat{b} =0, \quad \Delta c = 0, \quad a - b = 0.
\end{equation}
Of course this also follows since this Lagrangian algebra is just related to the algebra \eqref{eq:LagrangianForZ2^2WithDT} by permutation symmetry, so the boundary conditions will also just be a permutation of \eqref{eq:BCsForZ2^2WithDT}.

That leaves us with only one lagrangian algebra to identify, $\mathcal{L} = 1 \oplus U_{abc} \oplus \hat{U}_{ab} \oplus \hat{U}_{ac} \oplus \hat{U}_{bc}$, which would correspond to gauging the full $D_4$ symmetry with a choice of discrete torsion. In addition we can also check that the boundary variation cancels under the following boundary conditions:  
\begin{equation}
    \begin{aligned}
        \hat{a} = \hat{b} = \hat{c} , \qquad a + b + c =0 \\
    \end{aligned}
\end{equation}
We can now complete the table \ref{tab:lag_alg_d4} by adding the symmetry category corresponding to each Lagrangian algebra in the following table:
\begin{table}[H]
    \centering
    \begin{tabular}{|c|c|c|}
        \hline
            Subgroup of $D_4$  & Corresponding Lagrangian Algebra & Quantum Dual symmetry   \\
        \hline
             Trivial subgroup  & $1\oplus U_a \oplus U_b \oplus U_{ab} \oplus 2\hat{U}_c$ & $D_4$ \\
        \hline
            $\hat{\Z}^{(c)}_2$ & $1 \oplus U_a \oplus U_b \oplus U_c \oplus U_{ab} \oplus U_{bc} \oplus U_{ac} \oplus U_{abc}$ & $\Z^{(c)}_2 \times \Z^{(a)}_2 \times \Z^{(b)}_2$ \\
        \hline 
            $\Z^{(a)}_2$ & $1 \oplus U_b \oplus \hat{U}_a \oplus \hat{U}_c \oplus \hat{U}_{ac} $ & $\text{Rep}(D_4)$ \\
        \hline
            $\Z^{(b)}_2$ & $1 \oplus U_a  \oplus \hat{U}_b \oplus \hat{U}_c \oplus \hat{U}_{bc}$ & $\text{Rep}(D_4)$ \\
        \hline
            $\hat{\Z}^{(c)}_2 \times \Z^{(b)}_2$ & $1 \oplus U_a \oplus U_c \oplus U_{ac} \oplus 2\hat{U}_b$ & $D_4$ \\
        \hline 
            $\hat{\Z}^{(c)}_2 \times \Z^{(a)}_2$ & $1 \oplus U_b \oplus U_c \oplus U_{bc} \oplus 2\hat{U}_a$ & $D_4$ \\
        \hline
            $(\hat{\Z}^{(c)}_2 \times \Z^{(b)}_2)_{\text{with d.t.}}$ & $1 \oplus U_a \oplus U_{bc} \oplus U_{abc} \oplus 2\hat{U}_{bc} $ & $D_4$\\
        \hline 
            $(\hat{\Z}^{(c)}_2 \times \Z^{(a)}_2)_{\text{with d.t.}}$ & $1 \oplus U_b \oplus U_{ac} \oplus U_{abc} \oplus 2\hat{U}_{ac} $ & $D_4$\\
        \hline
            $\Z_4$ & $1 \oplus U_c \oplus U_{ab} \oplus U_{abc} \oplus 2\hat{U}_{ab}$ & $D_4$ \\
        \hline 
            $D_4$  & $ 1 \oplus U_c \oplus \hat{U}_a \oplus \hat{U}_b \oplus \hat{U}_{ab}$ & $\text{Rep}(D_4)$ \\
        \hline      
            $(D_4)_{\text{with d.t.}}$ & $1 \oplus U_{abc} \oplus \hat{U}_{ab} \oplus \hat{U}_{ac} \oplus \hat{U}_{bc}$ & $\text{Rep}(D_4)$ \\
        \hline
    \end{tabular}
    \caption{we are viewing $D_4$ as $0 \rightarrow \hat{\Z}^{(c)}_2 \rightarrow D_4 \rightarrow \Z^{(a)}_2 \times \Z^{(b)}_2 \rightarrow 0$). Starting with a Lagrangian algebra of $\mathcal{L} = 1\oplus U_a \oplus U_b \oplus U_{ab} \oplus 2\hat{U}_c$ that corresponds to $D_4$, we have listed the different gaugings and the quantum dual symmetry. $\hat{\Z}_2$ appearing in the second row is the Pontraygin dual of $\Z_2$. }
    \label{tab:d4_gauging}
\end{table}

\subsubsection*{$D_4$ symmetry web in the space of $c=1$ theories}

For a particular realization of a theory with $\operatorname{Vec}(D_4)$ symmetry we can take a compact free boson at radius $R$ (throughout this discussion we will ignore other symmetries of this theory, and for the various gaugings we will only discuss the symmetries which are directly related to the original $D_4$ symmetry).  This theory has a $\Z_4$ translation symmetry along with a $\Z_2$ reflection symmetry which together generate a non-anomalous global $D_4$ symmetry.  We know that if we gauge the $\Z_4$ translation symmetry we land on the free boson theory at radius $R/4$, where we still have the $\Z_2$ reflection symmetry as well as a quantum $\Z_4$ symmetry which acts as an order four translation in the dual coordinate.  These two symmetries again combine into a $D_4$ symmetry.  On the other hand, if we only gauge a $\Z_2$ subgroup of the translation symmetry, arriving at the free boson at radius $R/2$, then we are left with the remaining $\Z_2$ translation symmetry along with the quantum $\Z_2$ dual translation symmetry.  These commute with each other but have a mixed anomaly between them.  They both commute with the $\Z_2$ reflection symmetry, so all together we have $\Z_2^3$ with a mixed anomaly between two of the factors.

Next we turn to gaugings involving the reflection, which will take us to the orbifold branch.  Simply gauging the reflection symmetry brings us to the orbifold theory $S^1_R/\Z_2$.  There is another $\Z_2$ subgroup of $D_4$ which is conjugate to the reflection symmetry and whose gauging is completely equivalent.  There are also two more $\Z_2$ subgroups related to these by an outer automorphism, and gauging these will also land us on $S^1_R/\Z_2$.  Another option is to gauge the full $D_4$ symmetry, either with or without discrete torsion turned on.  This is equivalent to first gauging the normal $\Z_4$ subgroup and then gauging the quotient symmetry which acts as reflection.  In this formulation, discrete torsion corresponds to choices of the $\Z_2$ reflection symmetry that are related by outer automorphism (this can be confirmed by looking at the orbifold partition function and noting that the partial traces whose signs get flipped by the discrete torsion vanish), and in either case we land on the $S^1_{R/4}/\Z_2$ orbifold theory.  We know that this theory should have a quantum $\operatorname{Rep}(D_4)$ symmetry, and this matches known non-invertible symmetries of the orbifold branch.  In fact, by gauging the $D_4$ symmetry of $S^1_{R/4}$, we can argue that the $S^1_R/\Z_2$ theories that we got by gauging $\Z_2$ reflection subgroups also must have $\operatorname{Rep}(D_4)$ symmetry.  Finally, there are two $\Z_2\times\Z_2$ subgroups of $D_4$ which can be gauged either with or without turning on discrete torsion.  Again these can be implemented by first gauging the $\Z_2$ translation symmetry and then gauging a reflection symmetry.  All the choices can be shown to be equivalent, landing us on $S^1_{R/2}/\Z_2$.  To identify the symmetry present after one of these gaugings, we refer to the work in the previous subsection and identify the symmetry as $D_4$, which is also known to be present for orbifold branch theories at any radius.

\begin{figure}[H]
    \centering
    \begin{tikzpicture}[scale= 1.0]
      \draw (-1,0) -- (7,0);
      \draw (0,0) -- (0,7);
      \foreach \x in {2,4,6} \fill (\x,0) circle (3pt);
      \foreach \y in {2,4,6} \fill (0,\y) circle (3pt);
      \node at (6,-0.4) {$(D_4)$};
      \node at (4,-0.4) {$(\Z^3_2)$};
      \node at (2,-0.4) {$(D_4)$};
         
       \node at (-0.8,2) {$\text{Rep}(D_4)$};
      \node at (-0.5,4) {$D_4$};
      \node at (-0.8,6) {$\text{Rep}(D_4)$};

      \draw[->, thick, red, shorten >=4pt, shorten <=4pt] (6,0) -- (0,6) node[midway, above, sloped, red] {\footnotesize{$\Z^{(a)}_2$ or $\Z^{(b)}_2$}};
      
     \draw[->, thick, blue, shorten >=4pt, shorten <=4pt] (6,0) -- (0,4)  node[midway, above, sloped, blue] {\footnotesize{$\hat{\Z}^{(c)}_2 \times \Z^{(i)}_2, i=a,b$}};
     
     \draw[->, thick,orange, shorten >=4pt, shorten <=4pt] (6,0) -- (0,2) node[midway, above, sloped, orange] {$D_4$};

      \draw[->, thick, gray, shorten >=4pt, shorten <=4pt, bend left=45] (6,0) to node[midway, above, sloped, gray] {\footnotesize{$\hat{\Z}^{(c)}_2$}} (4,0);
      
      \draw[->, thick, purple, shorten >=4pt, shorten <=4pt, bend left=45] (6,0)  to node[midway, below, sloped, purple] {$\Z_4$} (2,0);
    \end{tikzpicture}
    \caption{The $D_4$ symmetry web in the space of $c=1$ theories. We are still viewing $D_4$ as an extension of $\Z^{(a)}_2 \times \Z^{(b)}_2$ by $\hat{\Z}^{(c)}_2$. The horizontal line represents the circle branch and the vertical line is the orbifold branch. The arrows represents different gaugings that connect the theories. Similar figures have also appeared in \cite{Perez-Lona:2023djo,Putrov:2024uor,Diatlyk:2023fwf,Thorngren:2021yso,Bartsch:2022ytj,Bhardwaj:2022maz}. }
    \label{fig:D4Web}
\end{figure}
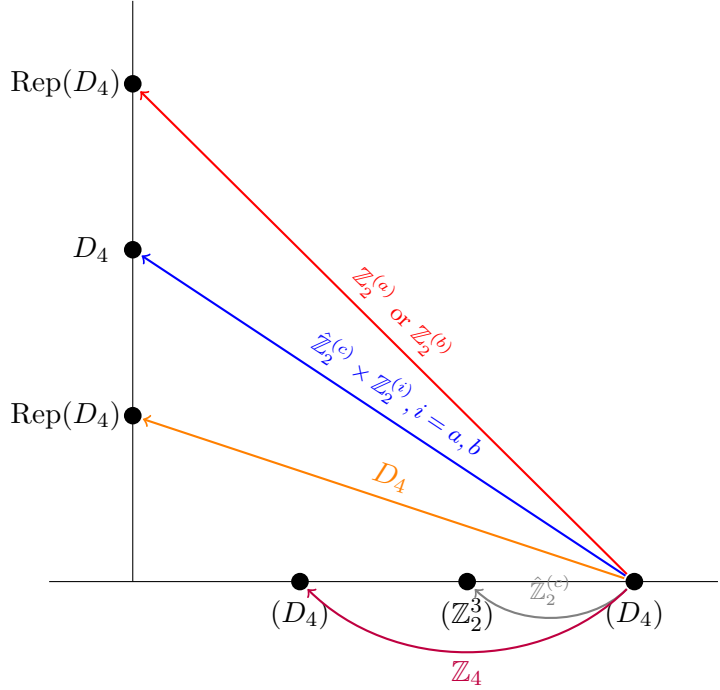

These relations between $c=1$ theories are summarized in Figure~\ref{fig:D4Web}.

\subsection{Club Sandwich, igSPT and anomaly resolution}

In this section, we discuss the non-Lagrangian condensable algebras associated with this SymTFT and their role in anomaly resolution\footnote{For a complete list of condensable algebras of $\text{Rep}(D_4)$, see \cite[Table III]{Bhardwaj:2024qrf}, \cite{Warman:2024lir}. Furthermore, the Lagrangian algebras of $\text{Rep}(H_8)$ have also appeared in \cite{Schafer-Nameki:2025fiy}. Finally, see \cite{Antinucci:2025fjp} for the condensable algebras of $\text{TY}(\Z_2 \times \Z_2, -)$}. A subset of these condensable algebras of $\mathcal{Z}(D_4)$ gives rise to gapless SPT phases\cite{Scaffidi_2017} depending on the Lagrangian algebra chosen on the symmetry boundary. Among these, there exists gapless phases that cannot be deformed into a gapped SPT phase --- these are the igSPT phases \cite{Bhardwaj:2024qrf,Thorngren:2020wet,Wen:2022tkg,Li:2023knf,Antinucci:2024ltv} \footnote{We thank Rui Wen for pointing us towards these references.}. From the perspective of missing charges, an igSPT phase can be defined as the phase with confined gapped charges of a symmetry that do not show up in any of the gapped SPT phases~\cite{Bhardwaj:2024qrf}\footnote{A notion of igSPTs for 1-form symmetries in $4d$ have recently been introduced in \cite{Antinucci:2024ltv}.}.
This idea is fundamental in the club sandwich construction \cite{Bhardwaj:2023bbf}. The notion of missing charges allows us to view the SymTFT of $D_4$ (or $\text{Rep}(D_4)$) symmetry with a trivially acting subgroup (or non-faithfully acting) as a club sandwich. The SymTFT for the effectively acting symmetry with an anomaly would show up once the left part of the club sandwich is reduced. The presence of this anomaly obstructs the symmetry preserving deformation of the gapless phase into a gapped SPT phase (hence the name intrinsically gapless SPT).

Trivially acting symmetries also play a fundamental role in anomaly resolution  \cite{Wang:2017loc, Tachikawa:2017gyf, Robbins:2020msp, Robbins:2021lry, Robbins:2021ibx,Robbins:2021xce,Robbins:2022wlr}\footnote{See \cite[section 3]{Robbins:2022wlr} for an introduction to the idea of trivially acting symmetries without using the SymTFT or Club Sandwich. A SymTFT interpretation of this construction was given in \cite[Section 4]{Lin:2025oml}. To understand the connection of trivially acting symmetries with decomposition, see \cite{Robbins:2021lry, Robbins:2021ibx, Robbins:2022wlr}. }. One can extend an anomalous symmetry of a theory by introducing trivially acting symmetries, thereby embedding it into a larger non-anomalous symmetry and resolving the anomaly.  A bit more explicitly, if we have a group $G$ with some anomaly $\omega\in H^3(G,U(1))$, and if we have a short exact sequence of groups $1\rightarrow K\rightarrow\Gamma\stackrel{\pi}{\rightarrow} G\rightarrow 1$, and if $\pi^\ast\omega=0\in H^3(\Gamma,U(1))$, then we say the anomaly is resolved.  Here the subgroup $K$ of $\Gamma$ are the trivially acting symmetries. The idea that the anomaly of any theory (general QFT) can be resolved by embedding this theory inside a larger non-anomalous fusion category symmetry namely an igSPT was laid out in \cite{Perez-Lona:2025ncg}. The effectively acting anomalous symmetry inside an igSPT phase was identified with the theory whose anomaly we are resolving. From the club sandwich this is more transparent; once we partially reduce the club sandwich, the anomalous symmetry or the anomaly that obstructs the deformation shows up.

Similarly, we can in principle write down short exact sequences of fusion categories, $1\rightarrow A\rightarrow B\rightarrow C\rightarrow 1$, where the arrows are now strong monoidal functors.  One natural interpretation of whether $C$ is ``anomalous'' is whether it has a fiber functor or not (no fiber functor means anomalous\cite{Thorngren:2019iar,Choi:2023xjw}), and we can look for sequences of this type where the fusion category $C$ does not have a fiber functor, but the category $B$ does.  This would be a notion of anomaly resolution in this setting~\cite{Perez-Lona:2025ncg}.

As a concrete example, there is an unique igSPT phase associated with $\text{Rep}(D_4)$ symmetry \cite{Bhardwaj:2024qrf}, associated with the condensable algebra $\mathcal{A} = 1 \oplus U_{ab} \oplus U_{bc} \oplus U_{ca}$. The reduced topological order for this condensable algebra was identified to be the Drinfeld center of $\text{Vec}(\Z^{\omega}_2)$. In other words, we have a $\text{Rep}(D_4)$ symmetry where only an anomalous $\Z^{\omega}_2$ acts effectively, and the rest of the symmetry acts trivially because the operators charged under a $\Z_2^2$ symmetry inside $\text{Rep}(D_4)$ will be missing from the theory. In terms of club sandwich we have, 
\begin{figure}[H]
    \centering
    \begin{tikzpicture}[scale=1]
    
    \fill[blue!30] (0,0) rectangle (2,3); 
    \fill[green!30] (2,0) rectangle (4,3); 
    
    \draw[thick, black] (2,0) -- (2,3);
    
    \draw[thick] (0,0) -- (0,3);
    \draw[thick] (4,0) -- (4,3);
    
    \node at (-0.5,-0.3) {\footnotesize{$\mathfrak{B}_\text{sym}$}}; 
    \node at (4.5,-0.3) {\footnotesize{$\mathfrak{B}_\text{phys}$}}; 
    \node at (2.3,3.3) {\footnotesize{$\mathcal{I}_\phi$}}; 
    \node at (1.0, 1.65) {\footnotesize{$\mathcal{Z}(\text{Rep}({D}_4)$)}};
    \node at (3.0, 1.65) {\footnotesize{$\mathcal{Z}({\Z}^\omega_2)$}};
    \end{tikzpicture}
    \caption{The club sandwich with the respective Drinfeld centers of $\text{Rep}({D}_4)$ and the reduced topological order, $\mathcal{Z}(\Z^\omega_2)$.}
    \label{fig:Anom_res_d4}
\end{figure}
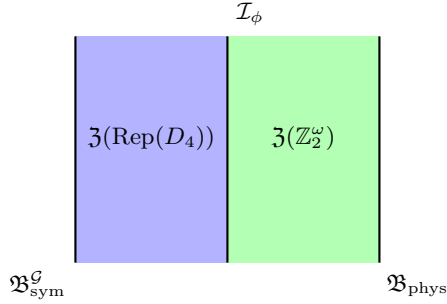
Following \cite{Perez-Lona:2025ncg}, we would identify this club sandwich with the following fusion categorical short exact sequence,
\begin{equation}
    0 \rightarrow \text{Vec}(\Z_2 \times \Z_2) \rightarrow \text{Rep}(D_4) \rightarrow \text{Vec}(\Z_2,\omega) \rightarrow 0
\end{equation} 
In other words, starting from a theory with an anomalous $\Z^\omega_2$ symmetry, we can resolve the anomaly by extending the anomalous $\Z_2$ symmetry by $\Z_2 \times \Z_2$ symmetry, leading us to a theory with $\text{Rep}(D_4)$ where the $\Z_2 \times \Z_2$ acts trivially. This fact can be understood by looking at the condensable algebra that leads us to the igSPT. The objects in the condensable algebra terminate on the interface of the club sandwich hence the corresponding charges associated with these anyons will be absent from the physical theory, telling us that these lines act trivially.  

\begin{figure}[ht]
    \centering
    \begin{tikzpicture}[scale=1.35]
    
    \fill[blue!30] (0,0) rectangle (2,3); 
    \fill[green!30] (2,0) rectangle (4,3); 
    
    \draw[thick, black] (2,0) -- (2,3);
    
    \draw[thick] (0,0) -- (0,3);
    \draw[thick] (4,0) -- (4,3);
    
    \node at (-0.5,-0.3) {$\mathfrak{B}_\text{sym}^{D_4}$}; 
    \node at (4.5,-0.3) {$\mathfrak{B}_\text{phys}$}; 
    \node at (2.1,3.3) {$\mathcal{I}_\phi$}; 
    \node at (1.0, 1.65) {$\mathcal{Z}(\text{Vec}_{{D}_4})$};
    \node at (3.0, 1.65) {$\mathcal{Z}({\Z}_4)$};
    
    \node at (5.5,1.5) {$\rightarrow$}; 

    \begin{scope}[shift={(7,0)}] 
        \fill[green!30] (0,0) rectangle (2,3); 
        
        \draw[thick, black] (2,0) -- (2,3);
        
        \draw[thick] (0,0) -- (0,3);
        \draw[thick] (2,0) -- (2,3); 
        
        \node at (-0.5,-0.3) {}; 
        \node at (2.5,-0.3) {$\mathfrak{B}_\text{phys}$}; 
        \node at (0,-0.3) {$\mathfrak{B}_\text{sym}^{(\Z_2 \times \Z_2)_\omega}$}; 
        \node at (1.1, 1.5) {$\mathcal{Z}({\Z}_4)$};
    \end{scope}
    \end{tikzpicture}
    
    \caption{Once we perform the interval compactification on the blue side of the club sandwich, we obtain the modified boundary condition for $\mathcal{Z}(\Z_4)$, which corresponds to gauging a $\Z_2$ subgroup of $\Z_4$. This gives rise to a theory with mixed anomaly. }
    \label{fig:combined_club_sandwich}
\end{figure}
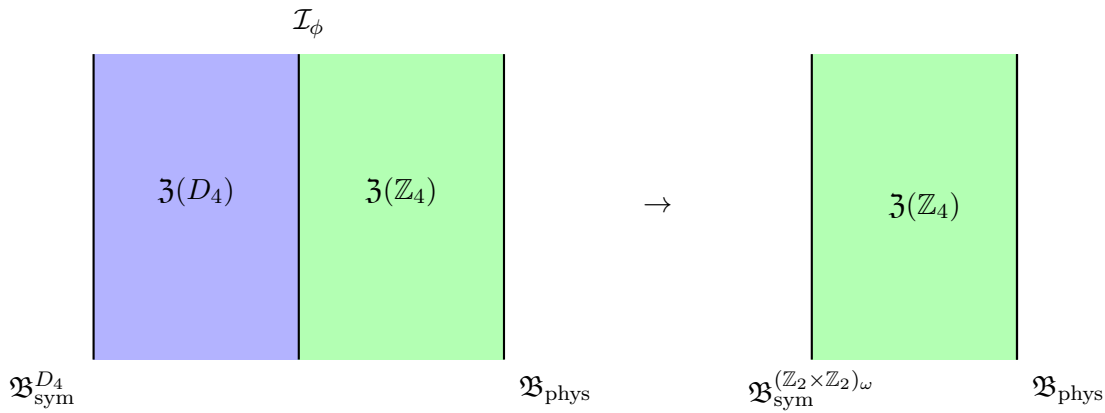

We can also talk about resolving anomalies by extending the anomalous symmetry to a dihedral group itself with a trivial acting symmetry. We have identified the condensable algebras of the Drinfeld center of the dihedral group that gives rise to an igSPT phase in appendix \ref{app:red_top_order_d4}. The condensable algebras are, 
\begin{equation}
    \mathcal{A}_1 = 1 \oplus U_{ac} , \qquad \mathcal{A}_2 = 1 \oplus U_{bc}
\end{equation}
The reduced topological order for both of them were found to be $\mathcal{Z}(\Z_4)$, with the following club sandwich picture, 
\begin{figure}[H]
    \centering
    \begin{tikzpicture}[scale=1]
    
    \fill[blue!30] (0,0) rectangle (2,3); 
    \fill[green!30] (2,0) rectangle (4,3); 
    
    \draw[thick, black] (2,0) -- (2,3);
    
    \draw[thick] (0,0) -- (0,3);
    \draw[thick] (4,0) -- (4,3);
    
    \node at (-0.5,-0.3) {\footnotesize{$\mathfrak{B}_\text{sym}$}}; 
    \node at (4.5,-0.3) {\footnotesize{$\mathfrak{B}_\text{phys}$}}; 
    \node at (2.3,3.3) {\footnotesize{$\mathcal{I}_\phi$}}; 
    \node at (1.0, 1.65) {\footnotesize{$\mathcal{Z}(\text{Vec}_{{D}_4})$}};
    \node at (3.0, 1.65) {\footnotesize{$\mathcal{Z}({\Z}_4)$}};
    \end{tikzpicture}
    \caption{The club sandwich with the respective Drinfeld centers of $\mathcal{Z}(\text{Vec}_{D_4})$ and the reduced topological order, $\mathcal{Z}(\Z_4)$.}
\end{figure}
Unlike in the previous case, the anomaly is a bit more obscure in this example. Things become clear once we reduce the blue part of the club sandwich to obtain the new boundary condition. To begin with the Lagrangian algebra on the symmetry boundary was $\mathcal{L} = 1 \oplus U_a \oplus U_b \oplus U_{ab} \oplus 2\hat{U}_c$, from our calculation in \ref{app:red_top_order_d4}, we can argue that the line $\widehat{U}_c$ cannot cross the interface\footnote{More precisely, if we consider any correlation function in which a $\hat{U}_c$ line penetrates into the $\mathfrak{Z}(\Z_4)$ part of the bulk, then the gauging of the condensable algebra on this side will lead to destructive interference from the lines of $\mathcal{A}$ fusing and braiding with $\widehat{U}_c$.  Effectively, we can say that the $\widehat{U}_c$ line is not invariant under this gauging and so gets projected out in this half of the bulk.}. Hence, the modified Lagrangian algebra on the reduced topological order is $\mathcal{L}' = 1 \oplus U_a \oplus U_b \oplus U_{ab}$. In terms of the lines of the reduced topological order in \eqref{eqn:lines_of_rto_d4}, it can be written as, $\mathcal{L}' = 1 \oplus L_1 \oplus L_2 \oplus L_3$. This is the boundary condition that corresponds to gauging a $\Z_2$ subgroup of $\Z_4$, which produces $\Z^2_2$ theory with a mixed anomaly (Fig.~\ref{fig:combined_club_sandwich}).
Now we are in a position to pin down the short exact sequence associated with this anomaly resolution\textcolor{blue}{\footnote{The mixed anomaly between the $\Z_2 \times \Z_2$ is given by $i\pi \int A_1 \cup A_2$. Under the extension, the gauge field of $\Z_2$ has a modified cocyle condition, \begin{equation}
    \delta B = A_1 \cup A_2
\end{equation}
which is the group extension condition for the dihedral group.}}, 
\begin{equation}
    0 \rightarrow \text{Vec}(\Z_2) \rightarrow \text{Vec}(D_4) \rightarrow \text{Vec}(\Z_2 \times \Z_2,\omega_{\text{mixed}}) \rightarrow 0
\end{equation}
This anomaly resolving sequence was first proposed in \cite[Section 5.2]{Robbins:2021lry}. Here we have derived this using the technology of club sandwich. 

\section{SymTFT for $\text{Rep}(Q_8)$}\label{sec:symtft_q4}

In this section, we extend the discussion of the previous section to the case of a group-like $Q_8$ symmetry and non-invertible $\text{Rep}(Q_8)$ symmetry. We begin by presenting the $3d$ SymTFT action that captures the $Q_8$ symmetry of a $2d$ QFT. We then discuss all possible gaugings of $Q_8$. Finally, we identify useful short exact sequences relevant for anomaly resolution involving $Q_8/\text{Rep}(Q_8)$ symmetry. 

\subsection{SymTFT Action}

The SymTFT action encoding $Q_8/\text{Rep}(Q_8)$ symmetry is given by three $\Z_2$ BF couplings with additional mixed anomalies:
\begin{equation}
    S_{bulk} = \pi\int_{X_3} \left(\hat{a}\delta a + \hat{b}\delta b + \hat{c}\delta c + abc + \frac{1}{2}c\delta b + \frac{1}{2} c\delta a\right)
\end{equation}
This action differs from the $D_4$ case only by the last two couplings, which is crucial to establish the group extension criteria \eqref{eqn:ext_q8} of $Q_8$ \cite{Kapustin:2014lwa,Kapustin:2014zva}. This Dijkgraaf-Witten action with twists have appeared in \cite{deWildPropitius:1995cf}. However, to the best of our knowledge, in the context of a $3d$ SymTFT for some $2d$ theory with $Q_8$ symmetry, such an action was not discussed before in the literature\footnote{Four dimensional SymTFTs capturing the $Q_8$ symmetry of $3d$ orthosymplectic gauge theories have been discussed previously in \cite{Bergman:2024its}. }. 

\subsubsection*{Gauge transformations} 

This action admits the same gauge transformation as the $D_4$ SymTFT, for convenience, we rewrite them here: 
\begin{equation}
    \begin{aligned}
         a \rightarrow a+ \delta \alpha, \qquad \hat{a} \rightarrow \hat{a} +\delta\hat{\alpha} - \left( \beta c - \gamma b + \beta \delta\gamma \right) \\
        b \rightarrow b+ \delta\beta, \qquad \hat{b} \rightarrow \hat{b} + \delta\hat{\beta} - \left( \gamma a - \alpha c + \gamma \delta \alpha \right)\\
        c \rightarrow c+ \delta\gamma, \qquad \hat{c} \rightarrow \hat{c} + \delta\hat{\gamma} - \left( \alpha b - \beta a + \alpha \delta\beta \right)
    \end{aligned}
\end{equation}

\subsubsection*{Topological operators}

The gauge transformations associated with this SymTFT are identical to the $D_4$ case\eqref{eqn:non_inv_op_d4}, the constraints enforced by the gauge bundle equations of motion differ.These differences ultimately modify the fusion rules of the line operators. The gauge bundle equations of motion are: 
\begin{equation}
\begin{aligned}
    2\oint a &\in \Z ,\quad 2\oint b \in \Z , \quad 2\oint c \in \Z \\
    \oint 2\hat{a} + c \in &\Z , \quad \oint 2\hat{b} + c \in \Z , \quad \oint 2\hat{c} + a + b \in \Z
\end{aligned}    
\end{equation}

\subsubsection*{Lagrangian Algebras for $Q_8$}

The SymTFT discussed in the last section captures $Q_8$ symmetry under a specific set of boundary conditions. However, there are many other possible boundary conditions for this SymTFT, which will give rise to different symmetry. Each such symmetry will be related to $Q_8$ via some discrete gauging. Each such boundary condition is associated with a distinct Lagrangian algebra; to understand the different gaugings of $Q_8$ we need to analyze all the Lagrangian algebras of this SymTFT, given in the table below. These algebras were derived in section (\ref{sec:condensablealgebraq8}), we have presented them in the  $\Z^{3}_2$ notation using the dictionary between the notations in Appendix \ref{app:dictionary}. 
\begin{table}[H]
    \centering
    \begin{tabular}{|c|c|}
    \hline
        & Lagrangian Algebras \\
        \hline
        1 & $1 \oplus U_a \oplus U_b \oplus U_c \oplus U_{ab} \oplus U_{bc} \oplus U_{ac} \oplus U_{abc}$ \\
        \hline
        2 & $1 \oplus U_a \oplus U_b \oplus U_{ab} \oplus 2\hat{U}_c$ \\
        \hline
        3 & $ 1 \oplus U_c \oplus \hat{U}_a \oplus \hat{U}_b \oplus \hat{U}_{ab}$ \\
        \hline
        4 & $1 \oplus U_c \oplus U_b \oplus U_{bc} \oplus 2\hat{U}_a$ \\
        \hline
        5 & $ 1 \oplus U_c \oplus U_a \oplus U_{ac} \oplus 2\hat{U}_b$ \\
        \hline
        6 & $ 1 \oplus U_{c} \oplus U_{ab} \oplus U_{abc} \oplus 2\hat{U}_{ab}$ \\
        \hline
    \end{tabular}
    \caption{Lagrangian algebras of $Q_8$ SymTFT in the $\Z^{3}_2$ notation.}
    \label{tab:lag_q8}
\end{table}

\subsection{Discrete gaugings and quantum symmetry}

We now identify the symmetry category associated with each Lagrangian algebra, as we did for $D_4$ in the last section. In order to realize a $Q_8$ symmetry, we need to impose Dirichlet boundary condition on $a,b$ and Neumann boundary condition on $c$. Viewing $Q_8$ as an extension of $\Z_2 \times \Z_2$ by $\Z_2$, 
\begin{equation}\label{eqn:q4_ses}
    0 \rightarrow \widehat{\Z}^{(c)}_2 \rightarrow Q_8 \rightarrow \Z^{(a)}_2 \times \Z^{(b)}_2 \rightarrow 0
\end{equation}
this choice enforces the group extension criteria. The corresponding Lagrangian is:
\begin{equation}
    \mathcal{L}_{Q_8} = 1\oplus U_a \oplus U_b \oplus U_{ab} \oplus 2\widehat{U}_c
\end{equation}
Now we would like to discuss different discrete gaugings of $Q_8$. One of the obvious choice is the non-anomalous central $\widehat{\Z}^{(c)}_2$ subgroup of $Q_8$, we can gauge that to obtain a theory with $\Z^{(c)}_2 \times \Z^{(a)}_2 \times \Z^{(b)}_2$ theory with a mixed anomaly. In terms of boundary conditions, this corresponds to imposing Dirichlet boundary condition to all of $a,b$ and $c$. The Lagrangian algebra associated with this boundary condition is given by, 
\begin{equation}
    \mathcal{L}_{\Z^{(c)}_2 \times \Z^{(a)}_2 \times \Z^{(b)}_2} = 1  \oplus U_a \oplus U_b \oplus U_c \oplus U_{ab}  \oplus U_{bc}  \oplus U_{ac}  \oplus U_{abc}
\end{equation}
Moreover, there is another obvious candidate, which is to gauge the entire $Q_8$ symmetry, which gives rise to the non-invertible $Rep(Q_8)$ symmetry. This requires exchanging the Neumann boundary conditions with Dirichlet and vice versa. The corresponding Lagrangian algebra is given by, 
\begin{equation}
    \mathcal{L}_{Rep(Q_8)} = 1 \oplus U_c \oplus \hat{U}_a \oplus \hat{U}_b \oplus \hat{U}_{ab}
\end{equation}
Beyond these, there are three $\Z_4$ subgroups of $Q_8$, hence the remaining Lagrangian algebras, 
\begin{equation}\label{eqn:lag_q4_z4}
    \begin{aligned}
        \mathcal{L}_1 &= 1 \oplus U_c \oplus U_b \oplus U_{bc} \oplus 2\hat{U}_{a} \\
        \mathcal{L}_2 &= 1 \oplus U_c \oplus U_a \oplus U_{ac} \oplus 2\hat{U}_{b} \\
        \mathcal{L}_3 &= 1 \oplus U_c \oplus U_{ab} \oplus U_{abc} \oplus 2\hat{U}_{ab}
    \end{aligned}
\end{equation}
must correspond to gauging these $\Z_4$ subgroups. This can be verified from the bulk anyon fusion rules. The $Q_8$ symmetry was realized upon choosing the Lagrangian algebra $\mathcal{L}_{Q_8}$, where we chose to condense $U_a$, $U_b$, $U_{ab}$ and $\widehat{U}_c$. The anyons, $\widehat{U}_a$, $\widehat{U}_b$ and $U_c$ remain topological or symmetry generators. In order to gauge a symmetry, we need to terminate/condense those anyons on the symmetry boundary; when we go from $\mathcal{L}_{Q_8}$ to $\mathcal{L}_i$, $i=1,2,3$, we are condensing $U_c$ and $\widehat{U}_a$ (or $\widehat{U}_b$, $\widehat{U}_{ab}$). The line operator $U_c$ ($([-1],\mathbf{1})$ in the Drinfeld double notation) gives rise to the central $\Z_2$ of $Q_8$ under the projection onto the symmetry boundary. In order to see the full $\Z_4$, we need to compute $\hat{U}_a \otimes \hat{U}_a$. In the Drinfeld double notation this is the following:
\begin{equation}
    ([j],\rho_0)^2=([1],\mathbf{1})+([1],\pi_b)+([-1],\mathbf{1})+([-1],\pi_b).
\end{equation}
In the $\Z_2$ language this translates to, 
\begin{equation}
    \hat{U}_a \otimes \hat{U}_a = 1 \oplus U_c \oplus U_b \oplus U_{bc},
\end{equation}
which can be checked explicitly, 
\begin{equation}
    \begin{aligned}
        \hat{U}_a \otimes \hat{U}_a = \int \mathcal{D}\phi \mathcal{D}\phi' \exp\left( i\pi \oint \left( 2\hat{a} + \phi_{b1} c - \phi_{c1}b - \phi_{b1}d\phi_{c1} + \phi_{b2}c - \phi_{c2}b - \phi_{b2}d\phi_{c2} \right) \right)
    \end{aligned}
\end{equation}
we do the following field redefinitions: 
\begin{equation}
    \begin{aligned}
        \phi_{b1} &= \phi'_{b} \qquad \phi_{c1} = \phi'_{c} \\
         (\phi_{b1} + \phi_{b2}) &= \phi_b \qquad (\phi_{c1} + \phi_{c2}) = \phi_c
    \end{aligned}
\end{equation}
\begin{equation}
\begin{aligned}
    \hat{U}_a \otimes \hat{U}_a &= \int \mathcal{D}\phi \mathcal{D}\phi' \exp\left( i \pi \oint \left( c + \phi_{b} c - \phi_{c}b - \phi_b d\phi_c + \phi_b d\phi_c' + \phi_b' d\phi_c - 2 \phi_b' d\phi_c' \right) \right) \\
    &= \sum_{\phi_b,\phi_c=0,1}\exp\left(i \pi \oint \left( (1+\phi_b)c-\phi_cb \right)\right)\\
    &= 1 \oplus U_b \oplus U_c \oplus U_{bc}.
\end{aligned}
\end{equation}
Projecting this configuration on the symmetry boundary, we see that $ \hat{U}^2_a = 2(1+U_c)$, which is consistent with a $\Z_4$ symmetry on the boundary if $\widehat{U}_a$ projects to the $\Z_4$ generator line plus its inverse, while $U_c$ corresponds to the generator line squared.  Alternatively, we can note that the $\Z_4$ subgroups of $Q_8$ are formed by extensions of $\Z_2^{(a)}$, $\Z_2^{(b)}$ or $\Z_2^{(ab)}$ by $\widehat{\Z}_2^{(c)}$.

In the same spirit, we can compute the following fusions: 
\begin{equation}
    \begin{aligned}
        ([i],\rho_0)^2 &= ([1],\mathbf{1})+([1],\pi_a)+([-1],\mathbf{1})+([-1],\pi_a) \\
        ([k],\rho_0)^2 &= ([1],\mathbf{1})+([1],\pi_c)+([-1],\mathbf{1})+([-1],\pi_c)
    \end{aligned}
\end{equation}
These would translate to the following in the $\Z^3_2$ notation: 
\begin{equation}
    \begin{aligned}
       \hat{U}_b \otimes \hat{U}_b &= 1 \oplus U_a \oplus U_c \oplus U_{ac} \\
       \hat{U}_{ab} \otimes \hat{U}_{ab} &=  1 \oplus U_c \oplus U_{ab} \oplus U_{abc}
    \end{aligned}
\end{equation}
we can argue like before that these are indeed the three $\Z_4$ subgroups of $Q_8$. Hence, changing the Lagrangian algebra from $\mathcal{L}_{Q_8}$ to \eqref{eqn:lag_q4_z4} corresponds to gauging one of the three $\Z_4$ subgroups.

\subsubsection*{Variation of the action}

We can check that the boundary variation of the action cancels under Lagrangian algebras specified above.
\begin{equation}
    \begin{aligned}
    \Delta S_{Q_8} = \int \left(\vphantom{\frac{1}{2}} \Delta \hat{a} \delta a + \hat{a} \delta \Delta a + \Delta \hat{b} \delta b + \hat{b} \delta \Delta b + \Delta \hat{c} \delta c + \hat{c} \delta \Delta c + \Delta a bc + a \Delta b c + a b \Delta c \right. \\ \left. + \frac{1}{2} \Delta c \delta b + \frac{1}{2} c \delta \Delta b + \frac{1}{2} \Delta c \delta a + \frac{1}{2} c \Delta \delta a \right).
    \end{aligned}
\end{equation}
We do an integration by parts, 
\begin{equation}
\begin{aligned}
    \Delta S_{Q_8} = \int  \left(\vphantom{\frac{1}{2}} \Delta \hat{a} \delta a + \Delta \hat{b} \delta b + \Delta \hat{c} \delta c -  \delta \hat{a} \Delta a -  \delta \hat{b} \Delta b -  \delta \hat{c} \Delta c + \Delta a bc + a \Delta b c + a b \Delta c \right. \\
    \left. + \frac{1}{2} \Delta c \delta b + \frac{1}{2} \Delta c \delta a - \frac{1}{2} \delta c  \Delta b - \frac{1}{2} \delta c  \Delta a \right) \\
    +\int_{\partial X_3} \left( \hat{a} \Delta a +  \hat{b} \Delta b +  \hat{c} \Delta c + \frac{1}{2} c \Delta b + \frac{1}{2} c \Delta a \right).
\end{aligned}    
\end{equation}
We want to impose boundary conditions that ensure the boundary variations vanish,
\begin{equation}\label{eqn:boun_var_q8}
    \partial S_{Q_8} = \int_{\partial X_3} \left( \hat{a} \Delta a +  \hat{b} \Delta b +  \hat{c} \Delta c + \frac{1}{2} c \Delta b + \frac{1}{2} c \Delta a \right).
\end{equation}
One obvious choice is the following: 
\begin{equation}
   \Delta a =0, \Delta b = 0, \Delta c =0, \qquad  a,b,c \rightarrow \text{Dirichlet}
\end{equation}
which corresponds to the $\mathcal{L}_{\Z^{(c)}_2 \times \Z^{(a)}_2 \times \Z^{(b)}_2}$. We can add the counterterm $c\hat{c}$ to the boundary variation in \eqref{eqn:boun_var_q8}, which would modify the boundary variation to the following: 
\begin{equation}
    \partial S_{Q_8} = \int_{\partial X_3} \left( \hat{a} \Delta a +  \hat{b} \Delta b +  c \Delta \hat{c} + \frac{1}{2} c \Delta b + \frac{1}{2} c \Delta a \right)
\end{equation}
which vanishes upon imposing Dirichlet boundary condition on $\hat{c}, b, a$ and Neumann boundary condition on $c, \hat{b}$ and $\hat{a}$, giving us another consistent set of boundary condition associated with $Q_8$ symmetry or trivial gauging. This boundary condition is associated with the Lagrangian algebra $\mathcal{L}_{Q_8}$.

The boundary condition associated with the Lagrangian algebra $\mathcal{L}_{\Z_{4,1}}$ correspond to imposing Dirichlet boundary condition on $c,b$ and $\hat{a}$, it is not too difficult to see that the boundary variation \eqref{eqn:boun_var_q8} vanishes once we add the counter term $a\hat{a}$ and impose this condition (the last term vanishes when we impose the Dirichlet condition $c=0$). The Lagrangian algebra $\mathcal{L}_{\Z_{4,2}}$ works out in the same manner.  For $\mathcal{L}_{\Z_{4,3}}$ we impose Dirichlet conditions on $c$ and $\hat{a}$, and then either impose the condition that $a=b$ at the boundary, or equivalently we make a field redefinition $b'=a+b$ and impose Dirichlet boundary conditions on $b'$ (along with Neumann conditions on $\hat{c}$, $\hat{b}'=\hat{a}+\hat{b}$, and $a$).  For $\mathcal{L}_{Rep(Q_8)}$, we impose Dirichlet conditions on $\hat{a}$, $\hat{b}$, and $c$ (setting $c=0$).  Along with an added $a\hat{a}+b\hat{b}$ term on the boundary, the variation can be seen to vanish.

We have identified the symmetry category associated with three of the Lagrangian algebras. In addition we have also emphasized how these are connected by some discrete gaugings. We did not identify the quantum symmetry for the three of these. It would be interesting to come back in the future and pin down the quantum symmetries associated with the gaugings of $\Z_4$ subgroups of $Q_8$. There is no $Q_8$ symmetry web in the space of $c=1$ theories. Although we can find $Q_8$ symmetry webs in 3d orthosymplectic gauge theories \cite{Bhardwaj:2022maz,Bergman:2024its}.

\subsection{Club Sandwich, igSPT and anomaly resolution}

In this section, we identify the igSPT phases associated with different Lagrangian algebras of the Drinfeld center of $Q_8$, which in turn will allow us to identify the categorical short exact sequences associated with anomaly resolution. The identification of gapped SPT phases and igSPT phases depend on the choice of Lagrangian algebra on the symmetry boundary. We have derived all the condensable algebras earlier, we tabulate them below. 
\begin{table}[H]
\centering
\caption{Condensable algebras in $\mathcal{Z}(Q_8)$}
\label{tab:Q8}
\rotatebox{90}{%
  \begin{minipage}{\textheight}
  \centering
  \begin{adjustbox}{max width=\textheight}
  \begin{tabular}{|c|c|c|c|}
  \hline
Dim. & Label & Con. Algebra in $\mathcal{Z}(Q_8)$ & in the $\Z^3_2$ notation \\
\hline
1 & $\mathcal{A}_1$ & $([1],\mathbf{1})$ & $1$ \\
2 & $\mathcal{A}_{2,1}$ & $([1],\mathbf{1})\oplus([1],\pi_a)$ & $1 \oplus U_a$ \\
2 & $\mathcal{A}_{2,2}$ & $([1],\mathbf{1})\oplus([1],\pi_b)$ & $1 \oplus U_b$ \\
2 & $\mathcal{A}_{2,3}$ & $([1],\mathbf{1})\oplus([1],\pi_c)$ & $1 \oplus U_{ab}$ \\
2 & $\mathcal{A}_{2,4}$ & $([1],\mathbf{1})\oplus([-1],\mathbf{1})$ & $1 \oplus U_c$ \\
2 & $\mathcal{A}_{2,5}$ & $([1],\mathbf{1})\oplus([-1],\pi_a)$ & $1 \oplus U_{ac}$ \\
2 & $\mathcal{A}_{2,6}$ & $([1],\mathbf{1})\oplus([-1],\pi_b)$ & $1 \oplus U_{bc}$ \\
2 & $\mathcal{A}_{2,7}$ & $([1],\mathbf{1})\oplus([-1],\pi_c)$ & $1 \oplus U_{abc}$ \\
4 & $\mathcal{A}_{4,1}$ & $([1],\mathbf{1})\oplus([1],\pi_a)\oplus([1],\pi_m)$ & $1 \oplus U_a \oplus \hat{U}_c$ \\
4 & $\mathcal{A}_{4,2}$ & $([1],\mathbf{1})\oplus([1],\pi_b)\oplus([1],\pi_m)$ & $1 \oplus U_b \oplus \hat{U}_c$ \\
4 & $\mathcal{A}_{4,3}$ & $([1],\mathbf{1})\oplus([1],\pi_c)\oplus([1],\pi_m)$ & $1 \oplus U_{ab} \oplus \hat{U}_c$ \\
4 & $\mathcal{A}_{4,4}$ & $([1],\mathbf{1})\oplus([1],\pi_a)\oplus([i],\rho_0)$ & $1 \oplus U_a \oplus \hat{U}_b$ \\
4 & $\mathcal{A}_{4,5}$ & $([1],\mathbf{1})\oplus([-1],\mathbf{1})\oplus([i],\rho_0)$ & $1 \oplus U_c \oplus \hat{U}_b$ \\
4 & $\mathcal{A}_{4,6}$ & $([1],\mathbf{1})\oplus([-1],\pi_a)\oplus([i],\rho_0)$ & $1 \oplus U_{ac} \oplus \hat{U}_b$ \\
4 & $\mathcal{A}_{4,7}$ & $([1],\mathbf{1})\oplus([1],\pi_b)\oplus([j],\rho_0)$ & $1 \oplus U_b \oplus \hat{U}_a$ \\
4 & $\mathcal{A}_{4,8}$ & $([1],\mathbf{1})\oplus([-1],\mathbf{1})\oplus([j],\rho_0)$ & $1 \oplus U_c \oplus \hat{U}_a$ \\
4 & $\mathcal{A}_{4,9}$ & $([1],\mathbf{1})\oplus([-1],\pi_b)\oplus([j],\rho_0)$ & $1 \oplus U_{bc} \oplus \hat{U}_a$ \\
4 & $\mathcal{A}_{4,10}$ & $([1],\mathbf{1})\oplus([1],\pi_c)\oplus([k],\rho_0)$ & $1 \oplus U_{ab} \oplus \hat{U}_{ab}$ \\
4 & $\mathcal{A}_{4,11}$ & $([1],\mathbf{1})\oplus([-1],\mathbf{1})\oplus([k],\rho_0)$ & $1 \oplus U_c \oplus \hat{U}_{ab}$ \\
4 & $\mathcal{A}_{4,12}$ & $([1],\mathbf{1})\oplus([-1],\pi_c)\oplus([k],\rho_0)$ & $1 \oplus U_{abc} \oplus \hat{U}_{ab}$ \\
4 & $\mathcal{A}_{4,13}$ & $([1],\mathbf{1})\oplus([1],\pi_a)\oplus([1],\pi_b)\oplus([1],\pi_c)$ & $1 \oplus U_a \oplus U_b \oplus U_{ab}$ \\
4 & $\mathcal{A}_{4,14}$ & $([1],\mathbf{1})\oplus([1],\pi_a)\oplus([-1],\mathbf{1})\oplus([-1],\pi_a)$ & $1 \oplus U_a \oplus U_c \oplus U_{ac}$ \\
4 & $\mathcal{A}_{4,15}$ & $([1],\mathbf{1})\oplus([1],\pi_b)\oplus([-1],\mathbf{1})\oplus([-1],\pi_b)$ & $1 \oplus U_b \oplus U_c \oplus U_{bc}$ \\
4 & $\mathcal{A}_{4,16}$ & $([1],\mathbf{1})\oplus([1],\pi_c)\oplus([-1],\mathbf{1})\oplus([-1],\pi_c)$ & $1 \oplus U_{ab} \oplus U_c \oplus U_{abc}$ \\
4 & $\mathcal{A}_{4,17}$ & $([1],\mathbf{1})\oplus([1],\pi_a)\oplus([-1],\pi_b)\oplus([-1],\pi_c)$ & $1 \oplus U_a \oplus U_{bc} \oplus U_{abc}$ \\
4 & $\mathcal{A}_{4,18}$ & $([1],\mathbf{1})\oplus([1],\pi_b)\oplus([-1],\pi_a)\oplus([-1],\pi_c)$ & $1 \oplus U_b \oplus U_{ac} \oplus U_{abc}$ \\
4 & $\mathcal{A}_{4,19}$ & $([1],\mathbf{1})\oplus([1],\pi_c)\oplus([-1],\pi_a)\oplus([-1],\pi_b)$ & $1 \oplus U_{ab} \oplus U_{ac} \oplus U_{bc}$ \\
8 & $\mathcal{A}_{8,1}$ & $([1],\mathbf{1})\oplus([1],\pi_a)\oplus([1],\pi_b)\oplus([1],\pi_c)\oplus2([1],\pi_m)$ & $1\oplus U_{a} \oplus U_{b} \oplus U_{ab} \oplus 2\hat{U}_{c}$ \\
8 & $\mathcal{A}_{8,2}$ & $([1],\mathbf{1})\oplus([1],\pi_a)\oplus([-1],\mathbf{1})\oplus([-1],\pi_a)\oplus2([i],\rho_0)$ & $1\oplus U_{a} \oplus U_{c} \oplus U_{ac} \oplus 2\hat{U}_{b}$ \\
8 & $\mathcal{A}_{8,3}$ & $([1],\mathbf{1})\oplus([1],\pi_b)\oplus([-1],\mathbf{1})\oplus([-1],\pi_b)\oplus2([j],\rho_0)$ & $1\oplus U_{b} \oplus U_{c} \oplus U_{bc} \oplus 2\hat{U}_{a}$ \\
8 & $\mathcal{A}_{8,4}$ & $([1],\mathbf{1})\oplus([1],\pi_c)\oplus([-1],\mathbf{1})\oplus([-1],\pi_c)\oplus2([k],\rho_0)$ & $1\oplus U_{ab} \oplus U_{c} \oplus U_{abc} \oplus 2\hat{U}_{ab}$ \\
8 & $\mathcal{A}_{8,5}$ & $([1],\mathbf{1})\oplus([-1],\mathbf{1})\oplus([i],\rho_0)\oplus([j],\rho_0)\oplus([k],\rho_0)$ & $1\oplus U_{c} \oplus \hat{U}_{a} \oplus \hat{U}_{b} \oplus \hat{U}_{ab}$ \\
8 & $\mathcal{A}_{8,6}$ & $([1],\mathbf{1})\oplus([1],\pi_a)\oplus([1],\pi_b)\oplus([1],\pi_c)\oplus([-1],\mathbf{1})\oplus([-1],\pi_a)\oplus([-1],\pi_b)\oplus([-1],\pi_c)$ & $1 \oplus U_a \oplus U_b \oplus U_c \oplus U_{ab} \oplus U_{bc} \oplus U_{ca} \oplus U_{abc}$ \\
\hline
\end{tabular}
\end{adjustbox}
\end{minipage}%
}
\end{table}
Gapped SPT phases can be identified by looking for Lagrangian algebras with trivial intersection with the symmetry Lagrangian algebra. Looking at our table, we can see that $\mathcal{A}_{8,6}$ does not have trivial overlap with any other Lagrangian algebra, hence there is no SPT phase associated with this symmetry lagragian algebra. Moving onto $\mathcal{A}_{8,5}$ as our symmetry lagrangian algebra, we see that this Lagrangian algebra has a trivial overlap with $\mathcal{A}_{8,1}$, hence this would lead us to an SPT phase. This is obvious, since $\mathcal{A}_{8,5}$ is the Lagrangian algebra associated with $\text{Rep}(Q_8)$ symmetry and $\mathcal{A}_{8,1}$ leads us to $Q_8$ symmetry. None of the other Lagrangian algebra will have trivial overlap among themselves, that leads us to the following table:
\begin{table}[H]
    \centering
    \begin{tabular}{|c|c|}
    \hline
    Symmetry Lagrangian algebra & SPT \\
    \hline
    $\mathcal{A}_{8,1}$ & $\mathcal{A}_{8,5}$ \\
    \hline
    $\mathcal{A}_{8,2}$ & -\\
    \hline
    $\mathcal{A}_{8,3}$ & -\\
    \hline
    $\mathcal{A}_{8,4}$ & -\\
    \hline
    $\mathcal{A}_{8,5}$ & $\mathcal{A}_{8,1}$ \\
    \hline
    $\mathcal{A}_{8,6}$ & - \\
    \hline
    \end{tabular}
    \caption{SPT phases and Lagrangian algebras}
\end{table}
Next, we detect the gapless SPT (gSPT) phases, which can be identified by hunting down the condensable algebras having trivial overlap with symmetry Lagrangian algebra.
\begin{table}[H]
    \centering
    \resizebox{\textwidth}{!}{%
    \begin{tabular}{|c|c|c|}
    \hline
    Symmetry Lagrangian algebra & SPT & gSPT \\
    \hline
    $\mathcal{A}_{8,1}$ & $\mathcal{A}_{8,5}$ & $\mathcal{A}_{2,4}, \mathcal{A}_{2,5} \mathcal{A}_{2,6},\mathcal{A}_{2,7}, \mathcal{A}_{4,5},\mathcal{A}_{4,6},\mathcal{A}_{4,8}, \mathcal{A}_{4,9}, \mathcal{A}_{4,11}, \mathcal{A}_{4,12}$ \\
    \hline
    $\mathcal{A}_{8,2}$ & - & $\mathcal{A}_{2,2},\mathcal{A}_{2,3},\mathcal{A}_{2,6},\mathcal{A}_{2,7}, \mathcal{A}_{4,2}, \mathcal{A}_{4,3}, \mathcal{A}_{4,7}, \mathcal{A}_{4,9}, \mathcal{A}_{4,10}, \mathcal{A}_{4,12} $\\
    \hline
    $\mathcal{A}_{8,3}$ & - & $\mathcal{A}_{2,1}, \mathcal{A}_{2,3}, \mathcal{A}_{2,5}, \mathcal{A}_{2,7}, \mathcal{A}_{4,1}, \mathcal{A}_{4,3}, \mathcal{A}_{4,4}, \mathcal{A}_{4,6},\mathcal{A}_{4,10},\mathcal{A}_{4,12}$\\
    \hline
    $\mathcal{A}_{8,4}$ & - & $\mathcal{A}_{2,1}, \mathcal{A}_{2,2}, \mathcal{A}_{2,5}, \mathcal{A}_{2,6}, \mathcal{A}_{4,1}, \mathcal{A}_{4,2}, \mathcal{A}_{4,4}, \mathcal{A}_{4,6}, \mathcal{A}_{4,7}, \mathcal{A}_{4,9}$\\
    \hline
    $\mathcal{A}_{8,5}$ & $\mathcal{A}_{8,1}$ & $ \mathcal{A}_{2,1}, \mathcal{A}_{2,2}, \mathcal{A}_{2,3}, \mathcal{A}_{2,5}, \mathcal{A}_{2,6}, \mathcal{A}_{2,7}, \mathcal{A}_{4,1},\mathcal{A}_{4,2}, \mathcal{A}_{4,3}, \mathcal{A}_{4,13}, \mathcal{A}_{4,17}, \mathcal{A}_{4,18},\mathcal{A}_{4,19} $\\
    \hline
    $\mathcal{A}_{8,6}$ & - & - \\
    \hline
    \end{tabular}}
\end{table}
Now we are in a position to identify the igSPT phases. These are associated with the condensable algebras that has trivial overlaps with the symmetry Lagrangian algebra and they are not subalgebras of the Lagrangian algebras leading to gapped SPT phases. The igSPT phase associated with $Q_8$ symmetry ($\mathcal{A}_{8,1}$) is given by $\mathcal{A}_{2,5}, \mathcal{A}_{2,6}, \mathcal{A}_{2,7}, \mathcal{A}_{4,6},\mathcal{A}_{4,9},\mathcal{A}_{4,12}$. Similarly, the igSPT phases associated with $\text{Rep}(Q_8)$ symmetry ($\mathcal{A}_{8,5}$) are $\mathcal{A}_{2,5}, \mathcal{A}_{2,6}, \mathcal{A}_{2,7}, \mathcal{A}_{4,17}, \mathcal{A}_{4,18},\mathcal{A}_{4,19}$. To the best of our knowledge, the igSPT phases for $Q_8$ or $\text{Rep}(Q_8)$ symmetry has not been discussed before. 

Our next target would be to associate these igSPT phases with the anomaly resolution story discussed earlier in the context of the dihedral group. In order to identify the categorical anomaly resolution sequences, we will need the information of reduced topological order associated with each condensable algebra, which we have derived explicitly in appendix \ref{app:reduced_topological_order}, we just list the results below.
\begin{table}[H]
    \centering
    \begin{tabular}{|c|c|}
    \hline
    Condensable Algebra & Reduced Topological order \\
    \hline
       $\mathcal{A}_{2,5}$  & $\Z^\omega_2 \times \Z^\omega_2$ \\
       \hline
        $ \mathcal{A}_{2,6}$ & $\Z^\omega_2 \times \Z^\omega_2$\\
        \hline
        $ \mathcal{A}_{2,7}$ &  $\Z^\omega_2 \times \Z^\omega_2$\\
        \hline
        $ \mathcal{A}_{4,6}$ & $\Z^\omega_2$ \\
        \hline
        $\mathcal{A}_{4,9}$ & $\Z^\omega_2$ \\
        \hline
        $\mathcal{A}_{4,12}$ & $\Z^\omega_2$ \\
        \hline
        $ \mathcal{A}_{4,17}$ & $\Z^\omega_2$ \\
        \hline
        $\mathcal{A}_{4,18}$ & $\Z^\omega_2$ \\
        \hline
        $\mathcal{A}_{4,19}$ & $\Z^\omega_2$ \\
        \hline
    \end{tabular}
    \caption{Reduced topological order associated with different condensable algebras of $\mathcal{Z}(Q_8)$ which leads to an igSPT.}
\end{table}
The information of the reduced topological order immediately allows us to associate a club sandwich with each condensable algebra. 
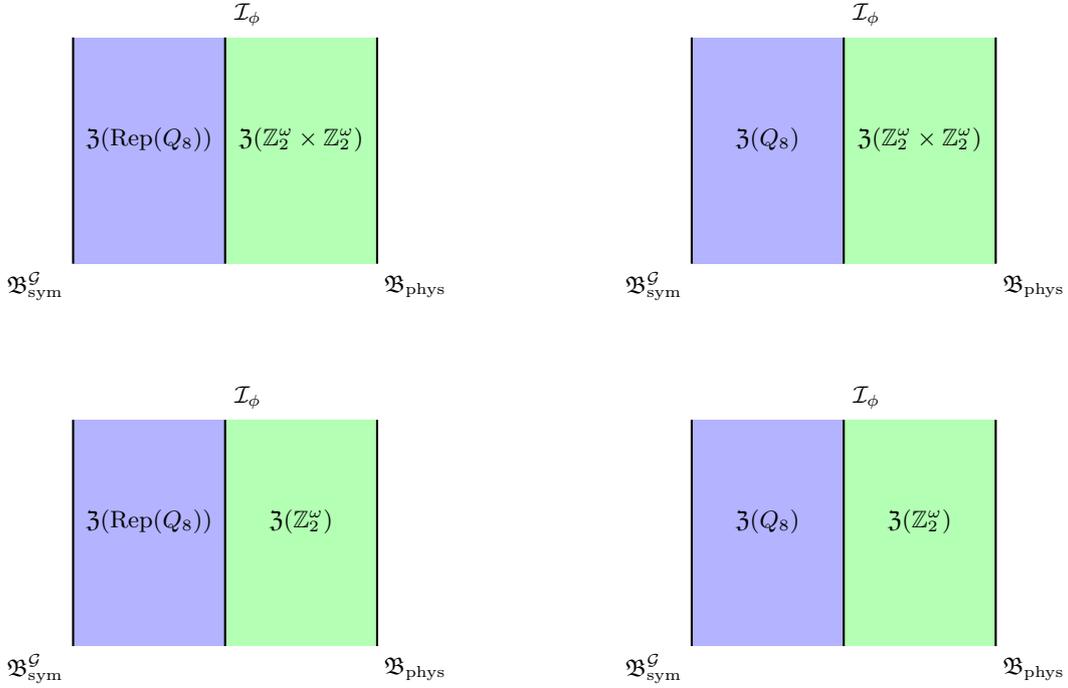
\begin{figure}[H]
    \centering
    \begin{minipage}[t]{0.48\textwidth}
        \centering
        \begin{tikzpicture}[scale=1]
            \fill[blue!30] (0,0) rectangle (2,3); 
            \fill[green!30] (2,0) rectangle (4,3); 
            
            \draw[thick, black] (2,0) -- (2,3);
            
            \draw[thick] (0,0) -- (0,3);
            \draw[thick] (4,0) -- (4,3);
            
            \node at (-0.5,-0.3) {\footnotesize{$\mathfrak{B}_\text{sym}$}};
            \node at (4.5,-0.3) {\footnotesize{$\mathfrak{B}_\text{phys}$}};
            \node at (2.3,3.3) {\footnotesize{$\mathcal{I}_\phi$}};
            \node at (1.0, 1.65) {\footnotesize{$\mathcal{Z}(\text{Rep}({Q}_8)$)}};
            \node at (3.0, 1.65) {\footnotesize{$\mathcal{Z}(\Z^\omega_2 \times \Z^\omega_2)$}};
        \end{tikzpicture}
    \end{minipage}
    \hfill
    \begin{minipage}[t]{0.48\textwidth}
        \centering
        \begin{tikzpicture}[scale=1]
            \fill[blue!30] (0,0) rectangle (2,3);
            \fill[green!30] (2,0) rectangle (4,3);
            
            \draw[thick, black] (2,0) -- (2,3);
            
            \draw[thick] (0,0) -- (0,3);
            \draw[thick] (4,0) -- (4,3);
            
            \node at (-0.5,-0.3) {\footnotesize{$\mathfrak{B}_\text{sym}$}};
            \node at (4.5,-0.3) {\footnotesize{$\mathfrak{B}_\text{phys}$}};
            \node at (2.3,3.3) {\footnotesize{$\mathcal{I}_\phi$}};
            \node at (1.0, 1.65) {\footnotesize{$\mathcal{Z}(\text{Vec}_{{Q}_8})$}};
            \node at (3.0, 1.65) {\footnotesize{$\mathcal{Z}({\mathbb{Z}}^\omega_2 \times \mathbb{Z}^\omega_2 )$}};
        \end{tikzpicture}
    \end{minipage}
\end{figure}
\begin{figure}[H]
    \centering
    \begin{minipage}[t]{0.48\textwidth}
        \centering
        \begin{tikzpicture}[scale=1]
            \fill[blue!30] (0,0) rectangle (2,3); 
            \fill[green!30] (2,0) rectangle (4,3); 
            
            \draw[thick, black] (2,0) -- (2,3);
            
            \draw[thick] (0,0) -- (0,3);
            \draw[thick] (4,0) -- (4,3);
            
            \node at (-0.5,-0.3) {\footnotesize{$\mathfrak{B}_\text{sym}$}};
            \node at (4.5,-0.3) {\footnotesize{$\mathfrak{B}_\text{phys}$}};
            \node at (2.3,3.3) {\footnotesize{$\mathcal{I}_\phi$}};
            \node at (1.0, 1.65) {\footnotesize{$\mathcal{Z}(\text{Rep}({Q}_8)$)}};
            \node at (3.0, 1.65) {\footnotesize{$\mathcal{Z}({\mathbb{Z}}^\omega_2)$}};
        \end{tikzpicture}
    \end{minipage}
    \hfill
    \begin{minipage}[t]{0.48\textwidth}
        \centering
        \begin{tikzpicture}[scale=1]
            \fill[blue!30] (0,0) rectangle (2,3);
            \fill[green!30] (2,0) rectangle (4,3);
            
            \draw[thick, black] (2,0) -- (2,3);
            
            \draw[thick] (0,0) -- (0,3);
            \draw[thick] (4,0) -- (4,3);
            
            \node at (-0.5,-0.3) {\footnotesize{$\mathfrak{B}_\text{sym}$}};
            \node at (4.5,-0.3) {\footnotesize{$\mathfrak{B}_\text{phys}$}};
            \node at (2.3,3.3) {\footnotesize{$\mathcal{I}_\phi$}};
            \node at (1.0, 1.65) {\footnotesize{$\mathcal{Z}(\text{Vec}_{{Q}_8})$}};
            \node at (3.0, 1.65) {\footnotesize{$\mathcal{Z}({\Z}^\omega_2)$}};
        \end{tikzpicture}
    \end{minipage}
\end{figure}
The club sandwich essentially captures the notion of missing charges from the anomalous theory. We can pin down the categorical short exact sequences associated with anomaly resolutions. There are six igSPT's associated with $Q_8$, three of which are resolving an anomalous $\Z_2$ and the remaining three corresponds to resolving an anomalous $\Z_2 \times \Z_2$. The dimension four condensable algebras lead us to the following short exact sequence:
\begin{equation}
    0 \rightarrow \text{Vec}(\Z_4) \rightarrow \text{Vec}(Q_8) \rightarrow \text{Vec}(\Z_2,\omega) \rightarrow 0.
\end{equation}
We have identified the $\Z_4$ by computing the fusion of the elements in the condensable algebra on the symmetry boundary. The club sandwich tells that the line operators associated with the $\Z_4 \subset Q_8$ terminates on the interface, hence the subgroup acts trivially. The three cases basically correspond to three different $\Z_4$ subgroups of $Q_8$. This anomaly resolution was discussed in the context of 2d orbifolds in \cite[Section 5]{Robbins:2021lry}. Moving onto dimension two condensable algebra, we have the following sequence:
\begin{equation}
    0 \rightarrow \text{Vec}(\Z_2)  \rightarrow \text{Vec}(Q_8) \rightarrow \text{Vec}(\Z_2 \times \Z_2,\omega') \rightarrow 0,
\end{equation}
where $\omega'=\pi_1^\ast\omega_1+\pi_2^\ast\omega_2$ and $\pi_i:\Z_2^2\rightarrow\Z_{2,i}^{\omega}$ is projection onto either of the $\Z_2$ factors.  We can also say that $\text{Vec}(\Z_2^2,\omega')\cong\text{Vec}(\Z_2,\omega)\boxtimes\text{Vec}(\Z_2,\omega)$.
A similar sequence was discussed in \cite[Section 4]{Robbins:2021xce} from the point of view of absolute theory. Moving onto the igSPTs associated with $\text{Rep}(Q_8)$. Once again we have six igSPTs, three (dimension four condensable algebras) of which corresponds to resolving an anomalous $\Z_2$, 
\begin{equation}
    0 \rightarrow \text{Vec}(\Z_2 \times \Z_2) \rightarrow \text{Rep}(Q_8) \rightarrow \text{Vec}(\Z_2,\omega) \rightarrow 0 
\end{equation}
once agian, we can identify the trivially acting subgroup by computing the fusion of elements participating in the condensable algebra. This anomaly resolution sequence has not been discussed before. The remaining three (dimension two condensable algebras) correspond to resolving an anomalous $\Z_2 \times \Z_2$. 
\begin{equation}
    0 \rightarrow \text{Vec}(\Z_2)  \rightarrow \text{Rep}(Q_8) \rightarrow \text{Vec}(\Z_2 \times \Z_2,\omega') \rightarrow 0 
\end{equation}
This sequence was described in \cite[Section 7]{Perez-Lona:2025ncg}.

\section{Conclusion and Outlook}\label{sec:conclusion}

In this note, we have analyzed the SymTFT for certain non-invertible symmetries. In the examples we have discussed, we have an explicit lagrangian description of the SymTFT. We have also reviewed the tools required to derive the list of condensable algebras of the Drinfeld center. Moreover, we have exemplified this with a detailed derivation of the condensable algebras of the Drinfeld center of $Q_8$. 

We have discussed the Symmetry TFT for $\text{Rep}(D_4)$ and $\text{Rep}(Q_8)$, beyond these, there is another candidate of the $\text{TY}(\Z_2 \times \Z_2)$ family -- $\text{Rep}(\mathcal{H}_8)$, which differs from the $\text{Rep}(D_4)$ and $\text{Rep}(Q_8)$ via the choice of associators and F-symbols\cite{TAMBARA1998692}. The next immediate task would be to derive the full list of condensable algebras and understand the different gaugings of $\text{Rep}(\mathcal{H}_8)$ from the SymTFT, connecting it with the results of \cite{Perez-Lona:2023djo,Choi:2023vgk}. This is very interesting because there is a symmetry web associated with $\text{Rep}(\mathcal{H}_8)$ in the space of $c=1$ theories. 

We have demonstrated that how we can associate the story of anomaly resolution by extending the symmetry is intimately related with the igSPTs and idea of missing charges. We have identified the fusion categorical short exact sequences associated with $D_4$ and $Q_8$. One of the obvious future directions would be to identify the igSPT phases for $\text{Rep}(\mathcal{H}_8)$. This would allow us to discover more fusion categorical anomaly resolutions.

Finally, as emphasized in the introduction the story of anomaly resolution and trivially acting symmetries fits nicely in the framework of renormalization group flows. 
It would be interesting to study the examples with trivially acting symmetries discussed in the text and construct models where the full symmetry acts effectively in the UV but where the operators charged under trivially acting symmetry go away  under renormalization group flow.

More recently, this whole picture has been extended to include emergent symmetries in \cite{Antinucci:2025fjp}. The anomaly of an UV symmetry is trivialized in the IR in the presence of an emergent 1-form symmetry. It would be interesting to incorporate trivially acting symmetries and emergent symmetries together and study the fate of the anomaly under the renormalization group flow. We leave this for future work.  

\subsubsection*{Acknowledgments}

We are grateful to Eric Sharpe, Xingyang Yu, Alonso Perez-Lona and Thomas Vandermuelen for reading the draft carefully.
We further thank Ling Lin for helpful discussions. This work was partially supported by SUNY Research Foundation.

\appendix

\section{Group extension basics}

In this appendix, we briefly discuss the cocyle conditions due to group extension, largely following \cite{Tachikawa:2017gyf}. Consider the following extension:
\begin{equation}\label{eqn:group_extension}
    0 \rightarrow A \rightarrow \Gamma \rightarrow G= \Gamma/A \rightarrow 0.
\end{equation}
$\Gamma$ is anomaly free. The extension we have is nontrivial, classified by $e\in H^2(G,A)$. As a set $\Gamma$ can be thought off as $A\times G$. Since the extension is nontrivial, 
\begin{equation}
    (0,g) \times (0,h) = \left(e(g,h),gh\right),
\end{equation}
where $g,h \in G$ and $e(g,h) \in A$, is the 2-cocycle that defines the extension. When the extension is trivial; $e = 0$, the background gauge field $a_1$ is a 1-cocyle in $H^1(X,A)$ ($X$ is some manifold). When $e \neq 0$, then $a_1$ is no longer a cocyle rather it is a cochain valued in $C^1 (X,A)$ i.e.\cite{Tachikawa:2017gyf} 
\begin{equation}\label{eqn:extension_criteria}
    \delta a_1 = e(g_1).
\end{equation}
Here $g_1$ is the background field of $G$ which we can think of as determining a map $f$ from $X$ to the classifying space $BG$.  Since $e\in H^2(G,A)=H^2(BG,A)$, we can use the map $f$ to pull back the cocycle $e$, thereby getting an element in $H^2(X,A)$ which is the $e(g_1)$ appearing on the right-hand side of \eqref{eqn:extension_criteria}. 

Consider the simplest case of $\Z_4$, 
\begin{equation}
    0 \rightarrow \Z^A_2 \rightarrow  ~\Z_4 \rightarrow \Z^G_2 \rightarrow 0.
\end{equation}
This extension is classified by $H^2(\Z_2, \Z_2) = \Z_2$. The trivial extension class corresponds to the direct product symmetry but the non trivial extension class results in a $\Z_4$ symmetry. In such case, we have a condition, 
\begin{equation}
    \delta a_1 = \frac{1}{2} \delta g_1 , \qquad a_1 \in C^1(X, \Z^A_2).
\end{equation}
Moving onto the case of $D_4$, which is an extension of $\Z_2 \times \Z_2$ by $\Z_2$, 
\begin{equation}
    0 \rightarrow \Z_2 \rightarrow  ~D_4 \rightarrow \Z_2\times \Z_2 \rightarrow 0,
\end{equation}
classified by $H^2(\Z_2 \times \Z_2, \Z_2)$. We have a condition\cite{Franco:2024mxa}, 
\begin{equation}
    \delta a_1 = bc.
\end{equation}
Finally the story of $Q_8$, which is a non trivial extension of $\Z_2 \times \Z_2$ by $\Z_2$, 
\begin{equation}
    0 \rightarrow \Z_2 \rightarrow Q_8 \rightarrow \Z_2 \times \Z_2 \rightarrow 0.
\end{equation}
The analogue of \eqref{eqn:extension_criteria} for $Q_8$ is given by the following\cite{Kapustin:2014lwa,Kapustin:2014zva}: 
\begin{equation}\label{eqn:ext_q8}
    \delta a_1 = bc + \frac{1}{2}\delta b + \frac{1}{2}\delta c.
\end{equation}

\section{Gauge invariance of $D_4$ SymTFT}\label{app:gauge_inv_d4}

The SymTFT action that captures $D_4$ symmetry is given by the following: 
\begin{equation}
    S_{3d} = \pi \int_{3d} \left(\hat{a}\delta a + \hat{b} \delta b + \hat{c} \delta c + abc\right).
\end{equation}
The action is gauge invariant under the following gauge transformations:
\begin{equation}
    \begin{aligned}
         a \rightarrow a+ \delta \alpha, \qquad \hat{a} \rightarrow \hat{a} +\delta\hat{\alpha} - \left( \beta c - \gamma b + \beta \delta\gamma \right), \\
        b \rightarrow b+ \delta\beta, \qquad \hat{b} \rightarrow \hat{b} + \delta\hat{\beta} - \left( \gamma a - \alpha c + \gamma \delta \alpha \right),\\
        c \rightarrow c+ \delta\gamma, \qquad \hat{c} \rightarrow \hat{c} + \delta\hat{\gamma} - \left( \alpha b - \beta a + \alpha \delta\beta \right).
    \end{aligned}
\end{equation}
These have already appeared in \cite{Kaidi:2023maf,He:2016xpi}. In this appendix, we check these explicitly. 
\begin{equation}
    \begin{aligned}
        S_{3d} = \pi \int_{3d} \left(\hat{a}\delta a + \delta \hat{\alpha} \delta a - \beta c \delta a + \gamma b \delta a - \beta \delta \gamma \delta a \right.\\
        + \hat{b}\delta b + \delta \hat{\beta} \delta b 
        - \gamma a \delta b + \alpha c \delta b - \gamma \delta \alpha \delta b \\
        + \hat{c}\delta c + \delta \hat{\gamma} \delta c - \alpha b \delta c 
        + \beta a \delta c - \alpha \delta \beta \delta c \\ 
        + abc +  \delta \alpha bc + a\delta \beta c  + ab \delta \gamma
        + \delta \alpha \delta \beta c  \\ 
        \left.+ a\delta \beta \delta \gamma + \delta \alpha b \delta \gamma + \delta \alpha \delta \beta \delta \gamma \right).
    \end{aligned}
\end{equation}
Some terms, like $\delta\hat{\alpha}\delta a$ and $\delta\alpha\delta\beta\delta\gamma$ are total derivatives by themselves and can be dropped.  Next, we can see that the following terms combine to form total derivatives: 
\begin{equation}
    \begin{aligned}        a\delta \beta c - \beta c \delta a + \beta a \delta c = \delta (c\beta a), \\
        \delta \alpha bc - \alpha b \delta c + \alpha c \delta b = \delta (\alpha bc), \\
        ab\delta \gamma - \gamma a \delta b + \gamma b \delta a = \delta (ab \gamma).
    \end{aligned}
\end{equation}
So, we can ignore these terms. That leaves us with the following terms, 
\begin{equation}
    \beta \delta \gamma \delta a + a \delta \beta \delta \gamma + \gamma\delta a \delta b + \delta \alpha b \delta \gamma + \alpha \delta \beta \delta c + \delta \alpha \delta \beta c.
\end{equation}
These also form total derivatives, 
\begin{equation}
    \begin{aligned}
        \beta \delta \gamma \delta a + a \delta \beta \delta \gamma = \delta (a\beta \delta \gamma), \\
         \gamma\delta a \delta b + \delta \alpha b \delta \gamma = \delta (\delta \alpha b \gamma ), \\
          \alpha \delta \beta \delta c + \delta \alpha \delta \beta c = \delta (\alpha \delta \beta c),
    \end{aligned}
\end{equation}
which ensures the fact that the action is gauge invariant under the gauge transformations.

\section{Dictionary for Anyons}\label{app:dictionary}

In this appendix, we compare different nomenclatures we have used for the anyons in the Drinfeld center of the $D_4$ and $Q_8$. We believe this will be a useful resource.  The dictionary between $\mathcal{Z}(D_4)$ anyon notation and RGB notation has appeared previously in~\cite{Bhardwaj:2024qrf,Iqbal:2023wvm}.
\subsection*{$D_4$:}
\begin{center}
\begin{tabular}{|c|c|c|c|c|}
    \hline
    anyon & $d$ & $s$ & $\Z^3_2$ Nomenclature & RGB notation\\
    \hline
    $([1],\mathbf{1})$ & $1$ & $1$ & $1$ & 1 \\
    \hline
    $([1],\pi_a)$ & $1$ & $1$ & $U_{ab}$ & $e_{RG}$ \\
    \hline
    $([1],\pi_b)$ & $1$ & $1$ & $U_b$ & $e_R$ \\
    \hline
    $([1],\pi_c)$ & $1$ & $1$ & $U_a$ & $e_G$ \\
    \hline
    $([1],\pi_m)$ & $2$ & $1$ & $\hat{U}_c$ & $m_B$ \\
    \hline
    $([x^2],\mathbf{1})$ & $1$ & $1$ & $U_{abc}$ & $e_{RGB}$ \\
    \hline
    $([x^2],\pi_a)$ & $1$ & $1$ & $U_c$ & $e_B$ \\
    \hline
    $([x^2],\pi_b)$ & $1$ & $1$ & $U_{ac}$ & $e_{GB}$ \\
    \hline
    $([x^2],\pi_c)$ & $1$ & $1$ & $U_{bc}$ & $e_{RB}$\\
    \hline
    $([x^2],\pi_m)$ & $2$ & $-1$ & $\hat{U}_c U_c$ & $f_B$\\
    \hline
    $([x],\rho_0)$ & $2$ & $1$ & $\hat{U}_{ab}$ & $m_{RG}$\\
    \hline
    $([x],\rho_1)$ & $2$ & $i$& $\hat{U}_{abc}$ & $s_{RGB}$ \\
    \hline
    $([x],\rho_2)$ & $2$ & $-1$ & $\hat{U}_{ab}U_a = \hat{U}_{ab}U_b$& $f_{RG}$\\
    \hline
    $([x],\rho_3)$ & $2$ & $-i$ & $\hat{U}_{abc}U_a=\hat{U}_{abc}U_b=\hat{U}_{abc}U_c$ &$\bar{s}_{RGB}$\\
    \hline
    $([y],\rho_{++})$ & $2$ & $1$ & $\hat{U}_{ac}$ & $m_{GB}$\\
    \hline
    $([y],\rho_{+-})$ & $2$ & $1$ & $\hat{U}_a$ & $m_G$\\
    \hline
    $([y],\rho_{-+})$ & $2$ & $-1$ & $\hat{U}_{ac}U_a=\hat{U}_{ac}U_c$ & $f_{GB}$ \\
    \hline
    $([y],\rho_{--})$ & $2$ & $-1$ & $\hat{U}_a U_a$& $f_G$\\
    \hline
    $([xy],\rho_{++})$ & $2$ & $1$ & $\hat{U}_{bc}$ & $m_{RB}$ \\
    \hline
    $([xy],\rho_{+-})$ & $2$ & $1$ & $\hat{U}_b$ & $m_R$\\
    \hline
    $([xy],\rho_{-+})$ & $2$ & $-1$ & $\hat{U}_{bc}U_b=\hat{U}_{bc}U_c$ & $f_{RB}$ \\
    \hline
    $([xy],\rho_{--})$ & $2$ & $-1$ & $\hat{U}_b U_b$ & $f_R$\\
    \hline
\end{tabular}
\end{center}

We'll note here that there is a little bit of uncertainty in identifying the labels in the $\Z_2^3$ nomenclature; in particular, we're not certain whether $\hat{U}_{abc}$ has topological spin $i$ or $-i$, and similarly whether $\hat{U}_{abc}U_a=\hat{U}_{abc}U_b=\hat{U}_{abc}U_c$ has topological spin $i$ or $-i$.  One of these will be the semion line and one is the anti-semion line, but we have only guessed which is which.  This uncertainty does not affect any of the other calculations in the paper, however.

\subsection*{$Q_8$:}

\begin{center}
\begin{tabular}{|c|c|c|c|}
    \hline
    anyon & $d$ & $s$ & $\Z^3_2$ Nomenclature \\
    \hline
    $([1],\mathbf{1})$ & $1$ & $1$ & $1$ \\
    \hline
    $([1],\pi_a)$ & $1$ & $1$ & $U_a$ \\
    \hline
    $([1],\pi_b)$ & $1$ & $1$ & $U_b$\\
    \hline
    $([1],\pi_c)$ & $1$ & $1$ & $U_{ab}$\\
    \hline
    $([1],\pi_m)$ & $2$ & $1$ & $\hat{U}_c$\\
    \hline
    $([-1],\mathbf{1})$ & $1$ & $1$ & $U_c$ \\
    \hline
    $([-1],\pi_a)$ & $1$ & $1$ & $U_{ac}$ \\
    \hline
    $([-1],\pi_b)$ & $1$ & $1$ & $U_{bc}$\\
    \hline
    $([-1],\pi_c)$ & $1$ & $1$ & $U_{abc}$\\
    \hline
    $([-1],\pi_m)$ & $2$ & $-1$ & $\hat{U}_c U_c$ \\
    \hline
    $([i],\rho_0)$ & $2$ & $1$ & $\hat{U}_b$\\
    \hline
    $([i],\rho_1)$ & $2$ & $i$ & $\hat{U}_{bc}$ \\
    \hline
    $([i],\rho_2)$ & $2$ & $-1$ & $\hat{U}_b U_b$\\
    \hline
    $([i],\rho_3)$ & $2$ & $-i$ & $\hat{U}_{bc}U_b=\hat{U}_{bc}U_c$\\
    \hline
    $([j],\rho_0)$ & $2$ & $1$ & $\hat{U}_a$ \\
    \hline
    $([j],\rho_1)$ & $2$ & $i$ & $\hat{U}_{ac}$ \\
    \hline
    $([j],\rho_2)$ & $2$ & $-1$ & $\hat{U}_a U_a$\\
    \hline
    $([j],\rho_3)$ & $2$ & $-i$ & $\hat{U}_{ac}U_a=\hat{U}_{ac}U_c$ \\
    \hline
    $([k],\rho_0)$ & $2$ & $1$ & $\hat{U}_{ab}$ \\
    \hline
    $([k],\rho_1)$ & $2$ & $i$ & $\hat{U}_{abc}$ \\
    \hline
    $([k],\rho_2)$ & $2$ & $-1$ & $\hat{U}_{ab}U_a = \hat{U}_{ab}U_b$\\
    \hline
    $([k],\rho_3)$ & $2$ & $-i$ & $\hat{U}_{abc}U_a=\hat{U}_{abc}U_b=\hat{U}_{abc}U_c$ \\
    \hline
\end{tabular}
\end{center}

As with the $D_4$ table, we're not certain how to rigorously identify the semion and anti-semion within each of the three semion-anti-semion pairs in the $\Z_2^3$ nomenclature, so those entries in the dictionary are not as certain as the rest of the table.

\section{Reduced Topological Orders}
In order to identify the reduced topological order we would need to find the simple objects on the interface, which we do using the following identity: 
\begin{equation}
    \text{Hom}_{\mathcal{F}}(a,b) := \text{Hom}_{\mathcal{Z}} (a, b\otimes \mathcal{A})
\end{equation}
where ${\mathcal{F}}$ is the fusion category of the condensed theory. Since two simple objects $L_i$ and $L_j$ in $\mathcal{F}$ should have
\begin{equation}
    \text{Hom}_{\mathcal{F}}(L_i,L_j)=\left\{\begin{matrix} \mathbb{C}, & \text{if\ }i=j, \\ 0, & \text{if\ }i\ne j,\end{matrix}\right.
\end{equation}
we can use this to identify how objects in $\mathcal{Z}$ correspond to simple objects in $\mathcal{F}$, so we can write, for each simple object $z$ in $\mathcal{Z}$,
\begin{equation}
    z=\sum_in_{z,i}L_i,
\end{equation}
where $n_{z,i}$ are non-negative integers.  Turning this around, we can define a lift of each simple object $L_i$ in $\mathcal{F}$ by
\begin{equation}
    L_i\longrightarrow \sum_zn_{z,i}z.
\end{equation}
And finally, we impose the condition that all of the lines $z$ appearing in the lift of $L_i$ must have the same topological spin, otherwise we discard them.  What remains is identified as the reduced topological order $\mathcal{Z}'$\cite{Bhardwaj:2023bbf,Chatterjee:2022tyg}.

We are only interested in computing the reduced topological order for the condensable algebras leading to igSPT's.

\subsection{Reduced Topological Order for $\mathcal{Z}(D_4)$}\label{app:red_top_order_d4}

As emphasized in the main text, there are eleven Lagrangian algebras associated with $\mathcal{Z}(D_4)$, these correspond to either $D_4, \text{Rep}(D_4)$ or $\Z^3_2$ (with mixed anomaly) symmetry. In \cite[Table III]{Bhardwaj:2024qrf}, the reduced topological orders and gapless SPT (both instrinsic and non-intrinsic) were identified for $\text{Rep}(D_4)$ symmetry. We can use that same table to pin down the igSPT phases for $D_4$ symmetry. Eventually, these would allow us to identify the known anomaly resolution sequences involving group like $D_4$ symmetry. In the table below, we list down the gapped SPT phases associated with each Lagrangian algebra of the Drinfeld center of $D_4$. In order to identify the gapped SPT phases, we have used the fact that gapped SPT phases correspond to the Lagrangian algebras with trivial overlap with the Symmetry Lagrangian algebra. On the other hand, the gapless phases can be identified by identifying the condensable algebras with trivial overlap with symmetry Lagrangian algebra.
\begin{table}[H]
    \centering
    \begin{tabular}{|c|c|c|}
    \hline
        & Symmetry Lagrangian Algebras & SPTs  \\
         \hline
    $\mathcal{L}_1$ & $1 \oplus e_G \oplus e_R \oplus e_B \oplus e_{GB} \oplus e_{RB} \oplus e_{RG} \oplus e_{RGB}$ & - \\
    \hline
    $\mathcal{L}_{2}$ &   $1 \oplus  e_G \oplus e_R \oplus e_{RG} \oplus 2m_{B}$  & $\mathcal{L}_5,\mathcal{L}_{11} $  \\
    \hline
    $\mathcal{L}_{3}$ &   $1 \oplus e_R \oplus e_B \oplus e_{RB} \oplus 2m_G$  & $\mathcal{L}_{6}, \mathcal{L}_{11} $ \\
    \hline
    $\mathcal{L}_{4}$ &   $ 1\oplus e_G \oplus e_B \oplus e_{GB} \oplus 2 m_{R}$  & $\mathcal{L}_7, \mathcal{L}_{11} $ \\
    \hline
    $\mathcal{L}_{5}$ &  $1 \oplus e_B \oplus m_G \oplus m_R \oplus m_{RG} $ &  $\mathcal{L}_2, \mathcal{L}_8, \mathcal{L}_9$ \\
    \hline  
    $\mathcal{L}_{6}$ &   $1 \oplus e_G \oplus m_R \oplus m_B \oplus m_{RB}$  & $\mathcal{L}_3, \mathcal{L}_9,\mathcal{L}_{10}$\\
    \hline 
    $\mathcal{L}_{7}$ &   $1 \oplus e_R \oplus m_G \oplus m_B \oplus m_{GB}$  & $\mathcal{L}_4,\mathcal{L}_8,\mathcal{L}_{10}$\\
    \hline  
    $\mathcal{L}_{8}$ &   $1 \oplus e_G \oplus e_{RB} \oplus e_{RGB} \oplus 2 m_{RB}$ &  $\mathcal{L}_{5},\mathcal{L}_{7}$ \\
    \hline 
    $\mathcal{L}_{9}$ &   $1 \oplus e_R \oplus e_{GB} \oplus e_{RGB} \oplus 2m_{GB}$ &  $\mathcal{L}_{5}, \mathcal{L}_{6}$ \\
    \hline 
    $\mathcal{L}_{10}$ &   $1 \oplus e_B \oplus e_{RG} \oplus e_{RGB} \oplus 2m_{RG}$  &  $\mathcal{L}_{6}, \mathcal{L}_{7}$ \\
    \hline
     $\mathcal{L}_{11}$ &  $1 \oplus e_{RGB} \oplus m_{GB} \oplus m_{RB} \oplus m_{RG}$ & $\mathcal{L}_{2}, \mathcal{L}_{3}, \mathcal{L}_{4}$ \\
    \hline
    \end{tabular}
   \caption{Lagrangian algebras for $\text{Rep}(D_4)$ \cite[Table III]{Bhardwaj:2024qrf}. The anyon $e_G, e_R$ and $e_B$ is $U_a, U_b$ and $U_c$ in $\Z^3_2$ notation.}
\end{table}
Among the Lagrangian algebras listed above\footnote{We have used this notation because we are going to use the fusion rules for $\mathcal{Z}(D_4)$ given in \cite{Bhardwaj:2024qrf,Iqbal:2023wvm}}, $\mathcal{L}_2,\mathcal{L}_3,\mathcal{L}_4, \mathcal{L}_8, \mathcal{L}_9, \mathcal{L}_{10}$ correspond to $D_4$ symmetry, so if we identify the igSPT for one of them, we will be done. We choose $\mathcal{L}_2$ as our symmetry Lagrangian algebra, then the gapless SPTs are the following: 
\begin{equation}
\begin{aligned}
     \mathcal{A}_1 = 1 \oplus e_{GB}, \quad \mathcal{A}_2 = 1\oplus e_{RB}, \quad \mathcal{A}_3 = 1\oplus e_B, \quad \mathcal{A}_4 = 1 \oplus e_{RGB} \\
     \mathcal{A}_5 = 1\oplus e_B \oplus m_{RG}, \quad \mathcal{A}_6= 1\oplus e_B \oplus m_R, \quad \mathcal{A}_7=1 \oplus e_B \oplus m_G,\\
     \mathcal{A}_8= 1\oplus e_{RGB} \oplus m_{RG}, \quad \mathcal{A}_9= 1 \oplus e_{RGB} \oplus m_{GB}, \quad  \mathcal{A}_{10}= 1 \oplus e_{RGB} \oplus m_{RB}
\end{aligned}
\end{equation}
Among these, only $\mathcal{A}_1$ and $\mathcal{A}_2$ are not subalgebras of the corresponding gapped SPT phases ($\mathcal{L}_5, \mathcal{L}_{11}$). Hence, the igSPT phase corresponding to $\text{Vec}(D_4)$ are given by the condensable algebras of dimension 2, $\mathcal{A}_1$ and $\mathcal{A}_2$. The reduced topological order corresponding to these is $\text{Vec}(\Z_4)$ \cite{Bhardwaj:2024qrf}. We will derive this result here. 
\begin{equation*}
    1 \otimes \left( 1 \oplus e_{GB} \right) = 1 \oplus e_{GB}, \qquad e_B \otimes \left( 1 \oplus e_{GB} \right) = e_G \oplus e_B
\end{equation*}
\begin{equation*}
    e_R \otimes \left( 1 \oplus e_{GB} \right) = e_R \oplus e_{RGB}, \qquad e_G \otimes \left( 1 \oplus e_{GB} \right) = e_G \oplus e_B
\end{equation*}
\begin{equation*}
     e_{RG} \otimes \left( 1 \oplus e_{GB} \right) =  e_{RG} \oplus e_{RB}, \qquad  e_{RB} \otimes \left( 1 \oplus e_{GB} \right) =  e_{RB} \oplus e_{RG}
\end{equation*}
\begin{equation*}
    e_{GB} \otimes \left( 1 \oplus e_{GB} \right)= 1 \oplus e_{GB}, \qquad e_{RGB} \otimes \left( 1 \oplus e_{GB} \right) = e_{RGB} \oplus e_R
\end{equation*}
\begin{equation*}
    m_G \otimes \left( 1 \oplus e_{GB} \right)= m_G \oplus f_B ,\qquad m_R  \otimes \left( 1 \oplus e_{GB} \right) = 2m_R
\end{equation*}
\begin{equation*}
    m_B \otimes \left( 1 \oplus e_{GB} \right) = m_B \oplus f_G, \qquad  m_{BG}  \otimes \left( 1 \oplus e_{GB} \right) = 2m_{BG}
\end{equation*}
\begin{equation*}
   m_{RB} \otimes \left( 1 \oplus e_{GB} \right) = m_{RB} \oplus f_{RB},  \qquad   m_{RG} \otimes \left( 1 \oplus e_{GB} \right) = m_{RG} \oplus f_{RG}
\end{equation*}
\begin{equation*}
   f_{R}  \otimes \left( 1 \oplus e_{GB} \right) = 2f_R,  \qquad  f_{GB}  \otimes \left( 1 \oplus e_{GB} \right) = 2f_{GB}
\end{equation*}
\begin{equation*}
   f_{B}  \otimes \left( 1 \oplus e_{GB} \right) = m_G \oplus f_B,  \qquad  f_{G}  \otimes \left( 1 \oplus e_{GB} \right) = f_G \oplus m_B,
\end{equation*}
\begin{equation*}
    f_{RG}  \otimes \left( 1 \oplus e_{GB} \right) = f_{RG} \oplus m_{RG} ,  \qquad  f_{RB}  \otimes \left( 1 \oplus e_{GB} \right) = f_{RB} \oplus m_{RB}
\end{equation*}
\begin{equation*}
   s_{RGB} \otimes \left( 1 \oplus e_{GB} \right) = 2 s_{RGB}, \qquad \bar{s}_{RGB}  \otimes \left( 1 \oplus e_{GB} \right) = 2\bar{s}_{RGB}
\end{equation*}
We observe that few simple objects of $\mathcal{Z}$ get identified with each other and few others split decompose into two simple objects. The objects that decompose are,
\begin{equation*}
    m_{R},  m_{GB}, f_{GB}, f_{R}, s_{RGB}, \bar{s}_{RGB}
\end{equation*}
Meanwhile, the following objects get identified, 
\begin{align*}
    1 \sim e_{GB},~ e_R \sim e_{RGB},~ e_G \sim e_B,~ e_{RG} \sim e_{RB} \\
    m_B \sim f_G, ~ m_G \sim f_B,~ m_{RG} \sim f_{RG},~ m_{RB} \sim f_{RB}
\end{align*}
We can write down the simple objects of the reduced topological order, 
\begin{equation}\label{eqn:lines_of_rto_d4}
\begin{aligned}
    L_0 &= 1 \oplus e_{GB}, \\
    L_1 &= e_G \oplus e_B, \\
    L_2 &= e_R \oplus e_{RGB}, \\
    L_3 &= e_{RG} \oplus e_{RB}, \\
    L_5 &= m_G \oplus f_B, \\
    L_6 &= m_B \oplus f_G, \\
    L_7 &= m_{RB} \oplus f_{RB}, \\
    L_8 &= m_{RG} \oplus f_{RG},
\end{aligned}
\end{equation}
twelve more dimension one object come from $m_{R},  m_{GB}, f_{GB}, f_{R}, s_{RGB}, \bar{s}_{RGB}$. We need to discard $L_5, L_6, L_7, L_8$ based on the topological spin rule, that leaves us with 16 dimension one objects, hence the reduced topological order is indeed $\mathcal{Z}(\Z_4)$. 

\subsection{Reduced Topological Orders for $\mathcal{Z}(Q_8)/\mathcal{Z}(\text{Rep}(Q_8))$}\label{app:reduced_topological_order}
Now we move onto the identifying the reduced topological order for $\mathcal{Z}(Q_8)/\mathcal{Z}(\text{Rep}(Q_8))$. We have already identified the condensable algebras leading to igSPT phases in the main text, in this appendix we derive the reduced topological order.
\subsubsection*{$\mathcal{A}_{2,5} = ([1],\mathbf{1}) \oplus ([-1],\pi_a )$:} We plug in $\mathcal{A} = \mathcal{A}_{2,5}$ and compute fusion of all the simple objects in the Drinfeld center of $Q_8$ to identify the simple objects of $\mathcal{F}$. 
\begin{equation*}
    ([1],\mathbf{1}) \otimes  (([1],\mathbf{1}) \oplus ([-1],\pi_a )) = ([1],\mathbf{1}) \oplus ([-1],\pi_a ),
\end{equation*}
\begin{equation*}
   ([1],\pi_a)\otimes (([1],\mathbf{1}) \oplus ([-1],\pi_a ))= ([1],\pi_a) \oplus ([-1],\mathbf{1} ),
\end{equation*}
\begin{equation*}
    ([1],\pi_b) \otimes (([1],\mathbf{1}) \oplus ([-1],\pi_a ))= ([1],\pi_b) \oplus ([-1],\pi_c ), 
\end{equation*}
\begin{equation*}
    ([1],\pi_c) \otimes (([1],\mathbf{1}) \oplus ([-1],\pi_a ))= ([1],\pi_c)  \oplus ([-1],\pi_b ), 
\end{equation*}
\begin{equation*}
   ([1],\pi_m) \otimes (([1],\mathbf{1}) \oplus ([-1],\pi_a ))= ([1],\pi_m) \oplus ([-1],\pi_m ),
\end{equation*}
\begin{equation*}
  ([-1],\mathbf{1}) \otimes (([1],\mathbf{1}) \oplus ([-1],\pi_a ))= ([-1],\mathbf{1}) \oplus ([1],\pi_a )
\end{equation*}
\begin{equation*}
   ([-1],\pi_a) \otimes (([1],\mathbf{1}) \oplus ([-1],\pi_a ))= ([-1],\pi_a) \oplus ([1], \mathbf{1}),
\end{equation*}
\begin{equation*}
  ([-1],\pi_b) \otimes (([1],\mathbf{1}) \oplus ([-1],\pi_a ))= ([-1],\pi_b) \oplus ([1], \pi_c),
\end{equation*}
\begin{equation*}
  ([-1],\pi_b) \otimes (([1],\mathbf{1}) \oplus ([-1],\pi_a ))= ([-1],\pi_b) \oplus ([1], \pi_c),
\end{equation*}
\begin{equation*}
 ([-1],\pi_c) \otimes (([1],\mathbf{1}) \oplus ([-1],\pi_a ))=([-1],\pi_c) \oplus ([1], \pi_b) ,
\end{equation*}
\begin{equation*}
 ([-1],\pi_m) \otimes (([1],\mathbf{1}) \oplus ([-1],\pi_a ))= ([-1],\pi_m)\oplus ([1], \pi_m),
\end{equation*}
\begin{equation*}
  ([i],\rho_0) \otimes (([1],\mathbf{1}) \oplus ([-1],\pi_a ))= ([i],\rho_0) \oplus ([i], \rho_0),
\end{equation*}
\begin{equation*}
  ([i],\rho_1) \otimes (([1],\mathbf{1}) \oplus ([-1],\pi_a ))= ([i],\rho_1) \oplus ([i], \rho_3) ,
\end{equation*}
\begin{equation*}
      ([i],\rho_2) \otimes (([1],\mathbf{1}) \oplus ([-1],\pi_a ))= ([i],\rho_2) \oplus ([i], \rho_2), 
\end{equation*}
\begin{equation*}
    ([i],\rho_3) \otimes (([1],\mathbf{1}) \oplus ([-1],\pi_a ))=([i],\rho_3) \oplus ([i], \rho_1)
\end{equation*}   
\begin{equation*}
     ([i],\rho_3) \otimes (([1],\mathbf{1}) \oplus ([-1],\pi_a ))=([i],\rho_3) \oplus ([i], \rho_1)
\end{equation*}       
\begin{equation*}
     ([j],\rho_0) \otimes (([1],\mathbf{1}) \oplus ([-1],\pi_a ))= ([j],\rho_0) \oplus, ([j], \rho_2)
\end{equation*}      
\begin{equation*}
     ([j],\rho_1) \otimes (([1],\mathbf{1}) \oplus ([-1],\pi_a ))= ([j],\rho_1) \oplus  ([j], \rho_1)
\end{equation*}           
\begin{equation*}
     ([j],\rho_2) \otimes (([1],\mathbf{1}) \oplus ([-1],\pi_a ))= ([j],\rho_2) \oplus ([j], \rho_0),
\end{equation*}    
\begin{equation*}
     ([j],\rho_3) \otimes (([1],\mathbf{1}) \oplus ([-1],\pi_a ))= ([j],\rho_3)\oplus ([j], \rho_3)
\end{equation*}    
\begin{equation*}
      ([k],\rho_0) \otimes (([1],\mathbf{1}) \oplus ([-1],\pi_a ))= ([k],\rho_0) \oplus ([k],\rho_2) ,
\end{equation*}  
\begin{equation*}
    ([k],\rho_1) \otimes (([1],\mathbf{1}) \oplus ([-1],\pi_a )) = ([k],\rho_1) \oplus ([k],\rho_1)
\end{equation*}  
\begin{equation*}
     ([k],\rho_2) \otimes (([1],\mathbf{1}) \oplus ([-1],\pi_a ))= ([k],\rho_2) \oplus ([k],\rho_0)
\end{equation*}  
\begin{equation*}
    ([k],\rho_3) \otimes (([1],\mathbf{1}) \oplus ([-1],\pi_a ))= ([k],\rho_3) \oplus ([k],\rho_3)
\end{equation*}  
We conclude that all the simple objects of $\mathcal{Z}$ remain simple in $\mathcal{F}$ except the following: 
\begin{equation*}
  ([i],\rho_0), ([i],\rho_2), ([j],\rho_1), ([j],\rho_3),([k],\rho_1), ([k],\rho_3).
\end{equation*}
These split into two simple objects in $\mathcal{F}$.  
Moreover, for the $\mathcal{Z}$ lines that remain simple in $\mathcal{F}$, we can read off the following identifications: 
\begin{align}
    ([1], \mathbf{1}) \sim ([-1], \pi_a),~ ([1], \pi_a) \sim ([-1], \mathbf{1}),~ ([1], \pi_b) \sim ([-1],\pi_c),~ ([1],\pi_c) \sim ([-1],\pi_b) \\
    ([1],\pi_m) \sim ([-1],\pi_m), ~ ([i],\rho_1) \sim ([i],\rho_3), ~ ([j],\rho_0) \sim ([j], \rho_2), ~  ([k], \rho_0) \sim ([k],\rho_2) 
\end{align}
Following these identifications, we can pin down the simple objects of the reduced topological order. Focusing on the objects that remain simple in $\mathcal{F}$, we can deduce the following:
\begin{equation*}
     \begin{aligned}
        L_0 &= ([1],\mathbf{1}) \oplus ([-1],\pi_a ), \quad L_1 = ([1],\pi_a) \oplus ([-1],\mathbf{1} ) \\
        L_2 &= ([1],\pi_b) \oplus ([-1],\pi_c ), \quad L_3 = ([1],\pi_c)  \oplus ([-1],\pi_b ) \\
        L_4 &=  ([1],\pi_m) \oplus ([-1],\pi_m ), \quad L_5 = ([i],\rho_1) \oplus ([i], \rho_3)  \\
        L_6 &= ([j],\rho_0) \oplus ([j], \rho_2), \quad L_7 = ([k],\rho_0) \oplus ([k],\rho_2) \\
    \end{aligned}
\end{equation*}
So we find four dimension one line and four dimension two lines. In addition to this, we will have 12 more dimension one lines arising from the objects that no longer remains simple in $\mathcal{F}$. These are given by the following: 
\begin{equation*}
\begin{aligned}
    ([i],\rho_0) = L_8 \oplus L_9, \quad ([i],\rho_2) = L_{10} \oplus L_{11}, \quad ([j],\rho_1)= L_{12} \oplus L_{13},\\ ([j],\rho_3)= L_{14} \oplus L_{15}, \quad ([k],\rho_1)= L_{16} \oplus L_{17}, \quad ([k],\rho_3)= L_{18} \oplus L_{19}
\end{aligned}
\end{equation*}
where $L_i, ~ \left(i= 8- 19\right)$ can be identified from the fusion relations. All together we have found 16 dimension one line and four dimension 2 lines, for a total dimension of 32. Now we apply the topological spin rule, we discard $L_4, L_5,L_6, L_7$. That leaves us with 16 dimension one lines.
\begin{equation*}
\begin{aligned}
   \left( L_0, L_1, L_2, L_3, L_8,L_9 \right) &\rightarrow \text{bosons} \\
   \left( L_{10}, L_{11}, \right) &\rightarrow \text{fermions}  \\
   \left( L_{12}, L_{13}, L_{16}, L_{17} \right) &\rightarrow \text{semions}   \\
   \left( L_{14}, L_{15}, L_{18}, L_{19} \right) &\rightarrow \text{antisemions} 
\end{aligned}  
\end{equation*}
We can identify this reduced topological order with $\Z^\omega_2 \times \Z^\omega_2$. We are going to repeat this method for the rest of the cases and identify the respective reduced topological order which is essential to identify the categorical short exact sequences related with the anomaly resolution. 
\subsubsection*{$ \mathcal{A}_{2,6} = ([1],\mathbf{1}) \oplus ([-1], \pi_b)$:}
\begin{equation*}
    ([1],\mathbf{1}) \otimes  (([1],\mathbf{1}) \oplus ([-1], \pi_b)) = ([1],\mathbf{1}) \oplus ([-1], \pi_b) ,
\end{equation*}
\begin{equation*}
     ([1],\pi_a)\otimes (([1],\mathbf{1}) \oplus ([-1], \pi_b))= ([1],\pi_a) \oplus ([-1], \pi_c) , 
\end{equation*}
\begin{equation*}
     ([1],\pi_b) \otimes (([1],\mathbf{1}) \oplus ([-1], \pi_b))= ([1],\pi_b) \oplus ([-1],\mathbf{1}),
\end{equation*}
\begin{equation*}
    ([1],\pi_c) \otimes (([1],\mathbf{1}) \oplus ([-1], \pi_b))= ([1],\pi_c) \oplus ([-1], \pi_a),
\end{equation*}
\begin{equation*}
    ([1],\pi_m) \otimes (([1],\mathbf{1}) \oplus ([-1], \pi_b))= ([1],\pi_m) \oplus ([-1],\pi_m),
\end{equation*}
\begin{equation*}
    ([-1],\mathbf{1}) \otimes (([1],\mathbf{1}) \oplus ([-1], \pi_b))= ([-1],\mathbf{1}) \oplus ([1], \pi_b),
\end{equation*}
\begin{equation*}
    ([-1],\pi_a) \otimes (([1],\mathbf{1}) \oplus ([-1], \pi_b))= ([-1],\pi_a) \oplus ([1],\pi_c) , 
\end{equation*}
\begin{equation*}
    ([-1],\pi_b) \otimes (([1],\mathbf{1}) \oplus ([-1], \pi_b))= ([-1],\pi_b) \oplus ([1],\mathbf{1}) , 
\end{equation*}
\begin{equation*}
   ([-1],\pi_c) \otimes (([1],\mathbf{1}) \oplus ([-1], \pi_b))= ([-1],\pi_c) \oplus ([1],\pi_a]) , 
\end{equation*}
\begin{equation*}
     ([-1],\pi_m) \otimes (([1],\mathbf{1}) \oplus ([-1], \pi_b))= ([-1],\pi_m) \oplus ([1],\pi_m), 
\end{equation*}
\begin{equation*}
    ([i],\rho_0) \otimes (([1],\mathbf{1}) \oplus ([-1], \pi_b))= ([i],\rho_0) \oplus ([i],\rho_2), 
\end{equation*}
\begin{equation*}
    ([i],\rho_1) \otimes (([1],\mathbf{1}) \oplus ([-1], \pi_b))= ([i],\rho_1) \oplus ([i],\rho_1), 
\end{equation*}
\begin{equation*}
   ([i],\rho_2) \otimes (([1],\mathbf{1}) \oplus ([-1], \pi_b))= ([i],\rho_2) \oplus ([i],\rho_0), 
\end{equation*}
\begin{equation*}
    ([i],\rho_3) \otimes (([1],\mathbf{1}) \oplus ([-1], \pi_b))= ([i],\rho_3) \oplus ([i],\rho_3),
\end{equation*}
\begin{equation*}
     ([j],\rho_0) \otimes (([1],\mathbf{1}) \oplus ([-1], \pi_b))= ([j],\rho_0) \oplus ([j],\rho_0),
\end{equation*}
\begin{equation*}
    ([j],\rho_1) \otimes (([1],\mathbf{1}) \oplus ([-1], \pi_b))= ([j],\rho_1) \oplus ([j],\rho_3) ,
\end{equation*}
\begin{equation*}
    ([j],\rho_2) \otimes (([1],\mathbf{1}) \oplus ([-1], \pi_b))= ([j],\rho_2) \oplus ([j],\rho_2),  
\end{equation*}
\begin{equation*}
       ([j],\rho_3) \otimes (([1],\mathbf{1}) \oplus ([-1], \pi_b))= ([j],\rho_3) \oplus ([j],\rho_1), 
\end{equation*}
\begin{equation*}
    ([k],\rho_0) \otimes (([1],\mathbf{1}) \oplus ([-1], \pi_b))= ([k],\rho_0) \oplus ([k],\rho_2)  
\end{equation*}
\begin{equation*}
     ([k],\rho_1) \otimes (([1],\mathbf{1}) \oplus ([-1], \pi_b)) = ([k],\rho_1) \oplus ([k],\rho_1) 
\end{equation*}      
\begin{equation*}
    ([k],\rho_2) \otimes (([1],\mathbf{1}) \oplus ([-1], \pi_b))= ([k],\rho_2) \oplus ([k],\rho_0)  
\end{equation*}
\begin{equation*}
    ([k],\rho_3) \otimes (([1],\mathbf{1}) \oplus ([-1], \pi_b))= ([k],\rho_3) \oplus ([k],\rho_3)
\end{equation*}      
We conclude that all the simple objects of $\mathcal{Z}$ remain simple in $\mathcal{F}$ except the following: 
\begin{equation*}
  ([i],\rho_1), ([i],\rho_3), ([j],\rho_0), ([j],\rho_2),([k],\rho_1), ([k],\rho_3).
\end{equation*}
These split into two simple objects in $\mathcal{F}$.
Moreover, for the $\mathcal{Z}$ lines that remain simple in $\mathcal{F}$, we can read off the following identifications: 
\begin{align}
    ([1],\mathbf{1}) \sim ([-1],\pi_b), ~ ([1],\pi_a) \sim ([-1],\pi_c), ~ ([1],\pi_m) \sim ([-1],\pi_m), ~ ([-1],\mathbf{1}) \sim ([1],\pi_b) \\
    ([-1],\pi_a) \sim ([1],\pi_c), ~ ([j], \rho_1) \sim ([j], \rho_3),~ ([i], \rho_0) \sim ([i], \rho_2) ,~ ([k], \rho_0) \sim ([k], \rho_2)
\end{align} 
Following these identifications, we identify the simple objects of the reduced topological order,
\begin{equation*}
     \begin{aligned}
        L_0 &= ([1],\mathbf{1}) \oplus ([-1],\pi_b ), \quad L_1 = ([1],\pi_b) \oplus ([-1],\mathbf{1} ) \\
        L_2 &= ([1],\pi_a) \oplus ([-1],\pi_c ), \quad L_3 = ([1],\pi_c)  \oplus ([-1],\pi_a ) \\
        L_4 &=  ([1],\pi_m) \oplus ([-1],\pi_m ), \quad L_5 = ([i],\rho_0) \oplus ([i], \rho_2)  \\
        L_6 &= ([j],\rho_1) \oplus ([j], \rho_3), \quad L_7 = ([k],\rho_0) \oplus ([k],\rho_2) \\
    \end{aligned}
\end{equation*}
In addition to this, we will have 12 more dimension one lines,  these are given by the following: 
\begin{equation*}
\begin{aligned}
    ([i],\rho_1) = L_8 \oplus L_9, \quad ([i],\rho_3) = L_{10} \oplus L_{11}, \quad ([j],\rho_0)= L_{12} \oplus L_{13},\\ ([j],\rho_2)= L_{14} \oplus L_{15}, \quad ([k],\rho_1)= L_{16} \oplus L_{17}, \quad ([k],\rho_3)= L_{18} \oplus L_{19}
\end{aligned}
\end{equation*}
All together we have found 16 dimension one line and four dimension 2 lines. We apply the topological spin rule and discard we discard $L_4, L_5,L_6, L_7$, 
\begin{equation*}
\begin{aligned}
   \left( L_0, L_1, L_2, L_3, L_{12},L_{13} \right) &\rightarrow \text{bosons} \\
   \left( L_{14}, L_{15} \right) &\rightarrow \text{fermions}  \\
   \left( L_{8}, L_{9}, L_{16}, L_{17} \right) &\rightarrow \text{semions}   \\
   \left( L_{10}, L_{11}, L_{18}, L_{19} \right) &\rightarrow \text{antisemions} 
\end{aligned}  
\end{equation*}
We can identify this reduced topological order with $\Z^\omega_2 \times \Z^\omega_2$. 
\subsubsection*{$\mathcal{A}_{2,7} = ([1], \mathbf{1}) \oplus ([-1], \pi_c)$:}
\begin{equation*}
       ([1],\mathbf{1}) \otimes  (([1],\mathbf{1}) \oplus ([-1], \pi_c)) = ([1],\mathbf{1}) \oplus ([-1], \pi_c) 
\end{equation*}       
\begin{equation*}
      ([1],\pi_a)\otimes (([1],\mathbf{1}) \oplus ([-1],\pi_c ))= ([1],\pi_a) \oplus ([-1],\pi_b), 
\end{equation*}          
\begin{equation*}
      ([1],\pi_b) \otimes (([1],\mathbf{1}) \oplus ([-1],\pi_c ))= ([1],\pi_b) \oplus ([-1],\pi_a),
\end{equation*}   
\begin{equation*}
       ([1],\pi_c) \otimes (([1],\mathbf{1}) \oplus ([-1],\pi_c ))= ([1],\pi_c) \oplus ([-1],\mathbf{1}),
\end{equation*}   
\begin{equation*}
      ([1],\pi_m) \otimes (([1],\mathbf{1}) \oplus ([-1],\pi_c ))= ([1],\pi_m) \oplus ([-1],\pi_m), 
\end{equation*}   
\begin{equation*}
       ([-1],\mathbf{1}) \otimes (([1],\mathbf{1}) \oplus ([-1],\pi_c ))= ([-1],\mathbf{1}) \oplus ([1],\pi_c ) , 
\end{equation*} 
\begin{equation*}
      ([-1],\pi_a) \otimes (([1],\mathbf{1}) \oplus ([-1],\pi_c ))= ([-1],\pi_a) \oplus ([1],\pi_b) ,
\end{equation*}   
\begin{equation*}
      ([-1],\pi_b) \otimes (([1],\mathbf{1}) \oplus ([-1],\pi_c ))= ([-1],\pi_b) \oplus ([1], \pi_a),  
\end{equation*}  
\begin{equation*}
     ([-1],\pi_c) \otimes (([1],\mathbf{1}) \oplus ([-1],\pi_c ))= ([-1],\pi_c) \oplus ([1] , \mathbf{1}) , 
\end{equation*}  
\begin{equation*}
     ([-1],\pi_m) \otimes (([1],\mathbf{1}) \oplus ([-1],\pi_c ))= ([-1],\pi_m) \oplus ([1], \pi_m) ,
\end{equation*}
\begin{equation*}
    ([i],\rho_0) \otimes (([1],\mathbf{1}) \oplus ([-1],\pi_c ))= ([i],\rho_0) \oplus ([i],\rho_2),
\end{equation*}
\begin{equation*}
     ([i],\rho_1) \otimes (([1],\mathbf{1}) \oplus ([-1],\pi_c ))= ([i],\rho_1) \oplus ([i],\rho_1) ,
\end{equation*}
\begin{equation*}
     ([i],\rho_2) \otimes (([1],\mathbf{1}) \oplus ([-1],\pi_c ))= ([i],\rho_2) \oplus ([i],\rho_0),
\end{equation*}
\begin{equation*}
     ([i],\rho_3) \otimes (([1],\mathbf{1}) \oplus ([-1],\pi_c ))= ([i],\rho_3) \oplus ([i],\rho_3),
\end{equation*}
\begin{equation*}
     ([j],\rho_0) \otimes (([1],\mathbf{1}) \oplus ([-1],\pi_c ))= ([j],\rho_0) \oplus ([j],\rho_2),
\end{equation*}
\begin{equation*}
     ([j],\rho_1) \otimes (([1],\mathbf{1}) \oplus ([-1],\pi_c ))= ([j],\rho_1) \oplus ([j],\rho_1),
\end{equation*}    
\begin{equation*}
    ([j],\rho_2) \otimes (([1],\mathbf{1}) \oplus ([-1],\pi_c ))= ([j],\rho_2) \oplus ([j],\rho_0),
\end{equation*}
\begin{equation*}
    ([j],\rho_3) \otimes (([1],\mathbf{1}) \oplus ([-1],\pi_c ))= ([j],\rho_3) \oplus ([j],\rho_3),
\end{equation*}
\begin{equation*}
    ([k],\rho_0) \otimes (([1],\mathbf{1}) \oplus ([-1],\pi_c ))= ([k],\rho_0) \oplus ([k],\rho_0),
\end{equation*}
\begin{equation*}
    ([k],\rho_1) \otimes (([1],\mathbf{1}) \oplus ([-1],\pi_c )) = ([k],\rho_1) \oplus ([k],\rho_3),
\end{equation*}
\begin{equation*}
    ([k],\rho_2) \otimes (([1],\mathbf{1}) \oplus ([-1],\pi_c ))= ([k],\rho_2) \oplus ([k],\rho_2),
\end{equation*}
\begin{equation*}
    ([k],\rho_3) \otimes (([1],\mathbf{1}) \oplus ([-1],\pi_c ))= ([k],\rho_3) \oplus ([k],\rho_1),
\end{equation*}
We conclude that all the simple objects of $\mathcal{Z}$ remain simple in $\mathcal{F}$ except the following: 
\begin{equation*}
  ([i],\rho_1), ([i],\rho_3), ([j],\rho_1), ([j],\rho_3),([k],\rho_0), ([k],\rho_2).
\end{equation*}
These split into two simple objects in $\mathcal{F}$.
Moreover, for the $\mathcal{Z}$ lines that remain simple in $\mathcal{F}$, we can read off the following identifications: :
\begin{align}
    ([1],\mathbf{1}) \sim ([-1],\pi_c),~ ([1],\pi_a) \sim ([-1],\pi_b),~ ~ ([1],\pi_b) \sim ([-1],\pi_a), ~ ([1],\pi_m) \sim ([-1],\pi_m) \\
    ([-1],\mathbf{1}) \sim ([1],\pi_c),~ ([k],\rho_1) \sim ([k],\rho_3), ~ ([j],\rho_0)\sim ([j],\rho_2), ~ ([i],\rho_0) \sim ([i],\rho_2) 
\end{align}
Following these identifications, we identify the simple objects of the reduced topological order,
\begin{equation*}
     \begin{aligned}
        L_0 &= ([1],\mathbf{1}) \oplus ([-1],\pi_c ), \quad L_1 = ([1],\pi_c) \oplus ([-1],\mathbf{1} ) \\
        L_2 &= ([1],\pi_a) \oplus ([-1],\pi_b ), \quad L_3 = ([1],\pi_b)  \oplus ([-1],\pi_a ) \\
        L_4 &=  ([1],\pi_m) \oplus ([-1],\pi_m ), \quad L_5 = ([i],\rho_0) \oplus ([i], \rho_2)  \\
        L_6 &= ([j],\rho_0) \oplus ([j], \rho_2), \quad L_7 = ([k],\rho_1) \oplus ([k],\rho_3) \\
    \end{aligned}
\end{equation*}
In addition to this, we will have 12 more dimension one lines,  these are given by the following: 
\begin{equation*}
\begin{aligned}
    ([i],\rho_1) = L_8 \oplus L_9, \quad ([i],\rho_3) = L_{10} \oplus L_{11}, \quad ([j],\rho_1)= L_{12} \oplus L_{13},\\ ([j],\rho_3)= L_{14} \oplus L_{15}, \quad ([k],\rho_0)= L_{16} \oplus L_{17}, \quad ([k],\rho_2)= L_{18} \oplus L_{19}
\end{aligned}
\end{equation*}
All together we have found 16 dimension one line and four dimension 2 lines. We apply the topological spin rule and discard we discard $L_4, L_5,L_6, L_7$, 
\begin{equation*}
\begin{aligned}
   \left( L_0, L_1, L_2, L_3, L_{16},L_{17} \right) &\rightarrow \text{bosons} \\
   \left( L_{18}, L_{19} \right) &\rightarrow \text{fermions}  \\
   \left( L_{8}, L_{9}, L_{12}, L_{13} \right) &\rightarrow \text{semions}   \\
   \left( L_{10}, L_{11}, L_{13}, L_{15} \right) &\rightarrow \text{antisemions} 
\end{aligned}  
\end{equation*}
We can identify this reduced topological order with $\Z^\omega_2 \times \Z^\omega_2$. 
\subsubsection*{$\mathcal{A}_{4,6}=([1],\mathbf{1})+([-1],\pi_a)+([i],\rho_0)$:}
The next three condensable algebras are associated purely with $\text{Vec}(Q_8)$. We start with $\mathcal{A}_{4,6}$. We plug this into our equation and compute the following fusions. 
\begin{equation*}    
       ([1],\mathbf{1}) \otimes  (([1],\mathbf{1})\oplus([-1],\pi_a)\oplus([i],\rho_0)) = ([1],\mathbf{1})\oplus([-1],\pi_a)\oplus([i],\rho_0)  ,  
\end{equation*}
\begin{equation*}
       ([1],\pi_a)\otimes (([1],\mathbf{1})\oplus([-1],\pi_a)\oplus([i],\rho_0)) = ([1],\pi_a) \oplus ([-1], \mathbf{1}) \oplus ([i],\rho_0) , 
\end{equation*}       
\begin{equation*}      
       ([1],\pi_b) \otimes (([1],\mathbf{1})\oplus([-1],\pi_a)\oplus([i],\rho_0)) = ([1],\pi_b) \oplus ([-1],\pi_c) \oplus ([i],\rho_2)  , 
\end{equation*}       
\begin{equation*}
       ([1],\pi_c) \otimes (([1],\mathbf{1})\oplus([-1],\pi_a)\oplus([i],\rho_0))= ([1],\pi_c) \oplus ([-1],\pi_b) \oplus ([i],\rho_2) , 
\end{equation*}       
\begin{equation*}
     ([1],\pi_m) \otimes (([1],\mathbf{1})\oplus([-1],\pi_a)\oplus([i],\rho_0))= ([1],\pi_m) \oplus ([-1],\pi_m) \oplus ([i],\rho_1) \oplus ([i],\rho_3) ,
\end{equation*} 
\begin{equation*}
    ([-1],\mathbf{1}) \otimes (([1],\mathbf{1})\oplus([-1],\pi_a)\oplus([i],\rho_0))= ([-1],\mathbf{1}) \oplus ([1],\pi_a) \oplus ([i],\rho_0)
\end{equation*} 
\begin{equation*}
    ([-1],\pi_a) \otimes (([1],\mathbf{1})\oplus([-1],\pi_a)\oplus([i],\rho_0))= ([-1],\pi_a) \oplus ([1],\mathbf{1}) \oplus ([i],\rho_0) ,
\end{equation*} 
\begin{equation*}
    ([-1],\pi_b) \otimes (([1],\mathbf{1})\oplus([-1],\pi_a)\oplus([i],\rho_0))= ([-1],\pi_b) \oplus ([1],\pi_c) \oplus ([i],\rho_2),
\end{equation*} 
\begin{equation*}
    ([-1],\pi_c) \otimes (([1],\mathbf{1})\oplus([-1],\pi_a)\oplus([i],\rho_0))= ([-1],\pi_c) \oplus ([1],\pi_b) \oplus ([i],\rho_2),
\end{equation*} 
\begin{equation*}
    ([-1],\pi_m) \otimes (([1],\mathbf{1})\oplus([-1],\pi_a)\oplus([i],\rho_0))= ([-1],\pi_m) \oplus ([1],\pi_m) \oplus ([i],\rho_1) \oplus ([i],\rho_3), 
\end{equation*} 
\begin{equation*}
    ([i],\rho_0) \otimes (([1],\mathbf{1})\oplus([-1],\pi_a)\oplus([i],\rho_0))= ([i],\rho_0) \oplus ([i],\rho_0) \oplus ([1],\mathbf{1}) \oplus ([1],\pi_a) \oplus ([-1],\mathbf{1}) \oplus ([-1],\pi_a),
\end{equation*}
\begin{equation*}
    ([i],\rho_1) \otimes (([1],\mathbf{1})\oplus([-1],\pi_a)\oplus([i],\rho_0))= ([i],\rho_1) \oplus ([i],\rho_3) \oplus ([1],\pi_m) \oplus ([-1], \pi_m) ,
\end{equation*}
\begin{equation*}
    ([i],\rho_2) \otimes (([1],\mathbf{1})\oplus([-1],\pi_a)\oplus([i],\rho_0))= ([i],\rho_2) \oplus ([i],\rho_2) \oplus ([1], \pi_b) \oplus ([1], \pi_c) \oplus ([-1], \pi_b) \oplus ([-1], \pi_c), 
\end{equation*}
\begin{equation*}
     ([i],\rho_3) \otimes (([1],\mathbf{1})\oplus([-1],\pi_a)\oplus([i],\rho_0))=([i],\rho_3) \oplus ([i],\rho_1) \oplus ([1],\pi_m) \oplus ([-1],\pi_m) ,
\end{equation*}
\begin{equation*}
    ([j],\rho_0) \otimes (([1],\mathbf{1})\oplus([-1],\pi_a)
       \oplus([i],\rho_0))= ([j],\rho_0) \oplus ([j],\rho_2) \oplus ([k],\rho_0) \oplus ([k],\rho_2)  ,
\end{equation*}
\begin{equation*}
    ([j],\rho_1) \otimes (([1],\mathbf{1})\oplus([-1],\pi_a)\oplus([i],\rho_0))= ([j],\rho_1) \oplus ([j],\rho_1) \oplus ([k],\rho_1) \oplus ([k],\rho_3),
\end{equation*}
\begin{equation*}
     ([j],\rho_2) \otimes (([1],\mathbf{1})\oplus([-1],\pi_a)\oplus([i],\rho_0))= ([j],\rho_2) \oplus ([j],\rho_0) \oplus ([k],\rho_2) \oplus ([k],\rho_0) , 
\end{equation*}
\begin{equation*}
    ([j],\rho_3) \otimes (([1],\mathbf{1})\oplus([-1],\pi_a)\oplus([i],\rho_0))=([j],\rho_3) \oplus ([j],\rho_3) \oplus ([k],\rho_3) \oplus ([k],\rho_1)  ,
\end{equation*}
\begin{equation*}
    ([k],\rho_0) \otimes (([1],\mathbf{1})\oplus([-1],\pi_a)\oplus([i],\rho_0))= ([k],\rho_0) \oplus ([k],\rho_2) \oplus ([j],\rho_0) \oplus ([j],\rho_2) ,
\end{equation*}
\begin{equation*}
     ([k],\rho_1) \otimes (([1],\mathbf{1})\oplus([-1],\pi_a)\oplus([i],\rho_0)) = ([k],\rho_1) \oplus ([k],\rho_1) \oplus ([j],\rho_1) \oplus ([j],\rho_3), 
\end{equation*}
\begin{equation*}
     ([k],\rho_2) \otimes (([1],\mathbf{1})\oplus([-1],\pi_a)\oplus([i],\rho_0))= ([k],\rho_2) \oplus ([k],\rho_0) \oplus ([j],\rho_2) \oplus ([j],\rho_0),
\end{equation*}
\begin{equation*}
       ([k],\rho_3) \otimes (([1],\mathbf{1})\oplus([-1],\pi_a)\oplus([i],\rho_0))= ([k],\rho_3) \oplus ([k],\rho_3) \oplus ([j],\rho_3) \oplus ([j],\rho_1)
\end{equation*}
We conclude that all the simple objects of $\mathcal{Z}$ remain simple in $\mathcal{F}$ except the following: 
\begin{equation*}
  ([i],\rho_0), ([i],\rho_2), ([j],\rho_1), ([j],\rho_3),([k],\rho_1), ([k],\rho_3).
\end{equation*}
These split into two simple objects in $\mathcal{F}$.
Now we can look for identifications, we find the following: 
\begin{align}
    ([1],\mathbf{1}) \sim ([-1],\pi_a),~ ([1],\pi_a) \sim ([-1],\mathbf{1}),~ ([1],\pi_b) \sim ([-1],\pi_c), ~ ([1],\pi_c) \sim ([-1],\pi_b) \\
    ([1],\pi_m) \sim ([-1],\pi_m)\sim ([i],\rho_1) \sim ([i],\rho_3), 
    ([j],\rho_0) \sim ([j],\rho_2) \sim ([k],\rho_0) \sim ([k],\rho_2),
\end{align}
Following these identifications, we identify the simple objects of the reduced topological order,
\begin{equation*}
     \begin{aligned}
        L_0 &= ([1],\mathbf{1})\oplus([-1],\pi_a)\oplus([i],\rho_0) \\
        L_1 &= ([1],\pi_a) \oplus ([-1], \mathbf{1}) \oplus ([i],\rho_0) \\
        L_2 &= ([1],\pi_b) \oplus ([-1],\pi_c) \oplus ([i],\rho_2), \\
        L_3 &= ([1],\pi_c) \oplus ([-1],\pi_b) \oplus ([i],\rho_2) \\
        L_4 &=  ([1],\pi_m) \oplus ([-1],\pi_m) \oplus ([i],\rho_1) \oplus ([i],\rho_3), \\
        L_5 &= ([j],\rho_0) \oplus ([j], \rho_2) \oplus ([k],\rho_0) \oplus ([k], \rho_2) 
    \end{aligned}
\end{equation*}
In addition to this, 
\begin{equation}
    ([i],\rho_0) = L_0 \oplus L_1, \quad ([i],\rho_2) = L_{2} \oplus L_{3}
\end{equation}
The remaining four dimension two lines that split into two simple objects in $\mathcal{F}$, can be written as the following: 
\begin{equation}
\begin{aligned}
    ([j],\rho_1)&= L_6 \oplus L_7, \\
    ([j],\rho_3)&= L_8 \oplus L_9, \\
    ([k],\rho_1)&= L_6 \oplus L_9 , \\
    ([k],\rho_3)&= L_8 \oplus L_7
\end{aligned}    
\end{equation}
where, 
\begin{equation}
    \begin{aligned}
        L_6 &= ([j],\rho_1) \oplus ([k],\rho_1), \quad L_7 = ([j],\rho_1) \oplus ([k],\rho_3) \\
        L_8 &= ([j],\rho_3) \oplus ([k],\rho_3), \quad L_9 = ([j],\rho_3) \oplus ([k],\rho_1)
    \end{aligned}
\end{equation}
we will have eight dimension one line and two dimension two lines, which gives us a total dimension of 16. We apply the topological spin rule and discard $L_1, L_3,L_4, L_5, L_7,L_9$. That leaves with,
\begin{equation*}
\begin{aligned}
   \left( L_0, L_2 \right) \rightarrow \text{bosons}, \quad
   \left( L_{6} \right) \rightarrow \text{semions}   \quad
   \left( L_{8} \right) \rightarrow \text{antisemions} 
\end{aligned}  
\end{equation*}
We can identify this reduced topological order with $\Z^\omega_2 $. 
\subsubsection*{$\mathcal{A}_{4,9} =([1],\mathbf{1})\oplus([-1],\pi_b)\oplus([j],\rho_0)$:}
\begin{equation*}
    ([1],\mathbf{1}) \otimes  (([1],\mathbf{1})\oplus([-1],\pi_b)\oplus([j],\rho_0)) = ([1],\mathbf{1})\oplus([-1],\pi_b)\oplus([j],\rho_0) ,
\end{equation*}
\begin{equation*}
    ([1],\pi_a)\otimes (([1],\mathbf{1})\oplus([-1],\pi_b)\oplus([j],\rho_0))= ([1],\pi_a) \oplus ([-1],\pi_c) \oplus ([j],\rho_2),
\end{equation*}
\begin{equation*}
    ([1],\pi_b) \otimes (([1],\mathbf{1})\oplus([-1],\pi_b)\oplus([j],\rho_0))= ([1],\pi_b) \oplus ([-1],\mathbf{1}) \oplus ([j],\rho_0),
\end{equation*}
\begin{equation*}
     ([1],\pi_c) \otimes (([1],\mathbf{1})\oplus([-1],\pi_b)\oplus([j],\rho_0))= ([1],\pi_c) \oplus ([-1],\pi_a) \oplus ([j],\rho_2),
\end{equation*}
\begin{equation*}
     ([1],\pi_m) \otimes (([1],\mathbf{1})\oplus([-1],\pi_b)\oplus([j],\rho_0))= ([1],\pi_m) \oplus ([-1],\pi_m) \oplus ([j],\rho_1) \oplus ([j],\rho_3),
\end{equation*}
\begin{equation*}
     ([-1],\mathbf{1}) \otimes (([1],\mathbf{1})\oplus([-1],\pi_b)\oplus([j],\rho_0))= ([-1],\mathbf{1}) \oplus ([1],\pi_b) \oplus ([j],\rho_0),
\end{equation*}
\begin{equation*}
    ([-1],\pi_a) \otimes (([1],\mathbf{1})\oplus([-1],\pi_b)\oplus([j],\rho_0)))= ([-1],\pi_a) \oplus ([1],\pi_c) \oplus ([j],\rho_2),
\end{equation*}   
\begin{equation*}
     ([-1],\pi_b) \otimes (([1],\mathbf{1})\oplus([-1],\pi_b)\oplus([j],\rho_0))= ([-1],\pi_b) \oplus ([1],\mathbf{1}) \oplus ([j],\rho_0),
\end{equation*}
\begin{equation*}
     ([-1],\pi_c) \otimes (([1],\mathbf{1})\oplus([-1],\pi_b)\oplus([j],\rho_0))= ([-1],\pi_c) \oplus ([1],\pi_a) \oplus ([j],\rho_2), 
\end{equation*}     
\begin{equation*}
     ([-1],\pi_m) \otimes (([1],\mathbf{1})\oplus([-1],\pi_b)\oplus([j],\rho_0))= ([-1],\pi_m) \oplus ([1],\pi_m) \oplus ([j],\rho_1) \oplus ([j],\rho_3), 
\end{equation*}
\begin{equation*}
    ([i],\rho_0) \otimes (([1],\mathbf{1})\oplus([-1],\pi_b)\oplus([j],\rho_0))= ([i],\rho_0) \oplus ([i],\rho_2) \oplus ([k],\rho_0) \oplus ([k],\rho_2) ,
\end{equation*} 
\begin{equation*}
     ([i],\rho_1) \otimes (([1],\mathbf{1})\oplus([-1],\pi_b)\oplus([j],\rho_0))= ([i],\rho_1) \oplus ([i],\rho_1) \oplus ([k],\rho_1)\oplus ([k],\rho_3),
\end{equation*}
\begin{equation*}
    ([i],\rho_2) \otimes (([1],\mathbf{1})\oplus([-1],\pi_b)\oplus([j],\rho_0))= ([i],\rho_2) \oplus ([i],\rho_0) \oplus ([k],\rho_0) \oplus ([k],\rho_2) , 
\end{equation*}
\begin{equation*}
    ([i],\rho_3) \otimes (([1],\mathbf{1})\oplus([-1],\pi_b)\oplus([j],\rho_0))= ([i],\rho_3) \oplus ([i],\rho_3) \oplus ([k],\rho_3) \oplus ([k],\rho_1),
\end{equation*}
\begin{equation*}
   ([j],\rho_0) \otimes (([1],\mathbf{1})\oplus([-1],\pi_b)\oplus([j],\rho_0))= ([j],\rho_0) \oplus ([j],\rho_0)\oplus ([1],\mathbf{1}) \oplus ([1],\pi_b) \oplus ([-1],\mathbf{1}) \oplus ([-1],\pi_b), 
\end{equation*}
\begin{equation*}
    ([j],\rho_1) \otimes (([1],\mathbf{1})\oplus([-1],\pi_b)\oplus([j],\rho_0))= ([j],\rho_1) \oplus ([j],\rho_3) \oplus ([1],\pi_m) \oplus ([-1],\pi_m),
\end{equation*}
\begin{equation*}
    ([j],\rho_2) \otimes (([1],\mathbf{1})\oplus([-1],\pi_b)\oplus([j],\rho_0))= ([j],\rho_2) \oplus ([j],\rho_2)  \oplus ([1],\pi_a)  \oplus ([1],\pi_c) \oplus ([-1],\pi_a)  \oplus ([-1],\pi_c),  
\end{equation*}
\begin{equation*}
    ([j],\rho_3) \otimes (([1],\mathbf{1})\oplus([-1],\pi_b)\oplus([j],\rho_0))= ([j],\rho_3) \oplus ([j],\rho_1) \oplus ([1],\pi_m) \oplus ([-1],\pi_m),
\end{equation*}
\begin{equation*}
     ([k],\rho_0) \otimes (([1],\mathbf{1})\oplus([-1],\pi_b)\oplus([j],\rho_0))= ([k],\rho_0) \oplus ([k],\rho_2) \oplus ([i], \rho_0) \oplus ([i],\rho_2),
\end{equation*}
\begin{equation*}
    ([k],\rho_1) \otimes (([1],\mathbf{1})\oplus([-1],\pi_b)\oplus([j],\rho_0)) = ([k],\rho_1) \oplus ([k],\rho_1) \oplus  ([i], \rho_1) \oplus  ([i], \rho_3)
\end{equation*}    
\begin{equation*}
    ([k],\rho_2) \otimes (([1],\mathbf{1})\oplus([-1],\pi_b)\oplus([j],\rho_0))= ([k],\rho_2) \oplus ([k],\rho_0) \oplus  ([i], \rho_2) \oplus  ([i], \rho_0),
\end{equation*}
\begin{equation*}
    ([k],\rho_3) \otimes (([1],\mathbf{1})\oplus([-1],\pi_b)\oplus([j],\rho_0))= ([k],\rho_3) \oplus ([k],\rho_3) \oplus  ([i], \rho_3) \oplus  ([i], \rho_1)
\end{equation*}
We conclude that all the simple objects of $\mathcal{Z}$ remain simple in $\mathcal{F}$ except the following: 
\begin{equation*}
  ([i],\rho_1), ([i],\rho_3), ([j],\rho_0), ([j],\rho_2),([k],\rho_1), ([k],\rho_3).
\end{equation*}
These split into two simple objects in $\mathcal{F}$.
Now we can look for identifications, we find the following: 
\begin{align}
    ([1],\mathbf{1}) \sim ([-1],\pi_b),~ ([1],\pi_a) \sim ([-1],\pi_c),~ ([1],\pi_b) \sim ([-1],\mathbf{1}), ~ ([1],\pi_c) \sim ([-1],\pi_a) \\
    ([1],\pi_m) \sim ([-1],\pi_m)\sim ([j],\rho_1) \sim ([j],\rho_3), 
    ([i],\rho_0) \sim ([i],\rho_2) \sim ([k],\rho_0) \sim ([k],\rho_2),
\end{align}
We identify the simple objects of the reduced topological order,
\begin{equation*}
     \begin{aligned}
        L_0 &= ([1],\mathbf{1})\oplus([-1],\pi_b)\oplus([j],\rho_0) \\
        L_1 &= ([1],\pi_a) \oplus ([-1],\pi_c) \oplus ([j],\rho_2) \\
        L_2 &= ([1],\pi_b) \oplus ([-1],\mathbf{1}) \oplus ([j],\rho_0), \\
        L_3 &= ([1],\pi_c) \oplus ([-1],\pi_a) \oplus ([j],\rho_2) \\
        L_4 &=  ([1],\pi_m) \oplus ([-1],\pi_m) \oplus ([j],\rho_1) \oplus ([j],\rho_3), \\
        L_5 &= ([i],\rho_0) \oplus ([i],\rho_2) \oplus ([k],\rho_0) \oplus ([k],\rho_2) 
    \end{aligned}
\end{equation*}
In addition to this, 
\begin{equation}
    ([j],\rho_0) = L_0 \oplus L_2, \quad ([j],\rho_2) = L_{1} \oplus L_{3}
\end{equation}
The remaining four dimension two lines that split into two simple objects in $\mathcal{F}$, can be written as the following: 
\begin{equation}
\begin{aligned}
    ([i],\rho_1)&= L_6 \oplus L_7, \\
    ([i],\rho_3)&= L_8 \oplus L_9, \\
    ([k],\rho_1)&= L_6 \oplus L_9 , \\
    ([k],\rho_3)&= L_8 \oplus L_7
\end{aligned}    
\end{equation}
where, 
\begin{equation}
    \begin{aligned}
        L_6 &= ([i],\rho_1) \oplus ([k],\rho_1), \quad L_7 = ([i],\rho_1) \oplus ([k],\rho_3) \\
        L_8 &= ([i],\rho_3) \oplus ([k],\rho_3), \quad L_9 = ([i],\rho_3) \oplus ([k],\rho_1)
    \end{aligned}
\end{equation}
We apply the topological spin rule and discard $L_2, L_3,L_4, L_5, L_7,L_9$. That leaves with,
\begin{equation*}
\begin{aligned}
   \left( L_0, L_1 \right) \rightarrow \text{bosons}, \quad
   \left( L_{6} \right) \rightarrow \text{semions}   \quad
   \left( L_{8} \right) \rightarrow \text{antisemions} 
\end{aligned}  
\end{equation*}
We can identify this reduced topological order with $\Z^\omega_2 $. 
\subsubsection*{$\mathcal{A}_{4,12} =([1],\mathbf{1})\oplus([-1],\pi_c)\oplus([k],\rho_0)$:}
\begin{equation*}
       ([1],\mathbf{1}) \otimes  (([1],\mathbf{1})\oplus([-1],\pi_c)\oplus([k],\rho_0)) = ([1],\mathbf{1})\oplus([-1],\pi_c)\oplus([k],\rho_0) ,  
\end{equation*}
\begin{equation*}      
       ([1],\pi_a)\otimes (([1],\mathbf{1})\oplus([-1],\pi_c)\oplus([k],\rho_0))= ([1],\pi_a) \oplus ([-1],\pi_b) 
       \oplus ([k],\rho_2),     
\end{equation*} 
\begin{equation*}
    ([1],\pi_b) \otimes (([1],\mathbf{1})\oplus([-1],\pi_c)\oplus([k],\rho_0))= ([1],\pi_b) \oplus ([-1],\pi_a) \oplus ([k],\rho_2),
\end{equation*}   
\begin{equation*}
    ([1],\pi_c) \otimes (([1],\mathbf{1})\oplus([-1],\pi_c)\oplus([k],\rho_0))= ([1],\pi_c) \oplus ([-1],\mathbf{1}) \oplus ([k],\rho_0) ,
\end{equation*}
\begin{equation*}
    ([1],\pi_m) \otimes (([1],\mathbf{1})\oplus([-1],\pi_c)\oplus([k],\rho_0))= ([1],\pi_m) \oplus ([-1],\pi_m) \oplus ([k],\rho_1) \oplus ([k],\rho_3) , 
\end{equation*}
\begin{equation*}
    ([-1],\mathbf{1}) \otimes (([1],\mathbf{1})\oplus([-1],\pi_c)\oplus([k],\rho_0))= ([-1],\mathbf{1}) \oplus ([1],\pi_c) \oplus ([k],\rho_0),
\end{equation*} 
\begin{equation*}
    ([-1],\pi_a) \otimes (([1],\mathbf{1})\oplus([-1],\pi_c)\oplus([k],\rho_0))= ([-1],\pi_a) \oplus ([1],\pi_b) \oplus ([k],\rho_2) , 
\end{equation*}       
\begin{equation*}
    ([-1],\pi_b) \otimes (([1],\mathbf{1})\oplus([-1],\pi_c)\oplus([k],\rho_0))= ([-1],\pi_b) \oplus ([1],\pi_a) \oplus ([k],\rho_2),
\end{equation*}
\begin{equation*}
       ([-1],\pi_c) \otimes (([1],\mathbf{1})\oplus([-1],\pi_c)\oplus([k],\rho_0))= ([-1],\pi_c) \oplus ([1],\mathbf{1}) \oplus ([k],\rho_0) ,
\end{equation*}       
\begin{equation*}
     ([-1],\pi_m) \otimes (([1],\mathbf{1})\oplus([-1],\pi_c)\oplus([k],\rho_0))= ([-1],\pi_m)  \oplus ([1],\pi_m) \oplus ([k],\rho_1) \oplus ([k],\rho_3),
\end{equation*}
\begin{equation*}
     ([i],\rho_0) \otimes (([1],\mathbf{1})\oplus([-1],\pi_c)\oplus([k],\rho_0))= ([i],\rho_0) \oplus  ([i],\rho_2) \oplus ([j],\rho_0) \oplus  ([j],\rho_2),
\end{equation*}       
\begin{equation*}
    ([i],\rho_1) \otimes (([1],\mathbf{1})\oplus([-1],\pi_c)\oplus([k],\rho_0))= ([i],\rho_1) \oplus ([i],\rho_1) \oplus ([j],\rho_1) \oplus  ([j],\rho_3) , 
\end{equation*}
\begin{equation*}
    ([i],\rho_2) \otimes (([1],\mathbf{1})\oplus([-1],\pi_c)\oplus([k],\rho_0))= ([i],\rho_2) \oplus ([i],\rho_0) \oplus ([j],\rho_0) \oplus  ([j],\rho_2),
\end{equation*}
\begin{equation*}
    ([i],\rho_3) \otimes (([1],\mathbf{1})\oplus([-1],\pi_c)\oplus([k],\rho_0))= ([i],\rho_3) \oplus ([i],\rho_3) \oplus ([j],\rho_3) \oplus  ([j],\rho_1) ,
\end{equation*}       
\begin{equation*}
     ([j],\rho_0) \otimes (([1],\mathbf{1})\oplus([-1],\pi_c)\oplus([k],\rho_0))= ([j],\rho_0) \oplus ([j],\rho_2) \oplus ([i],\rho_0) \oplus  ([i],\rho_2),
\end{equation*}
\begin{equation*}
     ([j],\rho_1) \otimes (([1],\mathbf{1})\oplus([-1],\pi_c)\oplus([k],\rho_0))= ([j],\rho_1) \oplus ([j],\rho_1) \oplus ([i],\rho_1) \oplus  ([i],\rho_3) ,
\end{equation*}      
\begin{equation*}
    ([j],\rho_2) \otimes (([1],\mathbf{1})\oplus([-1],\pi_c)\oplus([k],\rho_0))= ([j],\rho_2) \oplus ([j],\rho_0) \oplus ([i],\rho_2) \oplus  ([i],\rho_0) ,
\end{equation*}       
\begin{equation*}
     ([j],\rho_3) \otimes (([1],\mathbf{1})\oplus([-1],\pi_c)\oplus([k],\rho_0))= ([j],\rho_3) \oplus ([j],\rho_3) \oplus ([i],\rho_3) \oplus  ([i],\rho_1), 
\end{equation*}   
\begin{equation*}
    ([k],\rho_0) \otimes (([1],\mathbf{1})+([-1],\pi_c)+([k],\rho_0))= ([k],\rho_0) \oplus ([k],\rho_0) \oplus ([1],\mathbf{1}) \oplus ([-1],\mathbf{1}) \oplus ([1],\pi_c) \oplus ([-1],\pi_c) ,
\end{equation*}
\begin{equation*}
     ([k],\rho_1) \otimes (([1],\mathbf{1})+([-1],\pi_c)+([k],\rho_0)) = ([k],\rho_1) \oplus ([k],\rho_3) \oplus ([1],\pi_m) \oplus ([-1],\pi_m) 
\end{equation*}
\begin{equation*}
    ([k],\rho_2) \otimes (([1],\mathbf{1})+([-1],\pi_c)+([k],\rho_0))=  ([k],\rho_2) \oplus ([k],\rho_2) \oplus ([1],\pi_b) \oplus ([-1],\pi_b) \oplus ([1],\pi_a) \oplus ([-1],\pi_a) ,
\end{equation*}
\begin{equation*}
    ([k],\rho_3) \otimes (([1],\mathbf{1})+([-1],\pi_c)+([k],\rho_0))= ([k],\rho_3) \oplus ([k],\rho_1) \oplus ([1],\pi_m) \oplus ([-1],\pi_m)
\end{equation*}
We conclude that all the simple objects of $\mathcal{Z}$ remain simple in $\mathcal{F}$ except the following: 
\begin{equation*}
  ([i],\rho_1), ([i],\rho_3), ([j],\rho_1), ([j],\rho_3),([k],\rho_0), ([k],\rho_2).
\end{equation*}
These split into two simple objects in $\mathcal{F}$. Now we can look for identifications, we find the following: 
\begin{align}
    ([1],\mathbf{1}) \sim ([-1],\pi_c),~ ([1],\pi_a) \sim ([-1],\pi_b),~ ([1],\pi_c) \sim ([-1],\mathbf{1}), ~ ([1],\pi_b) \sim ([-1],\pi_a) \\
    ([1],\pi_m) \sim ([-1],\pi_m)\sim ([k],\rho_1) \sim ([k],\rho_3), 
    ([i],\rho_0) \sim ([i],\rho_2) \sim ([j],\rho_0) \sim ([j],\rho_2),
\end{align}
We identify the simple objects of the reduced topological order,
\begin{equation*}
     \begin{aligned}
        L_0 &= ([1],\mathbf{1})\oplus([-1],\pi_c)\oplus([k],\rho_0) \\
        L_1 &= ([1],\pi_a) \oplus ([-1],\pi_b) \oplus ([k],\rho_2) \\
        L_2 &= ([1],\pi_b) \oplus ([-1],\pi_a) \oplus ([k],\rho_2), \\
        L_3 &= ([1],\pi_c) \oplus  ([-1],\mathbf{1}) \oplus ([k],\rho_0) \\
        L_4 &=  ([1],\pi_m) \oplus ([-1],\pi_m) \oplus ([k],\rho_1) \oplus ([k],\rho_3), \\
        L_5 &= ([i],\rho_0) \oplus ([i],\rho_2) \oplus ([j],\rho_0) \oplus ([j],\rho_2) 
    \end{aligned}
\end{equation*}
In addition to this, 
\begin{equation}
    ([k],\rho_0) = L_0 \oplus L_3, \quad ([k],\rho_2) = L_{1} \oplus L_{2}
\end{equation}
The remaining four dimension two lines that split into two simple objects in $\mathcal{F}$, can be written as the following: 
\begin{equation}
\begin{aligned}
    ([i],\rho_1)&= L_6 \oplus L_7, \\
    ([i],\rho_3)&= L_8 \oplus L_9, \\
    ([j],\rho_1)&= L_6 \oplus L_9 , \\
    ([j],\rho_3)&= L_8 \oplus L_7
\end{aligned}    
\end{equation}
where, 
\begin{equation}
    \begin{aligned}
        L_6 &= ([i],\rho_1) \oplus ([j],\rho_1), \quad L_7 = ([i],\rho_1) \oplus ([j],\rho_3) \\
        L_8 &= ([i],\rho_3) \oplus ([j],\rho_3), \quad L_9 = ([i],\rho_3) \oplus ([j],\rho_1)
    \end{aligned}
\end{equation}
We apply the topological spin rule and discard $L_1, L_2,L_4, L_5, L_7,L_9$. That leaves with,
\begin{equation*}
\begin{aligned}
   \left( L_0, L_3 \right) \rightarrow \text{bosons}, \quad
   \left( L_{6} \right) \rightarrow \text{semions}   \quad
   \left( L_{8} \right) \rightarrow \text{antisemions} 
\end{aligned}  
\end{equation*}
We can identify this reduced topological order with $\Z^\omega_2 $. Now we move on to identifying the reduced topological order associated with the $\text{Rep}(Q_8)$ symmetry.
\subsubsection*{$\mathcal{A}_{4,17} = ([1],\mathbf{1})\oplus([1],\pi_a)\oplus([-1],\pi_b)\oplus([-1],\pi_c)$:}
\begin{equation*}
    ([1],\mathbf{1}) \otimes  (([1],\mathbf{1})\oplus([1],\pi_a)\oplus([-1],\pi_b)\oplus([-1],\pi_c)) = ([1],\mathbf{1})\oplus([1],\pi_a)\oplus([-1],\pi_b)\oplus([-1],\pi_c),
\end{equation*}
\begin{equation*}
     ([1],\pi_a)\otimes (([1],\mathbf{1})\oplus([1],\pi_a)\oplus([-1],\pi_b)\oplus([-1],\pi_c)) = ([1],\pi_a)\oplus  ([1],\mathbf{1})  \oplus ([-1],\pi_c)\oplus ([-1],\pi_b) ,
\end{equation*}
\begin{equation*}
    ([1],\pi_b) \otimes (([1],\mathbf{1})\oplus([1],\pi_a)\oplus([-1],\pi_b)\oplus([-1],\pi_c)) = ([1],\pi_b) \oplus ([1],\pi_c) \oplus ([-1],\mathbf{1}) \oplus ([-1],\pi_a),
\end{equation*}
\begin{equation*}
    ([1],\pi_c) \otimes (([1],\mathbf{1})\oplus([1],\pi_a)\oplus([-1],\pi_b)\oplus([-1],\pi_c))= ([1],\pi_c) \oplus ([1],\pi_b) \oplus ([-1],\pi_a) \oplus ([-1],\mathbf{1}),
\end{equation*}
\begin{equation*}
    ([1],\pi_m) \otimes (([1],\mathbf{1})\oplus([1],\pi_a)\oplus([-1],\pi_b)\oplus([-1],\pi_c))= ([1],\pi_m) \oplus ([1],\pi_m) \oplus ([-1],\pi_m) \oplus ([-1],\pi_m) ,
\end{equation*}
\begin{equation*}
     ([-1],\mathbf{1}) \otimes (([1],\mathbf{1})\oplus([1],\pi_a)\oplus([-1],\pi_b)\oplus([-1],\pi_c))= ([-1],\mathbf{1}) \oplus ([-1],\pi_a)  \oplus([1],\pi_b)\oplus([1],\pi_c)),
\end{equation*}
\begin{equation*}
    ([-1],\pi_a) \otimes (([1],\mathbf{1})\oplus([1],\pi_a)\oplus([-1],\pi_b)\oplus([-1],\pi_c))=([-1],\pi_a) \oplus ([1],\mathbf{1}) \oplus([1],\pi_c) \oplus([1],\pi_b) ,
\end{equation*}
\begin{equation*}
    ([-1],\pi_b) \otimes (([1],\mathbf{1})\oplus([1],\pi_a)\oplus([-1],\pi_b)\oplus([-1],\pi_c))= ([-1],\pi_b) \oplus ([-1],\pi_c) \oplus ([1],\mathbf{1}) \oplus([1],\pi_a),
\end{equation*}
\begin{equation*}
    ([-1],\pi_c) \otimes (([1],\mathbf{1})\oplus([1],\pi_a)\oplus([-1],\pi_b)\oplus([-1],\pi_c))= ([-1],\pi_c) \oplus([-1],\pi_b) \oplus([1],\pi_a) \oplus ([1],\mathbf{1})  , 
\end{equation*}
\begin{equation*}
    ([-1],\pi_m) \otimes (([1],\mathbf{1})\oplus([1],\pi_a)\oplus([-1],\pi_b)\oplus([-1],\pi_c))= ([-1],\pi_m) \oplus ([-1],\pi_m) \oplus ([1],\pi_m) \oplus ([1],\pi_m),
\end{equation*}
\begin{equation*}
    ([i],\rho_0) \otimes (([1],\mathbf{1})\oplus([1],\pi_a)\oplus([-1],\pi_b)\oplus([-1],\pi_c))= ([i],\rho_0) \oplus   ([i],\rho_0) \oplus  ([i],\rho_2) \oplus ([i],\rho_2)  ,
\end{equation*}
\begin{equation*}
    ([i],\rho_1) \otimes (([1],\mathbf{1})\oplus([1],\pi_a)\oplus([-1],\pi_b)\oplus([-1],\pi_c))= ([i],\rho_1) \oplus ([i],\rho_1) \oplus ([i],\rho_1) \oplus ([i],\rho_1),
\end{equation*}
\begin{equation*}
    ([i],\rho_2) \otimes (([1],\mathbf{1})\oplus([1],\pi_a)\oplus([-1],\pi_b)\oplus([-1],\pi_c)) =([i],\rho_2) \oplus ([i],\rho_2) \oplus ([i],\rho_0) \oplus   ([i],\rho_0),
\end{equation*}
\begin{equation*}
    ([i],\rho_3) \otimes (([1],\mathbf{1})\oplus([1],\pi_a)\oplus([-1],\pi_b)\oplus([-1],\pi_c))= ([i],\rho_3) \oplus ([i],\rho_3)\oplus ([i],\rho_3) \oplus ([i],\rho_3) ,
\end{equation*}
\begin{equation*}
    ([j],\rho_0) \otimes (([1],\mathbf{1})\oplus([1],\pi_a)\oplus([-1],\pi_b)\oplus([-1],\pi_c))= ([j],\rho_0) \oplus ([j],\rho_2) \oplus ([j],\rho_0) \oplus ([j],\rho_2) ,
\end{equation*}
\begin{equation*}
    ([j],\rho_1) \otimes (([1],\mathbf{1})\oplus([1],\pi_a)\oplus([-1],\pi_b)\oplus([-1],\pi_c))= ([j],\rho_1)  \oplus ([j],\rho_3) \oplus ([j],\rho_3) \oplus ([j],\rho_1) 
\end{equation*}
\begin{equation*}
    ([j],\rho_2) \otimes (([1],\mathbf{1})\oplus([1],\pi_a)\oplus([-1],\pi_b)\oplus([-1],\pi_c))= ([j],\rho_2)  \oplus ([j],\rho_0) \oplus ([j],\rho_2) \oplus ([j],\rho_0) 
\end{equation*}
\begin{equation*}
    ([j],\rho_3) \otimes (([1],\mathbf{1})\oplus([1],\pi_a)\oplus([-1],\pi_b)\oplus([-1],\pi_c))= ([j],\rho_3)  \oplus ([j],\rho_1) \oplus ([j],\rho_1) \oplus ([j],\rho_3),
\end{equation*}
\begin{equation*}
    ([k],\rho_0) \otimes (([1],\mathbf{1})\oplus([1],\pi_a)\oplus([-1],\pi_b)\oplus([-1],\pi_c))=  ([k],\rho_0) \oplus ([k],\rho_2) \oplus ([k],\rho_2) \oplus ([k],\rho_0), 
\end{equation*}
\begin{equation*}
     ([k],\rho_1) \otimes (([1],\mathbf{1})\oplus([1],\pi_a)\oplus([-1],\pi_b)\oplus([-1],\pi_c)) = ([k],\rho_1) \oplus ([k],\rho_3) \oplus ([k],\rho_1) \oplus ([k],\rho_3) ,
\end{equation*}
\begin{equation*}
    ([k],\rho_2) \otimes (([1],\mathbf{1})\oplus([1],\pi_a)\oplus([-1],\pi_b)\oplus([-1],\pi_c))= ([k],\rho_2) \oplus ([k],\rho_0) \oplus ([k],\rho_0) \oplus ([k],\rho_2)  , 
\end{equation*}
\begin{equation*}
   ([k],\rho_3) \otimes (([1],\mathbf{1})\oplus([1],\pi_a)\oplus([-1],\pi_b)\oplus([-1],\pi_c))= ([k],\rho_3) \oplus ([k],\rho_1) \oplus ([k],\rho_3) \oplus ([k],\rho_1) 
\end{equation*}     
We conclude that all the simple objects of $\mathcal{Z}$ remain simple in $\mathcal{F}$ except the following:
\begin{equation*}
  ([1], \pi_m), ([-1], \pi_m), ([i],\rho_0), ([i],\rho_2), ([j],\rho_i), ([k],\rho_i) , \quad i=(0,1,2,3)
\end{equation*}
in addition, $([i],\rho_1), ([i],\rho_3)$ decomposes into two copies of the same simple object in $\mathcal{F}$. Now we can look for identifications, we find the following: 
\begin{align*}
    ([1],\mathbf{1}) \sim ([1],\pi_a) \sim ([-1],\pi_b) \sim ([-1],\pi_c),\\ ([1],\pi_b) \sim ([1],\pi_c) \sim ([-1],\pi_a) \sim ([-1],\mathbf{1})\\
     ([1],\pi_m) \sim ([-1],\pi_m),~ ([i],\rho_0) \sim ([i],\rho_2), ~  ([j],\rho_0) \sim ([j],\rho_2),\\ ([j],\rho_1) \sim ([j],\rho_3),~([k],\rho_0) \sim ([k],\rho_2), ~([k],\rho_1) \sim ([k],\rho_3)
\end{align*}  
We identify the simple objects of the reduced topological order,
\begin{equation*}
     \begin{aligned}
        L_0 &= ([1],\mathbf{1})\oplus([1],\pi_a)\oplus([-1],\pi_b)\oplus([-1],\pi_c) \\
        L_1 &= ([1],\pi_b) \oplus ([1],\pi_c) \oplus ([-1],\mathbf{1}) \oplus ([-1],\pi_a) 
\end{aligned}
\end{equation*}
Moreover, we will have anyons that decompose in $\mathcal{F}$ that can be written as the following:
\begin{equation*}
    \begin{aligned}
        ([1],\pi_m) \sim ([-1],\pi_m) = L_2 \oplus L_3 \\
        ([i],\rho_0) \sim  ([i],\rho_2) = L_4 \oplus L_5 \\
        ([i],\rho_1) \sim 2L_6 , \quad ([i],\rho_3) \sim 2 L_7 \\
        ([j],\rho_0) \sim  ([j],\rho_2) = L_8 \oplus L_9 \\
        ([j],\rho_1) \sim  ([j],\rho_3) = L_{10} \oplus L_{11} \\
        ([k],\rho_0) \sim  ([k],\rho_2) =  L_{12} \oplus L_{13} \\
        ([k],\rho_1) \sim  ([k],\rho_3) = L_{14} \oplus L_{15} \\
    \end{aligned}
\end{equation*}
where, 
\begin{equation*}
    \begin{aligned}
        L_2 = L_3 &= ([1],\pi_m) \oplus ([-1],\pi_m) ,\\
        L_4 = L_5 &= ([i],\rho_0) \oplus  ([i],\rho_2) ,\\
        L_6 &= 2 ([i],\rho_1), \\
        L_7 &= 2([i],\rho_3), \\
        L_8 = L_9 &= ([j],\rho_0) \oplus  ([j],\rho_2), \\
        L_{10} = L_{11} &= ([j],\rho_1) \oplus  ([j],\rho_3) \\
        L_{12} = L_{13} &= ([k],\rho_0) \oplus  ([k],\rho_2) \\
        L_{14}= L_{15} &=([k],\rho_1) \oplus  ([k],\rho_3)
    \end{aligned}
\end{equation*}
Now we apply the topological spin rule to the lifts of the simple objects, we only keep $L_0,L_1,L_6,L_7$. Among them, we have two bosons corresponding to $L_0$ and $L_1$. The remaining ones are semion and anti semion. Hence the reduced topological order is $\Z^\omega_2$.
\subsubsection*{$\mathcal{A}_{4,18}=([1],\mathbf{1})\oplus([1],\pi_b)\oplus([-1],\pi_a)\oplus([-1]\,\pi_c)$:}
\begin{equation*}
    ([1],\mathbf{1}) \otimes  (([1],\mathbf{1})\oplus([1],\pi_b)\oplus([-1],\pi_a)\oplus([-1]\,\pi_c)) = ([1],\mathbf{1})\oplus([1],\pi_b)\oplus([-1],\pi_a)\oplus([-1]\,\pi_c) ,
\end{equation*}
\begin{equation*}
    ([1],\pi_a)\otimes (([1],\mathbf{1})\oplus([1],\pi_b)\oplus([-1],\pi_a)\oplus([-1]\,\pi_c))= ([1],\pi_a) \oplus([1],\pi_c) \oplus ([-1],\mathbf{1}) \oplus ([-1],\pi_b), 
\end{equation*}
\begin{equation*}
    ([1],\pi_b) \otimes (([1],\mathbf{1})\oplus([1],\pi_b)\oplus([-1],\pi_a)\oplus([-1]\,\pi_c))= ([1],\pi_b) \oplus ([1],\mathbf{1}) \oplus ([-1],\pi_c) \oplus ([-1],\pi_a),
\end{equation*}
\begin{equation*}
    ([1],\pi_c) \otimes (([1],\mathbf{1})\oplus([1],\pi_b)\oplus([-1],\pi_a)\oplus([-1]\,\pi_c))= ([1],\pi_c) \oplus ([1],\pi_a) \oplus ([-1],\pi_b) \oplus ([-1],\mathbf{1}), 
\end{equation*}
\begin{equation*}
    ([1],\pi_m) \otimes (([1],\mathbf{1})\oplus([1],\pi_b)\oplus([-1],\pi_a)\oplus([-1]\,\pi_c))= ([1],\pi_m) \oplus ([1],\pi_m) \oplus ([-1],\pi_m) \oplus ([-1],\pi_m) ,
\end{equation*}       
\begin{equation*}
    ([-1],\mathbf{1}) \otimes (([1],\mathbf{1})\oplus([1],\pi_b)\oplus([-1],\pi_a)\oplus([-1]\,\pi_c))= ([-1],\mathbf{1}) \oplus ([-1],\pi_b)\oplus ([1],\pi_a)\oplus ([1],\pi_c) ,
\end{equation*}       
\begin{equation*}
     ([-1],\pi_a) \otimes (([1],\mathbf{1})\oplus([1],\pi_b)\oplus([-1],\pi_a)\oplus([-1]\,\pi_c))= ([-1],\pi_a) \oplus ([-1],\pi_c)  \oplus ([1],\mathbf{1}) \oplus ([1],\pi_b) ,
\end{equation*}       
\begin{equation*}
    ([-1],\pi_b) \otimes (([1],\mathbf{1})\oplus([1],\pi_b)\oplus([-1],\pi_a)\oplus([-1]\,\pi_c))= ([-1],\pi_b) \oplus ([-1],\mathbf{1}) \oplus  ([1],\pi_c)\oplus ([1],\pi_a) ,
\end{equation*}       
\begin{equation*}
    ([-1],\pi_c) \otimes (([1],\mathbf{1})\oplus([1],\pi_b)\oplus([-1],\pi_a)\oplus([-1]\,\pi_c))= ([-1],\pi_c) \oplus ([-1],\pi_a)\oplus ([1],\pi_b)\oplus ([1],\mathbf{1}), 
\end{equation*}
\begin{equation*}
     ([-1],\pi_m) \otimes (([1],\mathbf{1})\oplus([1],\pi_b)\oplus([-1],\pi_a)\oplus([-1]\,\pi_c))= ([-1],\pi_m) \oplus  ([-1],\pi_m) \oplus  ([1],\pi_m) \oplus ([1],\pi_m) , 
\end{equation*}
\begin{equation*}
     ([i],\rho_0) \otimes (([1],\mathbf{1})\oplus([1],\pi_b)\oplus([-1],\pi_a)\oplus([-1]\,\pi_c))= ([i],\rho_0) \oplus ([i],\rho_2) \oplus ([i],\rho_0) \oplus ([i],\rho_2),
\end{equation*}
\begin{equation*}
    ([i],\rho_1) \otimes (([1],\mathbf{1})\oplus([1],\pi_b)\oplus([-1],\pi_a)\oplus([-1]\,\pi_c))= ([i],\rho_1) \oplus ([i],\rho_3)\oplus ([i],\rho_3) \oplus ([i],\rho_1)  , 
\end{equation*}
\begin{equation*}
    ([i],\rho_2) \otimes (([1],\mathbf{1})\oplus([1],\pi_b)\oplus([-1],\pi_a)\oplus([-1]\,\pi_c))= ([i],\rho_2) \oplus ([i],\rho_0) \oplus ([i],\rho_2) \oplus  ([i],\rho_0), 
\end{equation*}
\begin{equation*}
    ([i],\rho_3) \otimes (([1],\mathbf{1})\oplus([1],\pi_b)\oplus([-1],\pi_a)\oplus([-1]\,\pi_c))= ([i],\rho_3) \oplus ([i],\rho_1) \oplus ([i],\rho_1) \oplus  ([i],\rho_3) ,
\end{equation*}
\begin{equation*}
    ([j],\rho_0) \otimes (([1],\mathbf{1})\oplus([1],\pi_b)\oplus([-1],\pi_a)\oplus([-1]\,\pi_c))= ([j],\rho_0) \oplus ([j],\rho_0) \oplus ([j],\rho_2) \oplus ([j],\rho_2) ,
\end{equation*}
\begin{equation*}
    ([j],\rho_1) \otimes (([1],\mathbf{1})\oplus([1],\pi_b)\oplus([-1],\pi_a)\oplus([-1]\,\pi_c))= ([j],\rho_1) \oplus ([j],\rho_1) \oplus ([j],\rho_1) \oplus ([j],\rho_1),
\end{equation*}
\begin{equation*}
    ([j],\rho_2) \otimes (([1],\mathbf{1})\oplus([1],\pi_b)\oplus([-1],\pi_a)\oplus([-1]\,\pi_c))= ([j],\rho_2) \oplus ([j],\rho_2) \oplus ([j],\rho_0) \oplus ([j],\rho_0), 
\end{equation*}
\begin{equation*}
    ([j],\rho_3) \otimes (([1],\mathbf{1})\oplus([1],\pi_b)\oplus([-1],\pi_a)\oplus([-1]\,\pi_c))= ([j],\rho_3) \oplus ([j],\rho_3) \oplus ([j],\rho_3) \oplus ([j],\rho_3),
\end{equation*}
\begin{equation*}
    ([k],\rho_0) \otimes (([1],\mathbf{1})\oplus([1],\pi_b)\oplus([-1],\pi_a)\oplus([-1]\,\pi_c))=  ([k],\rho_0) \oplus ([k],\rho_2) \oplus ([k],\rho_2) \oplus ([k],\rho_0) ,
\end{equation*}    
\begin{equation*}
    ([k],\rho_1) \otimes (([1],\mathbf{1})\oplus([1],\pi_b)\oplus([-1],\pi_a)\oplus([-1]\,\pi_c)) = ([k],\rho_1)  \oplus ([k],\rho_3) \oplus ([k],\rho_1) \oplus ([k],\rho_3) ,
\end{equation*}       
\begin{equation*}
    ([k],\rho_2) \otimes (([1],\mathbf{1})\oplus([1],\pi_b)\oplus([-1],\pi_a)\oplus([-1]\,\pi_c))= ([k],\rho_2)  \oplus ([k],\rho_0) \oplus ([k],\rho_0) \oplus ([k],\rho_2),
\end{equation*}
\begin{equation*}
     ([k],\rho_3) \otimes (([1],\mathbf{1})\oplus([1],\pi_b)\oplus([-1],\pi_a)\oplus([-1]\,\pi_c))= ([k],\rho_3)  \oplus ([k],\rho_1) \oplus ([k],\rho_3) \oplus ([k],\rho_1)
\end{equation*}
We conclude that all the simple objects of $\mathcal{Z}$ remain simple in $\mathcal{F}$ except the following:
\begin{equation*}
  ([1], \pi_m), ([-1], \pi_m), ([i],\rho_i), ([j],\rho_0), ([j],\rho_2), ([k],\rho_i) , \quad i=(0,1,2,3)
\end{equation*}
in addition, $([j],\rho_1), ([j],\rho_3)$ splits into two copies of the same object. Now we can look for identifications, we find the following: 
\begin{align*}
    ([1],\mathbf{1}) \sim ([1],\pi_b) \sim ([-1],\pi_a) \sim ([-1],\pi_c),\\ ([1],\pi_a) \sim ([1],\pi_c) \sim ([-1],\pi_b) \sim ([-1],\mathbf{1})\\
     ([1],\pi_m) \sim ([-1],\pi_m),~ ([i],\rho_0) \sim ([i],\rho_2), ~  ([j],\rho_0) \sim ([j],\rho_2),\\ ([i],\rho_1) \sim ([i],\rho_3),~([k],\rho_0) \sim ([k],\rho_2), ~([k],\rho_1) \sim ([k],\rho_3)
\end{align*}  
We identify the simple objects of the reduced topological order,
\begin{equation*}
     \begin{aligned}
        L_0 &= ([1],\mathbf{1})\oplus([1],\pi_b)\oplus([-1],\pi_a)\oplus([-1],\pi_c) \\
        L_1 &= ([1],\pi_a) \oplus ([1],\pi_c) \oplus ([-1],\mathbf{1}) \oplus ([-1],\pi_b)
\end{aligned}
\end{equation*}
Moreover, we will have anyons that decompose in $\mathcal{F}$ that can be written as the following:
\begin{equation*}
    \begin{aligned}
        ([1],\pi_m) \sim ([-1],\pi_m) = L_2 \oplus L_3 \\
        ([i],\rho_0) \sim  ([i],\rho_2) = L_4 \oplus L_5 \\
        ([i],\rho_1) \sim  ([i],\rho_3) = L_6 \oplus L_7 \\
        ([j],\rho_0) \sim  ([j],\rho_2) = L_8 \oplus L_9 \\
        ([j],\rho_1) \sim 2 L_{10} , \quad ([j],\rho_3) \sim 2L_{11} \\
        ([k],\rho_0) \sim  ([k],\rho_2) =  L_{12} \oplus L_{13} \\
        ([k],\rho_1) \sim  ([k],\rho_3) = L_{14} \oplus L_{15} \\
    \end{aligned}
\end{equation*}
where, 
\begin{equation*}
    \begin{aligned}
        L_2 = L_3 &= ([1],\pi_m) \oplus ([-1],\pi_m) ,\\
        L_4 = L_5 &= ([i],\rho_0) \oplus  ([i],\rho_2) ,\\
        L_4 = L_5 &= ([i],\rho_1) \oplus  ([i],\rho_3) ,\\
        L_8 = L_9 &= ([j],\rho_0) \oplus  ([j],\rho_2), \\
        L_{10} &= 2 ([j],\rho_1), \\
        L_{11} &= 2([j],\rho_3), \\
        L_{12} = L_{13} &= ([k],\rho_0) \oplus  ([k],\rho_2) \\
        L_{14}= L_{15} &=([k],\rho_1) \oplus  ([k],\rho_3)
    \end{aligned}
\end{equation*}
Now we apply the topological spin rule to the lifts of the simple objects, we only keep $L_0,L_1,L_{10},L_{11}$. Among them, we have two bosons corresponding to $L_0$ and $L_1$. The remaining ones are semion and anti semion. Hence the reduced topological order is $\Z^\omega_2$.
\subsubsection*{$\mathcal{A}_{4,19}=([1],\mathbf{1})\oplus([1],\pi_c)\oplus([-1],\pi_a)\oplus([-1],\pi_b)$:}
\begin{equation*}
     ([1],\mathbf{1}) \otimes  (([1],\mathbf{1})\oplus([1],\pi_c)\oplus([-1],\pi_a)\oplus([-1],\pi_b)) = ([1],\mathbf{1})\oplus([1],\pi_c)\oplus([-1],\pi_a)\oplus([-1],\pi_b) ,
\end{equation*}
\begin{equation*}
    ([1],\pi_a)\otimes (([1],\mathbf{1})\oplus([1],\pi_c)\oplus([-1],\pi_a)\oplus([-1],\pi_b))= ([1],\pi_a) \oplus  ([1],\pi_b) \oplus ([-1],\mathbf{1})\oplus ([-1],\pi_c) ,
\end{equation*}
\begin{equation*}
    ([1],\pi_b) \otimes (([1],\mathbf{1})\oplus([1],\pi_c)\oplus([-1],\pi_a)\oplus([-1],\pi_b))= ([1],\pi_b) \oplus  ([1],\pi_a) \oplus ([-1],\pi_c)\oplus ([-1],\mathbf{1}),
\end{equation*}
\begin{equation*}
    ([1],\pi_c) \otimes (([1],\mathbf{1})\oplus([1],\pi_c)\oplus([-1],\pi_a)\oplus([-1],\pi_b))= ([1],\pi_c) \oplus  ([1],\mathbf{1}) \oplus ([-1],\pi_b)\oplus ([-1],\pi_a),
\end{equation*}
\begin{equation*}
    ([1],\pi_m) \otimes (([1],\mathbf{1})\oplus([1],\pi_c)\oplus([-1],\pi_a)\oplus([-1],\pi_b))= ([1],\pi_m) \oplus  ([1],\pi_m) \oplus ([-1],\pi_m)\oplus ([-1],\pi_m), 
\end{equation*}
\begin{equation*}
    ([-1],\mathbf{1}) \otimes (([1],\mathbf{1})\oplus([1],\pi_c)\oplus([-1],\pi_a)\oplus([-1],\pi_b))= ([-1],\mathbf{1}) \oplus  ([-1],\pi_c) \oplus ([1],\pi_a)\oplus ([1],\pi_b),
\end{equation*}
\begin{equation*}
    ([-1],\pi_a) \otimes (([1],\mathbf{1})\oplus([1],\pi_c)\oplus([-1],\pi_a)\oplus([-1],\pi_b))= ([-1],\pi_a) \oplus  ([-1],\pi_b) \oplus ([1],\mathbf{1})\oplus ([1],\pi_c),
\end{equation*}
\begin{equation*}
    ([-1],\pi_b) \otimes (([1],\mathbf{1})\oplus([1],\pi_c)\oplus([-1],\pi_a)\oplus([-1],\pi_b))= ([-1],\pi_b) \oplus  ([-1],\pi_a) \oplus ([1],\pi_c)\oplus ([1],\mathbf{1}) ,
\end{equation*}
\begin{equation*}
    ([-1],\pi_c) \otimes (([1],\mathbf{1})\oplus([1],\pi_c)\oplus([-1],\pi_a)\oplus([-1],\pi_b))= ([-1],\pi_c) \oplus  ([-1],\mathbf{1}) \oplus ([1],\pi_b)\oplus  ([1],\pi_a),
\end{equation*}
\begin{equation*}
    ([-1],\pi_m) \otimes (([1],\mathbf{1})\oplus([1],\pi_c)\oplus([-1],\pi_a)\oplus([-1],\pi_b))=  ([-1],\pi_m) \oplus  ([-1],\pi_m) \oplus ([1],\pi_m)\oplus ([1],\pi_m)  , 
\end{equation*}
\begin{equation*}
     ([i],\rho_0) \otimes (([1],\mathbf{1})\oplus([1],\pi_c)\oplus([-1],\pi_a)\oplus([-1],\pi_b))= ([i],\rho_0) \oplus ([i],\rho_2) \oplus ([i],\rho_0) \oplus ([i],\rho_2), 
\end{equation*}
\begin{equation*}
     ([i],\rho_1) \otimes (([1],\mathbf{1})\oplus([1],\pi_c)\oplus([-1],\pi_a)\oplus([-1],\pi_b))= ([i],\rho_1) \oplus ([i],\rho_3) \oplus ([i],\rho_3) \oplus ([i],\rho_1), 
\end{equation*}
\begin{equation*}
     ([i],\rho_2) \otimes (([1],\mathbf{1})\oplus([1],\pi_c)\oplus([-1],\pi_a)\oplus([-1],\pi_b))= ([i],\rho_2) \oplus ([i],\rho_0) \oplus ([i],\rho_2) \oplus ([i],\rho_0),
\end{equation*}
\begin{equation*}
    ([i],\rho_3) \otimes (([1],\mathbf{1})\oplus([1],\pi_c)\oplus([-1],\pi_a)\oplus([-1],\pi_b))= ([i],\rho_3) \oplus ([i],\rho_1) \oplus ([i],\rho_1) \oplus ([i],\rho_3),
\end{equation*}
\begin{equation*}
     ([j],\rho_0) \otimes (([1],\mathbf{1})\oplus([1],\pi_c)\oplus([-1],\pi_a)\oplus([-1],\pi_b))= ([j],\rho_0) \oplus ([j],\rho_2) \oplus ([j],\rho_2) \oplus ([j],\rho_0) ,
\end{equation*}
\begin{equation*}
    ([j],\rho_1) \otimes (([1],\mathbf{1})\oplus([1],\pi_c)\oplus([-1],\pi_a)\oplus([-1],\pi_b))= ([j],\rho_1)  \oplus ([j],\rho_3) \oplus ([j],\rho_1) \oplus ([j],\rho_3),
\end{equation*}
\begin{equation*}
    ([j],\rho_2) \otimes (([1],\mathbf{1})\oplus([1],\pi_c)\oplus([-1],\pi_a)\oplus([-1],\pi_b))= ([j],\rho_2)  \oplus ([j],\rho_0) \oplus ([j],\rho_0) \oplus ([j],\rho_0), 
\end{equation*}
\begin{equation*}
     ([j],\rho_3) \otimes (([1],\mathbf{1})\oplus([1],\pi_c)\oplus([-1],\pi_a)\oplus([-1],\pi_b))= ([j],\rho_3)  \oplus ([j],\rho_1) \oplus ([j],\rho_3) \oplus ([j],\rho_1),
\end{equation*}
\begin{equation*}
    ([k],\rho_0) \otimes (([1],\mathbf{1})\oplus([1],\pi_c)\oplus([-1],\pi_a)\oplus([-1],\pi_b))=  ([k],\rho_0)  \oplus ([k],\rho_0) \oplus ([k],\rho_2) \oplus ([k],\rho_2),
\end{equation*}
\begin{equation*}
     ([k],\rho_1) \otimes (([1],\mathbf{1})\oplus([1],\pi_c)\oplus([-1],\pi_a)\oplus([-1],\pi_b)) = ([k],\rho_1)  \oplus ([k],\rho_1) \oplus ([k],\rho_1) \oplus ([k],\rho_1),
\end{equation*}
\begin{equation*}
     ([k],\rho_2) \otimes (([1],\mathbf{1})\oplus([1],\pi_c)\oplus([-1],\pi_a)\oplus([-1],\pi_b))= ([k],\rho_2)  \oplus ([k],\rho_2) \oplus ([k],\rho_0) \oplus ([k],\rho_0) ,
\end{equation*}
\begin{equation*}
     ([k],\rho_3) \otimes (([1],\mathbf{1})\oplus([1],\pi_c)\oplus([-1],\pi_a)\oplus([-1],\pi_b))= ([k],\rho_3)  \oplus ([k],\rho_3) \oplus ([k],\rho_3) \oplus ([k],\rho_3)
\end{equation*}
We conclude that all the simple objects of $\mathcal{Z}$ remain simple in $\mathcal{F}$ except the following:
\begin{equation*}
  ([1], \pi_m), ([-1], \pi_m), ([i],\rho_i), ([j],\rho_i), ([k],\rho_0), ([k],\rho_2) , \quad i=(0,1,2,3)
\end{equation*}
in addition, $([k],\rho_1), ([k],\rho_3)$ splits into two copies of the same simple object. Now we can look for identifications, we find the following: 
\begin{align*}
    ([1],\mathbf{1}) \sim ([1],\pi_c) \sim ([-1],\pi_a) \sim ([-1],\pi_b),\\ ([1],\pi_a) \sim ([1],\pi_b) \sim ([-1],\pi_c) \sim ([-1],\mathbf{1})\\
     ([1],\pi_m) \sim ([-1],\pi_m),~ ([i],\rho_0) \sim ([i],\rho_2), ~  ([j],\rho_0) \sim ([j],\rho_2),\\ ([i],\rho_1) \sim ([i],\rho_3),~([k],\rho_0) \sim ([k],\rho_2), ~([j],\rho_1) \sim ([j],\rho_3)
\end{align*}  
We identify the simple objects of the reduced topological order,
\begin{equation*}
     \begin{aligned}
        L_0 &= ([1],\mathbf{1})\oplus([1],\pi_c)\oplus([-1],\pi_a)\oplus([-1],\pi_b) \\
        L_1 &= ([1],\pi_a) \oplus  ([1],\pi_b) \oplus ([-1],\mathbf{1})\oplus ([-1],\pi_c)
\end{aligned}
\end{equation*}
Moreover, we will have anyons that decompose in $\mathcal{F}$ that can be written as the following:
\begin{equation*}
    \begin{aligned}
        ([1],\pi_m) \sim ([-1],\pi_m) = L_2 \oplus L_3 \\
         ([i],\rho_0) \sim  ([i],\rho_2) = L_4 \oplus L_5 \\
        ([i],\rho_1) \sim  ([i],\rho_3) = L_6 \oplus L_7 \\
        ([j],\rho_0) \sim  ([j],\rho_2) = L_8 \oplus L_9 \\
        ([j],\rho_1) \sim  ([j],\rho_3) = L_{10} \oplus L_{11} \\
        ([k],\rho_0) \sim  ([k],\rho_2) =  L_{12} \oplus L_{13} \\
        ([j],\rho_1) \sim 2 L_{14} , \quad ([j],\rho_3) \sim 2L_{15}
    \end{aligned}
\end{equation*}
where, 
\begin{equation*}
    \begin{aligned}
        L_2 = L_3 &= ([1],\pi_m) \oplus ([-1],\pi_m) ,\\
        L_4 = L_5 &= ([i],\rho_0) \oplus  ([i],\rho_2) ,\\
        L_6 =  L_7 &=  ([i],\rho_1) \oplus ([i],\rho_3), \\
        L_8 = L_9 &= ([j],\rho_0) \oplus  ([j],\rho_2), \\
        L_{10} = L_{11} &= ([j],\rho_1) \oplus  ([j],\rho_3) \\
        L_{12} = L_{13} &= ([k],\rho_0) \oplus  ([k],\rho_2) \\
        L_{14} &= 2([k],\rho_1), \\ 
        L_{15} &= 2 ([k],\rho_3)
    \end{aligned}
\end{equation*}
Now we apply the topological spin rule to the lifts of the simple objects, we only keep $L_0,L_1,L_{14},L_{15}$. Among them, we have two bosons corresponding to $L_0$ and $L_1$. The remaining ones are semion and anti semion. Hence the reduced topological order is $\Z^\omega_2$.

\providecommand{\href}[2]{#2}\begingroup\raggedright\endgroup


\begin{thebibliography}{100}

\bibitem{Noether_1971}
E.~Noether, \emph{Invariant variation problems}, \href{https://doi.org/10.1080/00411457108231446}{\emph{Transport Theory and Statistical Physics} {\bfseries 1} (Jan., 1971) 186–207}.

\bibitem{Gaiotto:2014kfa}
D.~Gaiotto, A.~Kapustin, N.~Seiberg and B.~Willett, \emph{{Generalized Global Symmetries}}, \href{https://doi.org/10.1007/JHEP02(2015)172}{\emph{JHEP} {\bfseries 02} (2015) 172}, [\href{https://arxiv.org/abs/1412.5148}{{\ttfamily 1412.5148}}].

\bibitem{Freed:2022iao}
D.~S. Freed, \emph{{Introduction to topological symmetry in QFT.}}, \href{https://doi.org/10.1090/pspum/107/01946}{\emph{Proc. Symp. Pure Math.} {\bfseries 107} (2024) 93--106}, [\href{https://arxiv.org/abs/2212.00195}{{\ttfamily 2212.00195}}].

\bibitem{Cordova:2022ruw}
C.~Cordova, T.~T. Dumitrescu, K.~Intriligator and S.-H. Shao, \emph{{Snowmass White Paper: Generalized Symmetries in Quantum Field Theory and Beyond}},  in \emph{{Snowmass 2021}}, 5, 2022, \href{https://arxiv.org/abs/2205.09545}{{\ttfamily 2205.09545}}.

\bibitem{Bah:2022xfv}
I.~Bah, D.~Freed, G.~W. Moore, N.~Nekrasov, S.~S. Razamat and S.~Sch{\"a}fer-Nameki, \emph{{Snowmass Whitepaper: Physical Mathematics 2021}},  \href{https://arxiv.org/abs/2203.05078}{{\ttfamily 2203.05078}}.

\bibitem{Schafer-Nameki:2023jdn}
S.~Schafer-Nameki, \emph{{ICTP lectures on (non-)invertible generalized symmetries}}, \href{https://doi.org/10.1016/j.physrep.2024.01.007}{\emph{Phys. Rept.} {\bfseries 1063} (2024) 1--55}, [\href{https://arxiv.org/abs/2305.18296}{{\ttfamily 2305.18296}}].

\bibitem{Bhardwaj:2023kri}
L.~Bhardwaj, L.~E. Bottini, L.~Fraser-Taliente, L.~Gladden, D.~S.~W. Gould, A.~Platschorre et~al., \emph{{Lectures on generalized symmetries}}, \href{https://doi.org/10.1016/j.physrep.2023.11.002}{\emph{Phys. Rept.} {\bfseries 1051} (2024) 1--87}, [\href{https://arxiv.org/abs/2307.07547}{{\ttfamily 2307.07547}}].

\bibitem{Shao:2023gho}
S.-H. Shao, \emph{{What's Done Cannot Be Undone: TASI Lectures on Non-Invertible Symmetries}},  \href{https://arxiv.org/abs/2308.00747}{{\ttfamily 2308.00747}}.

\bibitem{Gomes:2023ahz}
P.~R.~S. Gomes, \emph{{An introduction to higher-form symmetries}}, \href{https://doi.org/10.21468/SciPostPhysLectNotes.74}{\emph{SciPost Phys. Lect. Notes} {\bfseries 74} (2023) 1}, [\href{https://arxiv.org/abs/2303.01817}{{\ttfamily 2303.01817}}].

\bibitem{Brennan:2023mmt}
T.~D. Brennan and S.~Hong, \emph{{Introduction to Generalized Global Symmetries in QFT and Particle Physics}},  \href{https://arxiv.org/abs/2306.00912}{{\ttfamily 2306.00912}}.

\bibitem{McGreevy:2022oyu}
J.~McGreevy, \emph{{Generalized Symmetries in Condensed Matter}}, \href{https://doi.org/10.1146/annurev-conmatphys-040721-021029}{\emph{Ann. Rev. Condensed Matter Phys.} {\bfseries 14} (2023) 57--82}, [\href{https://arxiv.org/abs/2204.03045}{{\ttfamily 2204.03045}}].

\bibitem{Chang:2018iay}
C.-M. Chang, Y.-H. Lin, S.-H. Shao, Y.~Wang and X.~Yin, \emph{{Topological Defect Lines and Renormalization Group Flows in Two Dimensions}}, \href{https://doi.org/10.1007/JHEP01(2019)026}{\emph{JHEP} {\bfseries 01} (2019) 026}, [\href{https://arxiv.org/abs/1802.04445}{{\ttfamily 1802.04445}}].

\bibitem{Bhardwaj:2022yxj}
L.~Bhardwaj, L.~E. Bottini, S.~Schafer-Nameki and A.~Tiwari, \emph{{Non-invertible higher-categorical symmetries}}, \href{https://doi.org/10.21468/SciPostPhys.14.1.007}{\emph{SciPost Phys.} {\bfseries 14} (2023) 007}, [\href{https://arxiv.org/abs/2204.06564}{{\ttfamily 2204.06564}}].

\bibitem{Iqbal:2024pee}
N.~Iqbal, \emph{{Jena lectures on generalized global symmetries: principles and applications}},  7, 2024, \href{https://arxiv.org/abs/2407.20815}{{\ttfamily 2407.20815}}.

\bibitem{Costa:2024wks}
D.~Costa et~al., \emph{{Simons Lectures on Categorical Symmetries}},  11, 2024, \href{https://arxiv.org/abs/2411.09082}{{\ttfamily 2411.09082}}.

\bibitem{Cordova:2019jnf}
C.~C{\' o}rdova, D.~S. Freed, H.~T. Lam and N.~Seiberg, \emph{{Anomalies in the Space of Coupling Constants and Their Dynamical Applications I}}, \href{https://doi.org/10.21468/SciPostPhys.8.1.001}{\emph{SciPost Phys.} {\bfseries 8} (2020) 001}, [\href{https://arxiv.org/abs/1905.09315}{{\ttfamily 1905.09315}}].

\bibitem{Cordova:2019uob}
C.~C{\' o}rdova, D.~S. Freed, H.~T. Lam and N.~Seiberg, \emph{{Anomalies in the Space of Coupling Constants and Their Dynamical Applications II}}, \href{https://doi.org/10.21468/SciPostPhys.8.1.002}{\emph{SciPost Phys.} {\bfseries 8} (2020) 002}, [\href{https://arxiv.org/abs/1905.13361}{{\ttfamily 1905.13361}}].

\bibitem{Aloni:2024jpb}
D.~Aloni, E.~Garc\'\i{}a-Valdecasas, M.~Reece and M.~Suzuki, \emph{{Spontaneously broken (-1)-form U(1) symmetries}}, \href{https://doi.org/10.21468/SciPostPhys.17.2.031}{\emph{SciPost Phys.} {\bfseries 17} (2024) 031}, [\href{https://arxiv.org/abs/2402.00117}{{\ttfamily 2402.00117}}].

\bibitem{Lin:2025oml}
L.~Lin, D.~Robbins and S.~Roy, \emph{{Decomposition and (Non-Invertible) (-1)-Form Symmetries from the Symmetry Topological Field Theory}},  \href{https://arxiv.org/abs/2503.21862}{{\ttfamily 2503.21862}}.

\bibitem{Yu:2024jtk}
X.~Yu, \emph{{Gauging in Parameter Space: A Top-Down Perspective}},  \href{https://arxiv.org/abs/2411.14997}{{\ttfamily 2411.14997}}.

\bibitem{Najjar:2024vmm}
M.~Najjar, L.~Santilli and Y.-N. Wang, \emph{{(\ensuremath{-}1)-form symmetries from M-theory and SymTFTs}}, \href{https://doi.org/10.1007/JHEP03(2025)134}{\emph{JHEP} {\bfseries 03} (2025) 134}, [\href{https://arxiv.org/abs/2411.19683}{{\ttfamily 2411.19683}}].

\bibitem{Heidenreich:2020pkc}
B.~Heidenreich, J.~McNamara, M.~Montero, M.~Reece, T.~Rudelius and I.~Valenzuela, \emph{{Chern-Weil global symmetries and how quantum gravity avoids them}}, \href{https://doi.org/10.1007/JHEP11(2021)053}{\emph{JHEP} {\bfseries 11} (2021) 053}, [\href{https://arxiv.org/abs/2012.00009}{{\ttfamily 2012.00009}}].

\bibitem{Robbins:2025apg}
D.~Robbins and S.~Roy, \emph{{(-1)-Form Symmetries and Anomaly Shifting from SymTFT}},  \href{https://arxiv.org/abs/2505.14807}{{\ttfamily 2505.14807}}.

\bibitem{Heckman:2022muc}
J.~J. Heckman, M.~H\"ubner, E.~Torres and H.~Y. Zhang, \emph{{The Branes Behind Generalized Symmetry Operators}}, \href{https://doi.org/10.1002/prop.202200180}{\emph{Fortsch. Phys.} {\bfseries 71} (2023) 2200180}, [\href{https://arxiv.org/abs/2209.03343}{{\ttfamily 2209.03343}}].

\bibitem{Apruzzi:2022rei}
F.~Apruzzi, I.~Bah, F.~Bonetti and S.~Schafer-Nameki, \emph{{Noninvertible Symmetries from Holography and Branes}}, \href{https://doi.org/10.1103/PhysRevLett.130.121601}{\emph{Phys. Rev. Lett.} {\bfseries 130} (2023) 121601}, [\href{https://arxiv.org/abs/2208.07373}{{\ttfamily 2208.07373}}].

\bibitem{Apruzzi:2021nmk}
F.~Apruzzi, F.~Bonetti, I.~Garc\'\i{}a~Etxebarria, S.~S. Hosseini and S.~Schafer-Nameki, \emph{{Symmetry TFTs from String Theory}}, \href{https://doi.org/10.1007/s00220-023-04737-2}{\emph{Commun. Math. Phys.} {\bfseries 402} (2023) 895--949}, [\href{https://arxiv.org/abs/2112.02092}{{\ttfamily 2112.02092}}].

\bibitem{Albertini:2020mdx}
F.~Albertini, M.~Del~Zotto, I.~Garc\'\i{}a~Etxebarria and S.~S. Hosseini, \emph{{Higher Form Symmetries and M-theory}}, \href{https://doi.org/10.1007/JHEP12(2020)203}{\emph{JHEP} {\bfseries 12} (2020) 203}, [\href{https://arxiv.org/abs/2005.12831}{{\ttfamily 2005.12831}}].

\bibitem{GarciaEtxebarria:2022vzq}
I.~Garc\'\i{}a~Etxebarria, \emph{{Branes and Non-Invertible Symmetries}}, \href{https://doi.org/10.1002/prop.202200154}{\emph{Fortsch. Phys.} {\bfseries 70} (2022) 2200154}, [\href{https://arxiv.org/abs/2208.07508}{{\ttfamily 2208.07508}}].

\bibitem{Freed:2012bs}
D.~S. Freed and C.~Teleman, \emph{{Relative quantum field theory}}, \href{https://doi.org/10.1007/s00220-013-1880-1}{\emph{Commun. Math. Phys.} {\bfseries 326} (2014) 459--476}, [\href{https://arxiv.org/abs/1212.1692}{{\ttfamily 1212.1692}}].

\bibitem{Freed:2022qnc}
D.~S. Freed, G.~W. Moore and C.~Teleman, \emph{{Topological symmetry in quantum field theory}},  \href{https://arxiv.org/abs/2209.07471}{{\ttfamily 2209.07471}}.

\bibitem{Gaiotto:2020iye}
D.~Gaiotto and J.~Kulp, \emph{{Orbifold groupoids}}, \href{https://doi.org/10.1007/JHEP02(2021)132}{\emph{JHEP} {\bfseries 02} (2021) 132}, [\href{https://arxiv.org/abs/2008.05960}{{\ttfamily 2008.05960}}].

\bibitem{Kaidi:2022cpf}
J.~Kaidi, K.~Ohmori and Y.~Zheng, \emph{{Symmetry TFTs for Non-invertible Defects}}, \href{https://doi.org/10.1007/s00220-023-04859-7}{\emph{Commun. Math. Phys.} {\bfseries 404} (2023) 1021--1124}, [\href{https://arxiv.org/abs/2209.11062}{{\ttfamily 2209.11062}}].

\bibitem{Kaidi:2023maf}
J.~Kaidi, E.~Nardoni, G.~Zafrir and Y.~Zheng, \emph{{Symmetry TFTs and anomalies of non-invertible symmetries}}, \href{https://doi.org/10.1007/JHEP10(2023)053}{\emph{JHEP} {\bfseries 10} (2023) 053}, [\href{https://arxiv.org/abs/2301.07112}{{\ttfamily 2301.07112}}].

\bibitem{Bhardwaj:2023ayw}
L.~Bhardwaj and S.~Schafer-Nameki, \emph{{Generalized Charges, Part II: Non-Invertible Symmetries and the Symmetry TFT}},  \href{https://arxiv.org/abs/2305.17159}{{\ttfamily 2305.17159}}.

\bibitem{Heckman:2022xgu}
J.~J. Heckman, M.~Hubner, E.~Torres, X.~Yu and H.~Y. Zhang, \emph{{Top down approach to topological duality defects}}, \href{https://doi.org/10.1103/PhysRevD.108.046015}{\emph{Phys. Rev. D} {\bfseries 108} (2023) 046015}, [\href{https://arxiv.org/abs/2212.09743}{{\ttfamily 2212.09743}}].

\bibitem{Baume:2023kkf}
F.~Baume, J.~J. Heckman, M.~H\"ubner, E.~Torres, A.~P. Turner and X.~Yu, \emph{{SymTrees and Multi-Sector QFTs}}, \href{https://doi.org/10.1103/PhysRevD.109.106013}{\emph{Phys. Rev. D} {\bfseries 109} (2024) 106013}, [\href{https://arxiv.org/abs/2310.12980}{{\ttfamily 2310.12980}}].

\bibitem{Heckman:2024oot}
J.~J. Heckman, M.~H\"ubner and C.~Murdia, \emph{{On the holographic dual of a topological symmetry operator}}, \href{https://doi.org/10.1103/PhysRevD.110.046007}{\emph{Phys. Rev. D} {\bfseries 110} (2024) 046007}, [\href{https://arxiv.org/abs/2401.09538}{{\ttfamily 2401.09538}}].

\bibitem{Heckman:2024zdo}
J.~J. Heckman and M.~H{\"u}bner, \emph{{Celestial Topology, Symmetry Theories, and Evidence for a NonSUSY D3-Brane CFT}}, \href{https://doi.org/10.1002/prop.202400270}{\emph{Fortsch. Phys.} {\bfseries 73} (2025) 2400270}, [\href{https://arxiv.org/abs/2406.08485}{{\ttfamily 2406.08485}}].

\bibitem{Heckman:2025lmw}
J.~J. Heckman, M.~H{\"u}bner and C.~Murdia, \emph{{Symmetry Theories, Wigner's Function, Compactification, and Holography}},  \href{https://arxiv.org/abs/2505.23887}{{\ttfamily 2505.23887}}.

\bibitem{Cvetic:2024dzu}
M.~Cveti{\v{c}}, R.~Donagi, J.~J. Heckman, M.~H{\"u}bner and E.~Torres, \emph{{Cornering relative symmetry theories}}, \href{https://doi.org/10.1103/PhysRevD.111.085026}{\emph{Phys. Rev. D} {\bfseries 111} (2025) 085026}, [\href{https://arxiv.org/abs/2408.12600}{{\ttfamily 2408.12600}}].

\bibitem{Bhardwaj:2023wzd}
L.~Bhardwaj and S.~Schafer-Nameki, \emph{{Generalized charges, part I: Invertible symmetries and higher representations}}, \href{https://doi.org/10.21468/SciPostPhys.16.4.093}{\emph{SciPost Phys.} {\bfseries 16} (2024) 093}, [\href{https://arxiv.org/abs/2304.02660}{{\ttfamily 2304.02660}}].

\bibitem{dijkgraaf1990topological}
R.~Dijkgraaf and E.~Witten, \emph{Topological gauge theories and group cohomology}, {\emph{Communications in Mathematical Physics} {\bfseries 129} (1990) 393--429}.

\bibitem{Bhardwaj:2023bbf}
L.~Bhardwaj, L.~E. Bottini, D.~Pajer and S.~Schafer-Nameki, \emph{{The Club Sandwich: Gapless Phases and Phase Transitions with Non-Invertible Symmetries}},  \href{https://arxiv.org/abs/2312.17322}{{\ttfamily 2312.17322}}.

\bibitem{Bhardwaj:2024qrf}
L.~Bhardwaj, D.~Pajer, S.~Schafer-Nameki and A.~Warman, \emph{{Hasse Diagrams for Gapless SPT and SSB Phases with Non-Invertible Symmetries}},  \href{https://arxiv.org/abs/2403.00905}{{\ttfamily 2403.00905}}.

\bibitem{Robbins:2022wlr}
D.~G. Robbins, E.~Sharpe and T.~Vandermeulen, \emph{{Decomposition, trivially-acting symmetries, and topological operators}}, \href{https://doi.org/10.1103/PhysRevD.107.085017}{\emph{Phys. Rev. D} {\bfseries 107} (2023) 085017}, [\href{https://arxiv.org/abs/2211.14332}{{\ttfamily 2211.14332}}].

\bibitem{Hellerman:2006zs}
S.~Hellerman, A.~Henriques, T.~Pantev, E.~Sharpe and M.~Ando, \emph{{Cluster decomposition, T-duality, and gerby CFT's}}, \href{https://doi.org/10.4310/ATMP.2007.v11.n5.a2}{\emph{Adv. Theor. Math. Phys.} {\bfseries 11} (2007) 751--818}, [\href{https://arxiv.org/abs/hep-th/0606034}{{\ttfamily hep-th/0606034}}].

\bibitem{Sharpe:2014tca}
E.~Sharpe, \emph{{Decomposition in diverse dimensions}}, \href{https://doi.org/10.1103/PhysRevD.90.025030}{\emph{Phys. Rev. D} {\bfseries 90} (2014) 025030}, [\href{https://arxiv.org/abs/1404.3986}{{\ttfamily 1404.3986}}].

\bibitem{Sharpe:2019ddn}
E.~Sharpe, \emph{{Undoing decomposition}}, \href{https://doi.org/10.1142/S0217751X19502336}{\emph{Int. J. Mod. Phys. A} {\bfseries 34} (2020) 1950233}, [\href{https://arxiv.org/abs/1911.05080}{{\ttfamily 1911.05080}}].

\bibitem{Tanizaki:2019rbk}
Y.~Tanizaki and M.~\"Unsal, \emph{{Modified instanton sum in QCD and higher-groups}}, \href{https://doi.org/10.1007/JHEP03(2020)123}{\emph{JHEP} {\bfseries 03} (2020) 123}, [\href{https://arxiv.org/abs/1912.01033}{{\ttfamily 1912.01033}}].

\bibitem{Komargodski:2020mxz}
Z.~Komargodski, K.~Ohmori, K.~Roumpedakis and S.~Seifnashri, \emph{{Symmetries and strings of adjoint QCD$_{2}$}}, \href{https://doi.org/10.1007/JHEP03(2021)103}{\emph{JHEP} {\bfseries 03} (2021) 103}, [\href{https://arxiv.org/abs/2008.07567}{{\ttfamily 2008.07567}}].

\bibitem{Cherman:2020cvw}
A.~Cherman and T.~Jacobson, \emph{{Lifetimes of near eternal false vacua}}, \href{https://doi.org/10.1103/PhysRevD.103.105012}{\emph{Phys. Rev. D} {\bfseries 103} (2021) 105012}, [\href{https://arxiv.org/abs/2012.10555}{{\ttfamily 2012.10555}}].

\bibitem{Robbins:2020msp}
D.~Robbins, E.~Sharpe and T.~Vandermeulen, \emph{{A generalization of decomposition in orbifolds}}, \href{https://doi.org/10.1007/JHEP10(2021)134}{\emph{JHEP} {\bfseries 21} (2020) 134}, [\href{https://arxiv.org/abs/2101.11619}{{\ttfamily 2101.11619}}].

\bibitem{Nguyen:2021naa}
M.~Nguyen, Y.~Tanizaki and M.~\"Unsal, \emph{{Noninvertible 1-form symmetry and Casimir scaling in 2D Yang-Mills theory}}, \href{https://doi.org/10.1103/PhysRevD.104.065003}{\emph{Phys. Rev. D} {\bfseries 104} (2021) 065003}, [\href{https://arxiv.org/abs/2104.01824}{{\ttfamily 2104.01824}}].

\bibitem{Sharpe:2021srf}
E.~Sharpe, \emph{{Topological operators, noninvertible symmetries and decomposition}}, \href{https://doi.org/10.4310/ATMP.2023.v27.n8.a2}{\emph{Adv. Theor. Math. Phys.} {\bfseries 27} (2023) 2319--2407}, [\href{https://arxiv.org/abs/2108.13423}{{\ttfamily 2108.13423}}].

\bibitem{Cherman:2021nox}
A.~Cherman, T.~Jacobson and M.~Neuzil, \emph{{Universal Deformations}}, \href{https://doi.org/10.21468/SciPostPhys.12.4.116}{\emph{SciPost Phys.} {\bfseries 12} (2022) 116}, [\href{https://arxiv.org/abs/2111.00078}{{\ttfamily 2111.00078}}].

\bibitem{Sharpe:2022ene}
E.~Sharpe, \emph{{An introduction to decomposition}},  \href{https://arxiv.org/abs/2204.09117}{{\ttfamily 2204.09117}}.

\bibitem{Lin:2022xod}
L.~Lin, D.~G. Robbins and E.~Sharpe, \emph{{Decomposition, Condensation Defects, and Fusion}}, \href{https://doi.org/10.1002/prop.202200130}{\emph{Fortsch. Phys.} {\bfseries 70} (2022) 2200130}, [\href{https://arxiv.org/abs/2208.05982}{{\ttfamily 2208.05982}}].

\bibitem{Wang:2017loc}
J.~Wang, X.-G. Wen and E.~Witten, \emph{{Symmetric Gapped Interfaces of SPT and SET States: Systematic Constructions}}, \href{https://doi.org/10.1103/PhysRevX.8.031048}{\emph{Phys. Rev. X} {\bfseries 8} (2018) 031048}, [\href{https://arxiv.org/abs/1705.06728}{{\ttfamily 1705.06728}}].

\bibitem{Tachikawa:2017gyf}
Y.~Tachikawa, \emph{{On gauging finite subgroups}}, \href{https://doi.org/10.21468/SciPostPhys.8.1.015}{\emph{SciPost Phys.} {\bfseries 8} (2020) 015}, [\href{https://arxiv.org/abs/1712.09542}{{\ttfamily 1712.09542}}].

\bibitem{Pantev:2005rh}
T.~Pantev and E.~Sharpe, \emph{{Notes on gauging noneffective group actions}},  \href{https://arxiv.org/abs/hep-th/0502027}{{\ttfamily hep-th/0502027}}.

\bibitem{Pantev:2022kpl}
T.~Pantev, D.~G. Robbins, E.~Sharpe and T.~Vandermeulen, \emph{{Orbifolds by 2-groups and decomposition}}, \href{https://doi.org/10.1007/JHEP09(2022)036}{\emph{JHEP} {\bfseries 09} (2022) 036}, [\href{https://arxiv.org/abs/2204.13708}{{\ttfamily 2204.13708}}].

\bibitem{Pantev:2022pbf}
T.~Pantev and E.~Sharpe, \emph{{Decomposition in Chern-Simons theories in three dimensions}}, \href{https://doi.org/10.1142/S0217751X2250227X}{\emph{Int. J. Mod. Phys. A} {\bfseries 37} (2022) 2250227}, [\href{https://arxiv.org/abs/2206.14824}{{\ttfamily 2206.14824}}].

\bibitem{Robbins:2021lry}
D.~G. Robbins, E.~Sharpe and T.~Vandermeulen, \emph{{Anomalies, extensions, and orbifolds}}, \href{https://doi.org/10.1103/PhysRevD.104.085009}{\emph{Phys. Rev. D} {\bfseries 104} (2021) 085009}, [\href{https://arxiv.org/abs/2106.00693}{{\ttfamily 2106.00693}}].

\bibitem{Robbins:2021ibx}
D.~G. Robbins, E.~Sharpe and T.~Vandermeulen, \emph{{Quantum symmetries in orbifolds and decomposition}}, \href{https://doi.org/10.1007/JHEP02(2022)108}{\emph{JHEP} {\bfseries 02} (2022) 108}, [\href{https://arxiv.org/abs/2107.12386}{{\ttfamily 2107.12386}}].

\bibitem{Robbins:2021xce}
D.~G. Robbins, E.~Sharpe and T.~Vandermeulen, \emph{{Anomaly resolution via decomposition}}, \href{https://doi.org/10.1142/S0217751X21502201}{\emph{Int. J. Mod. Phys. A} {\bfseries 36} (2021) 2150220}, [\href{https://arxiv.org/abs/2107.13552}{{\ttfamily 2107.13552}}].

\bibitem{Perez-Lona:2025ncg}
A.~Perez-Lona, D.~Robbins, S.~Roy, E.~Sharpe, T.~Vandermeulen and X.~Yu, \emph{{Anomaly resolution by non-invertible symmetries}},  \href{https://arxiv.org/abs/2504.06333}{{\ttfamily 2504.06333}}.

\bibitem{Scaffidi_2017}
T.~Scaffidi, D.~E. Parker and R.~Vasseur, \emph{Gapless symmetry-protected topological order}, \href{https://doi.org/10.1103/physrevx.7.041048}{\emph{Physical Review X} {\bfseries 7} (Nov., 2017) }.

\bibitem{Thorngren:2020wet}
R.~Thorngren, A.~Vishwanath and R.~Verresen, \emph{{Intrinsically gapless topological phases}}, \href{https://doi.org/10.1103/PhysRevB.104.075132}{\emph{Phys. Rev. B} {\bfseries 104} (2021) 075132}, [\href{https://arxiv.org/abs/2008.06638}{{\ttfamily 2008.06638}}].

\bibitem{Wen:2022tkg}
R.~Wen and A.~C. Potter, \emph{{Bulk-boundary correspondence for intrinsically gapless symmetry-protected topological phases from group cohomology}}, \href{https://doi.org/10.1103/PhysRevB.107.245127}{\emph{Phys. Rev. B} {\bfseries 107} (2023) 245127}, [\href{https://arxiv.org/abs/2208.09001}{{\ttfamily 2208.09001}}].

\bibitem{Li:2023knf}
L.~Li, M.~Oshikawa and Y.~Zheng, \emph{{Intrinsically/purely gapless-SPT from non-invertible duality transformations}}, \href{https://doi.org/10.21468/SciPostPhys.18.5.153}{\emph{SciPost Phys.} {\bfseries 18} (2025) 153}, [\href{https://arxiv.org/abs/2307.04788}{{\ttfamily 2307.04788}}].

\bibitem{Antinucci:2025fjp}
A.~Antinucci, C.~Copetti, Y.~Gai and S.~Schafer-Nameki, \emph{{Categorical Anomaly Matching}},  \href{https://arxiv.org/abs/2508.00982}{{\ttfamily 2508.00982}}.

\bibitem{Beigi:2010htr}
S.~Beigi, P.~W. Shor and D.~Whalen, \emph{{The Quantum Double Model with Boundary: Condensations and Symmetries}}, \href{https://doi.org/10.1007/s00220-011-1294-x}{\emph{Commun. Math. Phys.} {\bfseries 306} (2011) 663--694}, [\href{https://arxiv.org/abs/1006.5479}{{\ttfamily 1006.5479}}].

\bibitem{beigi2012indistinguishablechargeonfluxionpairsquantum}
S.~Beigi, P.~W. Shor and D.~Whalen, \emph{Indistinguishable chargeon-fluxion pairs in the quantum double of finite groups},  2012.

\bibitem{Kitaev:2005hzj}
A.~Kitaev, \emph{{Anyons in an exactly solved model and beyond}}, \href{https://doi.org/10.1016/j.aop.2005.10.005}{\emph{Annals Phys.} {\bfseries 321} (2006) 2--111}, [\href{https://arxiv.org/abs/cond-mat/0506438}{{\ttfamily cond-mat/0506438}}].

\bibitem{Fuchs:2002cm}
J.~Fuchs, I.~Runkel and C.~Schweigert, \emph{{TFT construction of RCFT correlators 1. Partition functions}}, \href{https://doi.org/10.1016/S0550-3213(02)00744-7}{\emph{Nucl. Phys. B} {\bfseries 646} (2002) 353--497}, [\href{https://arxiv.org/abs/hep-th/0204148}{{\ttfamily hep-th/0204148}}].

\bibitem{kong2021anyoncondensationtensorcategories}
L.~Kong, \emph{Anyon condensation and tensor categories},  2021.
\newblock https://doi.org/10.1016/j.nuclphysb.2014.07.003; 10.1016/j.nuclphysb.2021.115607.

\bibitem{Wen:2023otf}
R.~Wen and A.~C. Potter, \emph{{Classification of 1+1D gapless symmetry protected phases via topological holography}}, \href{https://doi.org/10.1103/PhysRevB.111.115161}{\emph{Phys. Rev. B} {\bfseries 111} (2025) 115161}, [\href{https://arxiv.org/abs/2311.00050}{{\ttfamily 2311.00050}}].

\bibitem{Benini:2022hzx}
F.~Benini, C.~Copetti and L.~Di~Pietro, \emph{{Factorization and global symmetries in holography}}, \href{https://doi.org/10.21468/SciPostPhys.14.2.019}{\emph{SciPost Phys.} {\bfseries 14} (2023) 019}, [\href{https://arxiv.org/abs/2203.09537}{{\ttfamily 2203.09537}}].

\bibitem{Chatterjee:2022tyg}
A.~Chatterjee and X.-G. Wen, \emph{{Holographic theory for continuous phase transitions: Emergence and symmetry protection of gaplessness}}, \href{https://doi.org/10.1103/PhysRevB.108.075105}{\emph{Phys. Rev. B} {\bfseries 108} (2023) 075105}, [\href{https://arxiv.org/abs/2205.06244}{{\ttfamily 2205.06244}}].

\bibitem{Zhang:2023wlu}
C.~Zhang and C.~C\'ordova, \emph{{Anomalies of $(1+1)D$ categorical symmetries}},  \href{https://arxiv.org/abs/2304.01262}{{\ttfamily 2304.01262}}.

\bibitem{davydov2011wittgroupnondegeneratebraided}
A.~Davydov, M.~Mueger, D.~Nikshych and V.~Ostrik, \emph{The witt group of non-degenerate braided fusion categories},  2011.

\bibitem{kong2014anyon}
L.~Kong, \emph{Anyon condensation and tensor categories}, {\emph{Nuclear Physics B} {\bfseries 886} (2014) 436--482}.

\bibitem{Cheng:2020rpl}
M.~Cheng and D.~J. Williamson, \emph{{Relative anomaly in ( 1+1 )d rational conformal field theory}}, \href{https://doi.org/10.1103/PhysRevResearch.2.043044}{\emph{Phys. Rev. Res.} {\bfseries 2} (2020) 043044}, [\href{https://arxiv.org/abs/2002.02984}{{\ttfamily 2002.02984}}].

\bibitem{Cong:2017ffh}
I.~Cong, M.~Cheng and Z.~Wang, \emph{{Hamiltonian and Algebraic Theories of Gapped Boundaries in Topological Phases of Matter}}, \href{https://doi.org/10.1007/s00220-017-2960-4}{\emph{Commun. Math. Phys.} {\bfseries 355} (2017) 645--689}, [\href{https://arxiv.org/abs/1707.04564}{{\ttfamily 1707.04564}}].

\bibitem{He:2016xpi}
H.~He, Y.~Zheng and C.~von Keyserlingk, \emph{{Field theories for gauged symmetry-protected topological phases: Non-Abelian anyons with Abelian gauge group $\mathbb Z_2^3$}}, \href{https://doi.org/10.1103/PhysRevB.95.035131}{\emph{Phys. Rev. B} {\bfseries 95} (2017) 035131}, [\href{https://arxiv.org/abs/1608.05393}{{\ttfamily 1608.05393}}].

\bibitem{Yu:2023nyn}
X.~Yu, \emph{{Noninvertible symmetries in 2D from type IIB string theory}}, \href{https://doi.org/10.1103/PhysRevD.110.065008}{\emph{Phys. Rev. D} {\bfseries 110} (2024) 065008}, [\href{https://arxiv.org/abs/2310.15339}{{\ttfamily 2310.15339}}].

\bibitem{Franco:2024mxa}
S.~Franco and X.~Yu, \emph{{Generalized symmetries in 2D from string theory: SymTFTs, intrinsic relativeness, and anomalies of non-invertible symmetries}}, \href{https://doi.org/10.1007/JHEP11(2024)004}{\emph{JHEP} {\bfseries 11} (2024) 004}, [\href{https://arxiv.org/abs/2404.19761}{{\ttfamily 2404.19761}}].

\bibitem{deWildPropitius:1995cf}
M.~D.~F. de~Wild~Propitius, \emph{{Topological interactions in broken gauge theories}}, Ph.D. thesis, Amsterdam U., 1995.
\newblock \href{https://arxiv.org/abs/hep-th/9511195}{{\ttfamily hep-th/9511195}}.

\bibitem{Kapustin:2014lwa}
A.~Kapustin and R.~Thorngren, \emph{{Anomalies of discrete symmetries in three dimensions and group cohomology}}, \href{https://doi.org/10.1103/PhysRevLett.112.231602}{\emph{Phys. Rev. Lett.} {\bfseries 112} (2014) 231602}, [\href{https://arxiv.org/abs/1403.0617}{{\ttfamily 1403.0617}}].

\bibitem{Kapustin:2014zva}
A.~Kapustin and R.~Thorngren, \emph{{Anomalies of discrete symmetries in various dimensions and group cohomology}},  \href{https://arxiv.org/abs/1404.3230}{{\ttfamily 1404.3230}}.

\bibitem{Bhardwaj:2023idu}
L.~Bhardwaj, L.~E. Bottini, D.~Pajer and S.~Sch\"afer-Nameki, \emph{{Gapped Phases with Non-Invertible Symmetries: (1+1)d}}, \href{https://doi.org/10.21468/SciPostPhys.18.1.032}{\emph{SciPost Phys.} {\bfseries 18} (2025) 032}, [\href{https://arxiv.org/abs/2310.03784}{{\ttfamily 2310.03784}}].

\bibitem{Perez-Lona:2023djo}
A.~Perez-Lona, D.~Robbins, E.~Sharpe, T.~Vandermeulen and X.~Yu, \emph{{Notes on gauging noninvertible symmetries. Part I. Multiplicity-free cases}}, \href{https://doi.org/10.1007/JHEP02(2024)154}{\emph{JHEP} {\bfseries 02} (2024) 154}, [\href{https://arxiv.org/abs/2311.16230}{{\ttfamily 2311.16230}}].

\bibitem{Putrov:2024uor}
P.~Putrov and R.~Radhakrishnan, \emph{{Non-anomalous non-invertible symmetries in 1+1D from gapped boundaries of SymTFTs}},  \href{https://arxiv.org/abs/2405.04619}{{\ttfamily 2405.04619}}.

\bibitem{Diatlyk:2023fwf}
O.~Diatlyk, C.~Luo, Y.~Wang and Q.~Weller, \emph{{Gauging non-invertible symmetries: topological interfaces and generalized orbifold groupoid in 2d QFT}}, \href{https://doi.org/10.1007/JHEP03(2024)127}{\emph{JHEP} {\bfseries 03} (2024) 127}, [\href{https://arxiv.org/abs/2311.17044}{{\ttfamily 2311.17044}}].

\bibitem{Thorngren:2021yso}
R.~Thorngren and Y.~Wang, \emph{{Fusion category symmetry. Part II. Categoriosities at c = 1 and beyond}}, \href{https://doi.org/10.1007/JHEP07(2024)051}{\emph{JHEP} {\bfseries 07} (2024) 051}, [\href{https://arxiv.org/abs/2106.12577}{{\ttfamily 2106.12577}}].

\bibitem{Bartsch:2022ytj}
T.~Bartsch, M.~Bullimore, A.~E.~V. Ferrari and J.~Pearson, \emph{{Non-invertible symmetries and higher representation theory II}}, \href{https://doi.org/10.21468/SciPostPhys.17.2.067}{\emph{SciPost Phys.} {\bfseries 17} (2024) 067}, [\href{https://arxiv.org/abs/2212.07393}{{\ttfamily 2212.07393}}].

\bibitem{Bhardwaj:2022maz}
L.~Bhardwaj, L.~E. Bottini, S.~Schafer-Nameki and A.~Tiwari, \emph{{Non-invertible symmetry webs}}, \href{https://doi.org/10.21468/SciPostPhys.15.4.160}{\emph{SciPost Phys.} {\bfseries 15} (2023) 160}, [\href{https://arxiv.org/abs/2212.06842}{{\ttfamily 2212.06842}}].

\bibitem{Warman:2024lir}
A.~Warman, F.~Yang, A.~Tiwari, H.~Pichler and S.~Schafer-Nameki, \emph{{Categorical Symmetries in Spin Models with Atom Arrays}},  \href{https://arxiv.org/abs/2412.15024}{{\ttfamily 2412.15024}}.

\bibitem{Schafer-Nameki:2025fiy}
S.~Schafer-Nameki, A.~Tiwari, A.~Warman and C.~Zhang, \emph{{SymTFT Approach for Mixed States with Non-Invertible Symmetries}},  \href{https://arxiv.org/abs/2507.05350}{{\ttfamily 2507.05350}}.

\bibitem{Antinucci:2024ltv}
A.~Antinucci, C.~Copetti and S.~Schafer-Nameki, \emph{{SymTFT for (3+1)d Gapless SPTs and Obstructions to Confinement}}, \href{https://doi.org/10.21468/SciPostPhys.18.3.114}{\emph{SciPost Phys.} {\bfseries 18} (2025) 114}, [\href{https://arxiv.org/abs/2408.05585}{{\ttfamily 2408.05585}}].

\bibitem{Thorngren:2019iar}
R.~Thorngren and Y.~Wang, \emph{{Fusion category symmetry. Part I. Anomaly in-flow and gapped phases}}, \href{https://doi.org/10.1007/JHEP04(2024)132}{\emph{JHEP} {\bfseries 04} (2024) 132}, [\href{https://arxiv.org/abs/1912.02817}{{\ttfamily 1912.02817}}].

\bibitem{Choi:2023xjw}
Y.~Choi, B.~C. Rayhaun, Y.~Sanghavi and S.-H. Shao, \emph{{Remarks on boundaries, anomalies, and noninvertible symmetries}}, \href{https://doi.org/10.1103/PhysRevD.108.125005}{\emph{Phys. Rev. D} {\bfseries 108} (2023) 125005}, [\href{https://arxiv.org/abs/2305.09713}{{\ttfamily 2305.09713}}].

\bibitem{Bergman:2024its}
O.~Bergman and F.~Mignosa, \emph{{String theory and the SymTFT of 3d orthosymplectic Chern-Simons theory}}, \href{https://doi.org/10.1007/JHEP04(2025)047}{\emph{JHEP} {\bfseries 04} (2025) 047}, [\href{https://arxiv.org/abs/2412.00184}{{\ttfamily 2412.00184}}].

\bibitem{TAMBARA1998692}
D.~Tambara and S.~Yamagami, \emph{Tensor categories with fusion rules of self-duality for finite abelian groups}, \href{https://doi.org/https://doi.org/10.1006/jabr.1998.7558}{\emph{Journal of Algebra} {\bfseries 209} (1998) 692--707}.

\bibitem{Choi:2023vgk}
Y.~Choi, D.-C. Lu and Z.~Sun, \emph{{Self-duality under gauging a non-invertible symmetry}}, \href{https://doi.org/10.1007/JHEP01(2024)142}{\emph{JHEP} {\bfseries 01} (2024) 142}, [\href{https://arxiv.org/abs/2310.19867}{{\ttfamily 2310.19867}}].

\bibitem{Iqbal:2023wvm}
M.~Iqbal et~al., \emph{{Non-Abelian topological order and anyons on a trapped-ion processor}}, \href{https://doi.org/10.1038/s41586-023-06934-4}{\emph{Nature} {\bfseries 626} (2024) 505--511}, [\href{https://arxiv.org/abs/2305.03766}{{\ttfamily 2305.03766}}].

\end{thebibliography}
\end{document}